\documentclass[ALICE,manyauthors]{cernphprep}
\usepackage[comma,square,numbers,sort&compress]{natbib}
\usepackage[T1]{fontenc}
\usepackage{hyperref}
\usepackage{lineno}
\usepackage{xspace}
\usepackage{xcolor} 
\usepackage{csquotes}
\usepackage{multirow}
\usepackage{upgreek}
\usepackage{booktabs}
\usepackage{amsmath}
\usepackage{orcidlink}

\begin{document}
%

\newcommand{\pp}           {pp\xspace}
\newcommand{\ppbar}        {\mbox{$\mathrm {p\overline{p}}$}\xspace}
\newcommand{\XeXe}         {\mbox{Xe--Xe}\xspace}
\newcommand{\PbPb}         {\mbox{Pb--Pb}\xspace}
\newcommand{\pA}           {\mbox{pA}\xspace}
\newcommand{\pPb}          {\mbox{p--Pb}\xspace}
\newcommand{\AuAu}         {\mbox{Au--Au}\xspace}
\newcommand{\dAu}          {\mbox{d--Au}\xspace}

\newcommand{\s}            {\ensuremath{\sqrt{s}}\xspace}
\newcommand{\snn}          {\ensuremath{\sqrt{s_{\mathrm{NN}}}}\xspace}
\newcommand{\pt}           {\ensuremath{p_{\rm T}}\xspace}
\newcommand{\meanpt}       {$\langle p_{\mathrm{T}}\rangle$\xspace}
\newcommand{\ycms}         {\ensuremath{y_{\rm CMS}}\xspace}
\newcommand{\ylab}         {\ensuremath{y_{\rm lab}}\xspace}
\newcommand{\etarange}[1]  {\mbox{$\left | \eta \right |~<~#1$}}
\newcommand{\yrange}[1]    {\mbox{$\left | y \right |~<~#1$}}
\newcommand{\dndy}         {\ensuremath{\mathrm{d}N_\mathrm{ch}/\mathrm{d}y}\xspace}
\newcommand{\dndeta}       {\ensuremath{\mathrm{d}N_\mathrm{ch}/\mathrm{d}\eta}\xspace}
\newcommand{\avdndeta}     {\ensuremath{\langle\dndeta\rangle}\xspace}
\newcommand{\dNdy}         {\ensuremath{\mathrm{d}N_\mathrm{ch}/\mathrm{d}y}\xspace}
\newcommand{\Npart}        {\ensuremath{N_\mathrm{part}}\xspace}
\newcommand{\Ncoll}        {\ensuremath{N_\mathrm{coll}}\xspace}
\newcommand{\dEdx}         {\ensuremath{\textrm{d}E/\textrm{d}x}\xspace}
\newcommand{\RpPb}         {\ensuremath{R_{\rm pPb}}\xspace}

\newcommand{\nineH}        {$\sqrt{s}~=~0.9$~Te\kern-.1emV\xspace}
\newcommand{\seven}        {$\sqrt{s}~=~7$~Te\kern-.1emV\xspace}
\newcommand{\twoH}         {$\sqrt{s}~=~0.2$~Te\kern-.1emV\xspace}
\newcommand{\twosevensix}  {$\sqrt{s}~=~2.76$~Te\kern-.1emV\xspace}
\newcommand{\five}         {$\sqrt{s}~=~5.02$~Te\kern-.1emV\xspace}
\newcommand{\twosevensixnn}{$\sqrt{s_{\mathrm{NN}}}~=~2.76$~Te\kern-.1emV\xspace}
\newcommand{\fivenn}       {$\sqrt{s_{\mathrm{NN}}}~=~5.02$~Te\kern-.1emV\xspace}
\newcommand{\LT}           {L{\'e}vy-Tsallis\xspace}
\newcommand{\GeVc}         {Ge\kern-.1emV/$c$\xspace}
\newcommand{\MeVc}         {Me\kern-.1emV/$c$\xspace}
\newcommand{\TeV}          {Te\kern-.1emV\xspace}
\newcommand{\GeV}          {Ge\kern-.1emV\xspace}
\newcommand{\MeV}          {Me\kern-.1emV\xspace}
\newcommand{\GeVmass}      {Ge\kern-.2emV/$c^2$\xspace}
\newcommand{\MeVmass}      {Me\kern-.2emV/$c^2$\xspace}
\newcommand{\lumi}         {\ensuremath{\mathcal{L}}\xspace}

\newcommand{\ITS}          {\rm{ITS}\xspace}
\newcommand{\TOF}          {\rm{TOF}\xspace}
\newcommand{\ZDC}          {\rm{ZDC}\xspace}
\newcommand{\ZDCs}         {\rm{ZDCs}\xspace}
\newcommand{\ZNA}          {\rm{ZNA}\xspace}
\newcommand{\ZNC}          {\rm{ZNC}\xspace}
\newcommand{\SPD}          {\rm{SPD}\xspace}
\newcommand{\SDD}          {\rm{SDD}\xspace}
\newcommand{\SSD}          {\rm{SSD}\xspace}
\newcommand{\TPC}          {\rm{TPC}\xspace}
\newcommand{\TRD}          {\rm{TRD}\xspace}
\newcommand{\VZERO}        {\rm{V0}\xspace}
\newcommand{\VZEROA}       {\rm{V0A}\xspace}
\newcommand{\VZEROC}       {\rm{V0C}\xspace}
\newcommand{\Vdecay} 	   {\ensuremath{V^{0}}\xspace}

\newcommand{\ee}           {\rm \ensuremath{e^{+}e^{-}}} 
\newcommand{\pip}          {\ensuremath{\pi^{+}}\xspace}
\newcommand{\pim}          {\ensuremath{\pi^{-}}\xspace}
\newcommand{\kap}          {\ensuremath{\rm{K}^{+}}\xspace}
\newcommand{\kam}          {\ensuremath{\rm{K}^{-}}\xspace}
\newcommand{\pbar}         {\ensuremath{\rm\overline{p}}\xspace}
\newcommand{\kzero}        {\ensuremath{{\rm K}^{0}_{\rm{S}}}\xspace}
\newcommand{\lmb}          {\ensuremath{\Lambda}\xspace}
\newcommand{\almb}         {\ensuremath{\overline{\Lambda}}\xspace}
\newcommand{\Om}           {\ensuremath{\Omega^-}\xspace}
\newcommand{\Mo}           {\ensuremath{\overline{\Omega}^+}\xspace}
\newcommand{\X}            {\ensuremath{\Xi^-}\xspace}
\newcommand{\Ix}           {\ensuremath{\overline{\Xi}^+}\xspace}
\newcommand{\Xis}          {\ensuremath{\Xi^{\pm}}\xspace}
\newcommand{\Oms}          {\ensuremath{\Omega^{\pm}}\xspace}
\newcommand{\degree}       {\ensuremath{^{\rm o}}\xspace}

\newcommand{\kT}{k_{\rm T}}
\newcommand{\OmegacZero}{\rm \Omega_{c}^{0}}
\newcommand{\XicZero}{\rm \Xi_{c}^{0}}
\newcommand{\XicPlus}{\rm \Xi_{c}^{+}}
\newcommand{\XicZeroPlus}{\rm \Xi_{c}^{0,+}}
\newcommand{\Jpsi}{{\rm J}/\psi}
\newcommand{\acceff}{\rm Acc.\times\varepsilon}
\newcommand{\Dzero}{\rm D^{0}}
\newcommand{\Dplus}{\rm D^+}
\newcommand{\Dstar}{\rm D^{\ast +}}
\newcommand{\Ds}{\rm D_s^+}
\newcommand{\DtoKpi}{\rm D^0\to K^-\pi^+}
\newcommand{\DplusToKpipi}{\Dplus \to \rm K^{-} \pi^{+} \pi^{+}}
\newcommand{\DstarToDzeropi}{\Dstar \to \Dzero \pi^{+} \to K^- \pi^+\pi^{+}}
\newcommand{\DsToPhipiToKKpi}{\Ds \to \phi \pi^{+} \to K^{+}K^{-}\pi^{+}}
\newcommand{\Sigmac}{\rm \Sigma_{c}}
\newcommand{\SigmacZeroPlusPlus}{{\rm \Sigma_{c}^{0,++}}}
\newcommand{\SigmacZero}{{\rm \Sigma_{c}^{0}}}
\newcommand{\SigmacPlusPlus}{{\rm \Sigma_{c}^{++}}}
\newcommand{\SigmacZeroPlusPlusPlus}{\rm \Sigma_{c}^{0,+,++}}
\newcommand{\Lambdac}{\rm \Lambda_{c}^{+}}
\newcommand{\LambdacTopKpi}{\rm \Lambda_{c}^+\rightarrow pK^{-}\pi^{+}}
\newcommand{\LambdacTopKzeroS}{\rm \Lambda_{c}^{+}\rightarrow pK^{0}_{S}}
\newcommand{\pKzeroS}{\rm pK^{0}_{S}}
\newcommand{\Lambdab}{\rm \Lambda_{b}^{0}}
\newcommand{\ep}{\rm ep}
\newcommand{\pKpi}{\rm pK^{-}\pi^{+}}
\newcommand{\pKzeros}{\rm pK^{0}_{S}}
\newcommand{\ccbar}{\rm c\overline{c}}
\newcommand{\bbbar}{\rm b\bar{b}}
\newcommand{\charm}{{\rm c}}
\newcommand{\noop}[1]{}
\newcommand{\hc}{\rm h_{c}\xspace}

\renewcommand{\pi}{\uppi}
\renewcommand{\phi}{\upphi}

\begin{titlepage}
\PHyear{2023}       
\PHnumber{162}      
\PHdate{05 August}  

\title{Charm production and fragmentation fractions at midrapidity in \pp collisions at \pmb{$\s=13$} TeV}
\ShortTitle{Charm production and fragmentation fractions in \pp at $\s=13$ TeV}   

\Collaboration{ALICE Collaboration\thanks{See Appendix~\ref{app:collab} for the list of collaboration members}}
\ShortAuthor{ALICE Collaboration} 

\begin{abstract}
Measurements of the production cross sections of prompt $\Dzero$, $\Dplus$, $\Dstar$, $\Ds$, $\Lambdac$, and $\XicPlus$ charm hadrons at  midrapidity in proton$-$proton collisions at $\s=13$ TeV with the ALICE detector are presented. The D-meson cross sections as a function of transverse momentum ($\pt$) are provided with improved precision and granularity. 
The ratios of $\pt$-differential meson production cross sections based on this publication and on measurements at different rapidity and collision energy provide a constraint on gluon parton distribution functions at low values of Bjorken-$x$ ($10^{-5}-10^{-4}$).
The measurements of $\Lambdac$ ($\XicPlus$) baryon production extend the measured $\pt$ intervals down to $\pt=0(3)$~GeV$/c$.
These measurements are used to determine the charm-quark fragmentation fractions and the $\ccbar$ production cross section at midrapidity ($|y|<0.5$) based on the sum of the cross sections of the weakly-decaying ground-state charm hadrons $\Dzero$, $\Dplus$, $\Ds$, $\Lambdac$, $\XicZero$ and, for the first time, $\XicPlus$, and of the strongly-decaying $\Jpsi$ mesons.
The first measurements of $\XicPlus$ and $\SigmacZeroPlusPlus$ fragmentation fractions at midrapidity are also reported.
A significantly larger fraction of charm quarks hadronising to baryons is found compared to $\ee$ and $\ep$ collisions. The $\ccbar$ production cross section at midrapidity is found to be at the upper bound of state-of-the-art perturbative QCD calculations.

\end{abstract}
\end{titlepage}

\setcounter{page}{2} 


\section{Introduction}
\label{sec:intro}

Measurements of heavy-flavour (i.e.~charm and beauty) hadron production in ultra-relativistic proton--proton (\pp) collisions provide fundamental tests of perturbative quantum chromodynamics (pQCD) calculations. The transverse momentum ($\pt$) differential production cross section of heavy-flavour hadrons is usually calculated in pQCD by the convolution of three ingredients employing a factorisation approach~\cite{Collins:1989gx}. The first ingredient corresponds to the parton distribution functions (PDFs), which describe the probability distributions of the parton momentum fractions in the proton. The second term is the partonic cross section, which defines the scattering probability calculated as a perturbative series expansion in the strong coupling constant ($\alpha_{\rm s}$). The third ingredient corresponds to the fragmentation function (FF) which parametrises the non-perturbative transition of a heavy quark into a hadron.

Factorisation can be implemented in pQCD-based calculations in different ways, for example
in terms of the transferred momentum squared $Q ^2$ (collinear factorisation)~\cite{Collins:1989gx} or of the parton transverse momentum $\kT$~\cite{CATANI1991135, Luszczak:2008je, Maciula:2013wg, PhysRevD.100.054001, Shabelski:2017kmy}. Calculations for LHC energies implementing the former approach, like the general-mass variable-flavour-number scheme (GM-VFNS)~\cite{Benzke:2017yjn, Kramer:2017gct, Kniehl:2004fy, Kniehl:2005mk, Kniehl:2012ti, Helenius:2018uul}
and the fixed order plus next-to-leading logarithms (FONLL) approach~\cite{Cacciari:1998it, Cacciari:2012ny} provide a next-to-leading order (NLO) accuracy with all-order resummation of next-to-leading logarithms. Calculations of heavy-flavour hadron production in the $\kT$ framework are also able to go beyond leading-order expansions in $\alpha_{\rm s}$~\cite{Shabelski:2017kmy, PhysRevD.100.054001}. The most recent implementation of $\kT$-factorisation has also employed the variable-flavour-number scheme approach~\cite{Guiot:2021vnp}. D-meson production down to low $\pt$ is calculable 
at $\pt$ scales far below the charm mass with transverse-momentum dependent factorisation approach (TMD~\cite{Boer:2012bt, Ma:2012hh}), and in the low-$x$ regime with the colour-glass condensate model (CGC~\cite{Ma:2014mri}).

All of these models describe the production of heavy-flavour  mesons within uncertainties, in different kinematic regions and at different energies in \pp and \ppbar collisions~\cite{ALICE:2019nxm, ALICE:2012inj, ALICE:2012gkr, ALICE:2017olh, ALICE:2021mgk, ATLAS:2015igt, CMS:2017qjw, LHCb:2013xam, LHCb:2015swx, LHCb:2016ikn, CMS:2021lab}.
In this article, the measured $\pt$-differential production cross sections of prompt $\Dzero$, $\Dplus$, $\Ds$, and $\Dstar$ mesons at midrapidity in \pp collisions at $\s=13$ TeV are compared with the predictions from these models, which are significantly less precise than the available experimental measurements. The main source of theoretical uncertainty is related to the choice of the energy scales for the validity of the perturbative regime (factorisation and renormalisation scales). However, as explained in Ref.~\cite{Cacciari:2015fta}, these uncertainties may become subdominant in the calculation of cross section ratios at different rapidities and energies. For this reason, precise measurements of charm meson-to-meson production ratios originating from charm-quark hadronisation (i.e.~prompt) down to low $\pt$ (i.e. $\pt \lesssim 5$ GeV$/c$) are sensitive to gluon PDFs at low Bjorken-$x$.

Calculations based on a collinear factorisation approach, and using fragmentation functions 
tuned on $\ee$ collision data, do not describe the production of charm baryons at midrapidity 
in \pp collisions at the LHC~\cite{ALICE:2022wpn}. The $\Lambdac$-baryon production cross section at low $\pt$ and midrapidity ($|y|<0.5$) in \pp collisions at $\s=5.02$ TeV~\cite{ALICE:2020wfu,ALICE:2020wla,CMS:2019uws, ALICE:2022ych} is underestimated by a factor 3$-$4 by GM-VFNS calculations adopting $\Lambdac$-baryon fragmentation functions derived from the fit of OPAL data~\cite{de35328685ef47f1807976f1f15f14be}, and by a factor 15 by the POWHEG predictions~\cite{Frixione_2007} matched with PYTHIA 6~\cite{Sjostrand:2006za} to generate the parton
shower.
The prediction from the PYTHIA 8 Monte Carlo (MC) generator~\cite{SJOSTRAND2015159} with the standard Monash tune~\cite{Skands:2014pea}, where the charm-quark fragmentation is constrained with $\ee$ and $\ep$ measurements, 
underestimates the measurement by a factor 2$-$10 depending on $\pt$ in the region $\pt<12$ GeV$/c$.
The baryon production observed in pp collisions at the LHC challenges the concept of \enquote{jet universality}, according to which the parton fragmentation is universal between collision systems and can be constrained from $\ee$ results. This implies the breakdown of multi-parton interaction (MPI)-based event generators implementing jet universality~\cite{SJOSTRAND2015159, Bellm:2015jjp, Sherpa:2019gpd}. 

The measured baryon production in pp collision can only be described by model calculations that account for novel hadronisation mechanisms.
The data description in PYTHIA improves when colour reconnection mechanisms beyond the leading-colour approximation (CR-BLC) are adopted in PYTHIA simulations~\cite{Christiansen:2015yqa}. In these cases, the predicted $\Lambdac$-baryon production is enhanced by the presence of new topologies, called \enquote{junctions}, which preferentially fragment into baryons.
Several other model calculations foresee a relative increase of charm-baryon production with respect to that of mesons in \pp collisions at the LHC. In the statistical hadronisation model with relativistic quark model (SHM+RQM)~\cite{HE2019117} the presence of a large set of mostly unobserved excited charm baryons, which decay strongly and enrich the abundance of ground-state charm baryons, is foreseen. These excited charm baryons are predicted by the RQM model~\cite{Ebert:2011kk}. Their abundances are assumed to follow the thermal densities derived from the SHM~\cite{BraunMunzinger:2003zd} depending only on their mass and spin degeneracy.
Different assumptions are made in the Catania model~\cite{Minissale:2020bif}, where the charm quark can hadronise either via fragmentation, 
or by recombining with surrounding light quarks already produced from the event underlying the hard scattering.
In this model, the recombination mechanisms enhances the production of baryons at intermediate $\pt$. 
In the quark (re)combination model (QCM)~\cite{Song:2018tpv} the charm quarks produced in the hard scatterings hadronise by recombining with surrounding equal-velocity light quarks. In this model, thermal weights regulate the relative abundances of different charm baryons.

Further information can be derived from the measurement of the relative production rates of different charm hadrons as a function of $\pt$. The measured $\Lambdac/\Dzero$ ratios at midrapidity in \pp collisions at $\s=5.02$~TeV and 13~TeV are described by the predictions provided by these model calculations. The measurements at the two collision energies are compatible within uncertainty, and they show an enhancement of about a factor 5 at low $\pt$ with respect to $\ee$ and $\ep$ collisions. The $\SigmacZeroPlusPlusPlus/\Dzero$ ratio at midrapidity in \pp collisions at $\s~=~13$~TeV~\cite{Acharya:2021vpo}
is described by the predictions of PYTHIA CR-BLC, SHM+RQM, and QCM models. This ratio is severely underestimated by the PYTHIA Monash tune simulations and is larger by around a factor 10 at $\pt < 6$ GeV$/c$ compared to that observed in $\ee$ collisions~\cite{Niiyama:2017wpp}. Given the available precision, the current experimental results are not sufficient to fully distinguish between the pictures described above. 
Finally, further measurements of $\XicZeroPlus$  and $\OmegacZero$ at midrapidity demonstrate that even the hadronisation mechanisms discussed above do not provide a complete picture, since significant tensions are observed when describing the relative abundance of charm baryons containing a strange valence quark~\cite{Acharya:2021dsq, Acharya:2021vjp, ALICE:2022cop}. In this article, the production of $\Lambdac$ and $\XicPlus$ baryons is measured down to lower transverse momenta, extending the comparison with the theoretical predictions in a kinematic region where the models calculations differ from each other.

The enhancement of the relative abundance of baryons compared to that of mesons at midrapidity in \pp collisions at the LHC has a strong impact on the charm-quark fragmentation fractions $f(\charm\to \rm{h_\charm})$. These describe the probability of a charm quark to produce a hadron of species $\rm{h_\charm}$. The measurement of $f(\charm\to \rm{h_\charm})$ for a given charm hadron implies integrating the $\pt$-differential production cross section and dividing it by the sum of the $\pt$-integrated production cross sections of all charm hadrons. It is crucial to extend the measurements of the $\pt$-differential production cross section down to $\pt = 0$~GeV$/c$, in order to minimise the uncertainties arising from extrapolations. The most recent measurement from the ALICE Collaboration in \pp collisions at $\s=5.02$ TeV~\cite{ALICE:2021dhb} shows a drop by about a factor 1.2$-$1.4 of the $\Dzero$-meson fragmentation fraction with respect to that observed in $\ee$ collisions at the B-factories and at LEP, as well as in deep-inelastic scattering measurements in $\ep$ collisions at HERA~\cite{Lisovyi:2015uqa}.
An increase by about a factor 3.3 is observed for the $\Lambdac$-baryon fragmentation fraction, while the relative fraction of the $\XicZero$ baryon in \pp collisions was measured to be similar to that of the $\Ds$ meson. In this article, the analogous result in \pp collisions at $\s=13$ TeV is proposed.

The fragmentation fractions also impact the determination of the $\ccbar$ production cross section. The first measurements of the $\ccbar$ cross sections at $\s=2.76$ TeV~\cite{ALICE:2012inj} and 7 TeV~\cite{ALICE:2017olh} at midrapidity ($|y|~<~0.5$) were derived only from D-meson cross section measurements using the fragmentation fractions from $\ee$ collisions. They were thus affected by an underestimation of the charm-baryon contribution, and they increased by about 40\% when applying updated D-meson fragmentation fractions taking the measured charm-baryon yields into account~\cite{ALICE:2021dhb}. Such precise measurements of the $\ccbar$ cross section are not only important tests of pQCD calculations, but they also provide a reference to study the charm dynamics in the quark$-$gluon plasma (QGP) produced in ultra-relativistic heavy-ion collisions. For example, the knowledge of the $\ccbar$ production cross section per nucleon$-$nucleon collision is a key ingredient to determine the production of charmonia and the influence of recombination effects in the QGP~\cite{BRAUNMUNZINGER2000196, ZHAO2011114, LIU200972}. Finally, precise results of $\pt$-differential production cross sections of different charm-hadron species are crucial as a reference for measurements of the nuclear modification factor ($R_{\rm AA}$)~\cite{ALICE:2021rxa}.

In this article, the $\pt$-differential production cross sections of prompt $\Dzero$, $\Dplus$, $\Ds$, and $\Dstar$ mesons at midrapidity in \pp collisions at $\s=13$ TeV are reported.
The production cross sections of prompt $\Dzero$  and $\Ds$ mesons in this article supersede those published in Refs.~\cite{Acharya:2021vpo, ALICE:2021npz}, and provide an extended $\pt$ coverage and a finer granularity in $\pt$ than the previous results.
In addition, the first results of the $\Lambdac$-baryon reconstruction in both $\pKpi$ and $\pKzeros$ decay channels down to $\pt=0$ at midrapidity at the LHC, as well as that of the $\XicPlus$ baryon down to $\pt=3$~GeV$/c$, are presented. These results extend the already published ones in Refs.~\cite{Acharya:2021vpo} and~\cite{Acharya:2021vjp}, respectively. This article then presents the first measurements of the $\ccbar$ production cross section per unit of rapidity at midrapidity and the charm-quark fragmentation fractions at midrapidity in \pp collisions at $\s=13$ TeV. Using similar methods as in Ref.~\cite{ALICE:2021dhb}, these measurements consider the sum of ground-state charm hadrons $\Dzero$, $\Dplus$, $\Ds$, $\Lambdac$, $\XicZero$ and, for the first time, $\XicPlus$, and of the strongly-decaying $\Jpsi$ mesons. Furthermore, the first measurement of the $\SigmacZeroPlusPlusPlus$-baryon fragmentation fraction is presented. The paper is organised as follows. The ALICE detector and the properties of the data samples used in these analyses are described in Section~\ref{sec:apparatus}. The details of the data analysis are provided in Section~\ref{sec:analysis} and the estimation of the systematic uncertainties is described in Section~\ref{sec:systematics}. The results of the analyses and their comparison with model calculations are discussed in Section~\ref{sec:results}. Finally, a brief summary is given in Section~\ref{sec:summary}.

\section{Experimental apparatus and data sample}
\label{sec:apparatus}
The ALICE apparatus and its performance are comprehensively described in Refs.~\cite{ALICE:2008ngc,ALICE:2014sbx}. 
The charm-hadron decays were reconstructed with the central barrel detectors, which cover the pseudorapidity interval $|\eta| < 0.9$ and are embedded in a cylindrical solenoid providing a magnetic field $ B \rm = 0.5$ T along the beam direction.
The trajectories of charged particles are reconstructed with the Inner Tracking System (\ITS) and the Time Projection Chamber (\TPC). The \ITS detector is the innermost ALICE subsystem. It is composed of six cylindrical layers of silicon detectors for precise measurements of track parameters in the vicinity of the interaction point (primary vertex). The \ITS detector provides a  precise determination of the track impact parameter (i.e.~the distance of closest approach of the track to the interaction point). For tracks with $\pt > 1$~GeV$/c$, a resolution better than 75 $\upmu$m is achieved in the plane orthogonal to the beam direction (transverse plane)~\cite{ALICE:2010tia}. Therefore, this detector is crucial to reconstruct the decay vertex of heavy-flavour hadrons (secondary vertex) and to distinguish it from the beam interaction point.
The \TPC provides track reconstruction with up
to 159 three-dimensional space points per track and charged-particle identification (PID) via the measurement of the specific ionisation energy loss ($\dEdx$).
The charm-hadron decay products are also identified with the Time-Of-Flight (\TOF) detector, which measures the flight time of the charged particles. For triggering and event selection, the \VZERO detector is used. It is composed of two scintillator arrays located on both sides of the nominal collision point covering the pseudorapidity intervals $-3.7 < \eta <-1.7$ and $2.8 < \eta < 5.1$.

The \pp collisions considered in the analyses presented in this article were collected at a centre-of-mass energy $\s=13$~TeV in 2016, 2017, and 2018 at the LHC using a minimum bias (MB) trigger, requiring coincident signals in the \VZERO scintillators on both sides. Background events coming from possible interactions between protons in the beam and residual gas inside the beam pipe were rejected offline exploiting the timing information of the \VZERO arrays and the correlation between the number of hits and tracks reconstructed in the two innermost layers of the ITS consisting of silicon pixel detectors (\SPD). Furthermore, all events with more than one reconstructed primary vertex were rejected in the analyses in order to exclude pile-up events within the same bunch crossing~\cite{ALICE:2014sbx}. Finally, only events with a position along the beam direction within $\pm 10$~cm from the centre of the apparatus were considered in this analysis to grant a uniform pseudorapidity acceptance. The number of analysed MB-triggered events corresponds to an integrated luminosity of $\mathcal{L}_{\rm int}= 31.9 \pm 0.5 \text{ nb}^{-1}$~\cite{ALICE-PUBLIC-2021-005}.

Monte Carlo samples of \pp collisions at the same centre-of-mass energy were used for efficiency and acceptance corrections as well as to train the machine learning algorithm to classify the signal and background candidates. This will be discussed in more detail in Section~\ref{sec:analysis}. 
These MC samples were generated by simulating \pp collisions with the Monash tune of the PYTHIA 8.243 event generator~\cite{Christiansen:2015yqa} requiring for each of them the production of at least a c$\overline{\rm c}$ or b$\overline{\rm b}$ pair. The produced charm hadrons  were forced to decay in the channels of interest for the analyses discussed in this paper. The produced particles were propagated through the detector using the GEANT3 transport code~\cite{Brun:1082634}. In the simulations, the conditions during the data taking were reproduced.

\section{Data analysis}
\label{sec:analysis}

\begin{table}[t]
    \centering
    \caption{Reconstructed decay channels for the measurement of $\Dzero$-, $\Dplus$-, $\Ds$-, $\Dstar$-, $\Lambdac$-, and $\XicPlus$-hadron signals. The branching ratios (BR) are taken from Ref.~\cite{10.1093/ptep/ptac097}, with the exception of $\rm BR(\XicPlus\to\X\pip\pip)$ which is taken from Ref.~\cite{Belle:2019bgi}.}
    \begin{tabular}{|c|c|c|c|}
         \hline
         Decay channel & Charm-hadron BR (\%) & Daughter decay & $\rm BR_{daughter}$ (\%) \\
         \hline
         $\Dzero\to \kam\pip$ & $\rm 3.95\pm0.03$ & - & -\\
         $\Dplus\to\kam\pip\pip$ & $\rm 9.38\pm0.16$ & - & - \\
         $\Ds\to\phi\pip\to\kap\kam\pip$ & $\rm 2.22\pm0.06$ & - & - \\
         $\Dstar\to\Dzero\pip$ & $\rm 67.7\pm0.5$ & $\Dzero\to \kam\pip$ & $\rm 3.95\pm0.03$ \\
         $\Lambdac\to\pKpi$ & $\rm 6.28\pm0.32$ & -  & -\\
         $\Lambdac\to\pKzeros$ & $\rm 1.59\pm0.08$ & $\kzero\to\pip\pim$ & $69.20\pm0.05$ \\
         \multirow{2}{*}{$\XicPlus\to\X\pip\pip$} & \multirow{2}{*}{$\rm 2.86\pm1.21 (stat.)\pm0.38(syst.)$~\cite{Belle:2019bgi}} & $\X\to\Lambda\pim$ & $99.887\pm 0.035$ \\
          & & $\Lambda\to \rm p\pim$ & $63.9 \pm 0.5$ \\
         \hline
    \end{tabular}
    \label{tab:BR}
\end{table}

%
%

The charm hadrons and their charge conjugates were reconstructed via the decay channels reported in Table~\ref{tab:BR}.
%
%
The $\Dzero$, $\Dplus$, $\Ds$, and $\Lambdac\to\pKpi$ signals were measured by combining pairs or triplets of tracks with $|\eta|<0.8$ and $\pt>0.3$~GeV$/c$. Only tracks crossing at least 70 pad rows in the TPC and reconstructed with at least one hit in the SPD detector and further selected with the track-quality criteria described in Ref.~\cite{ALICE:2019nxm} were considered. The $\Dstar$-meson candidates were formed by combining pion tracks selected with the criteria described in Ref.~\cite{ALICE:2019nxm} with $\Dzero$-meson candidates within about 3 standard deviations from the PDG mass~\cite{10.1093/ptep/ptac097}. Pions and kaons were identified by requiring the $\dEdx$ and time-of-flight signals measured respectively by the TPC and TOF detectors to be compatible with the expected values within 3 times the detector resolution.

For the measurement of the $\Lambdac\to\pKzeroS$ signal, the reconstruction of the $\kzero\to\pip\pim$ candidates was performed by pairing opposite-sign tracks, selected as discussed in Ref.~\cite{ALICE:2020wla}, into a neutral decay vertex displaced from the primary vertex. An additional primary proton~\cite{ALICE-PUBLIC-2017-005} was considered to reconstruct the $\Lambdac$-baryon decay.

The tracks used to reconstruct the $\Lambda$- and $\X$-baryon signals from the $\XicPlus$-baryon cascades were selected as described in Ref.~\cite{Acharya:2021vjp} and references therein. For this analysis, pion and proton tracks were identified requiring that the measured signals were compatible with the expected values within 3 and 5 times the detector resolution in the TPC and TOF, respectively.
The $\lmb$-baryon candidates, reconstructed as pairs of pion and proton tracks, were combined with pion tracks with a transverse momentum larger than 0.15~GeV$/c$ to form the $\X$-baryon decay vertex. The masses of the reconstructed $\lmb$ and $\X$ particles were constrained to not deviate more than $1\%$ from their respective PDG masses~\cite{10.1093/ptep/ptac097}. Positively charged pion tracks with at least 3 hits in the ITS detector and a transverse momentum larger than 0.4~GeV$/c$ were selected and combined with the reconstructed $\X$ baryon to form the $\XicPlus$-baryon candidates.

%
%
As in Ref.~\cite{ALICE:2019nxm}, the applied track selections affect the D-meson and $\Lambdac$-baryon acceptance in rapidity, which decreases rapidly for $|y|>0.5$ at low $\pt$ and $|y|>0.8$ for $\pt > 5$~GeV$/c$. Therefore, a fiducial acceptance selection was applied on the D-meson and $\Lambdac$-baryon rapidity, $|y| < y_{\rm fid} (\pt)$. The $y_{\rm fid} (\pt)$ factor increases with a polynomial form from 0.5 to 0.8 in the range $0 < \pt < 5$~GeV$/c$ and $y_{\rm fid} = 0.8$ above $5$~GeV$/c$. For the reconstruction of the $\XicPlus$-baryon the acceptance criterion $|y|<0.8$ was applied.

%
%
\subsection{Analyses with charm-hadron secondary vertex reconstruction}
The reconstruction of charm-hadron signals in exclusive hadronic channels is characterised by a substantial combinatorial background arising from the erroneous association of charged tracks to the charm-hadron decays.
Given the typical proper lifetime of a few hundred micrometres ($c\tau\approx 120-300$ $\upmu$m~\cite{10.1093/ptep/ptac097})
the measurement of $\Dzero$-
, $\Dplus$-, $\Ds$-meson, and $\XicPlus$-baryon signals was based on the reconstruction of the decay-vertex topologies displaced from the primary vertex. The spatial separation between the production and the decay vertices, allowed the signal-to-background separation to be improved exploiting variables related to the displaced vertex decay topology. 

The topological variables considered in the analyses reported in this article are similar to those already used in previous works~\cite{ALICE:2021mgk, ALICE:2019nxm, ALICE:2012inj, Acharya:2021vpo, Acharya:2021vjp}. Some of them are the track impact parameter in the transverse plane, the distance between the primary and decay vertices (decay length, $L$), and the pointing angle of the reconstructed charm-hadron momentum to the primary vertex ($\theta_{\rm pointing}$). As the $\Dstar$ meson decays strongly, its decay vertex cannot be resolved from the primary one, and so topological selections are only used on the reconstructed $\Dzero$ daughter.

The $\XicPlus\to\X\pip\pip$ secondary vertex was reconstructed from the decay channels $\rm \X\rightarrow \lmb\pi^{-}$ and $\rm \lmb\rightarrow p\pi^{-}$ together with their charge conjugates using a Kalman Filter (KF) vertexing algorithm~\cite{KF2012}. The algorithm provides a full description of the decay particle both at its production and decay vertex. As discussed in Ref.~\cite{Acharya:2021vjp}, the KF software gives the possibility to set constraints to the mass and the production point of the reconstructed particles using information about the uncertainties of the daughter particle trajectories. Setting the mass constraint to the reconstructed decay products of a decay chain helps to improve the mass and momentum resolution of the reconstructed mother particle, while the production point constraint can be used to determine whether the particle emerges either from the primary or from a displaced vertex. A $\chi^2$ is calculated for the reconstructed particle, which quantifies the probability of the hypothesis that the particle truly emerges from the assigned vertex. In this analysis the mass constraint was applied to the $\lmb$- and $\X$-baryon candidates. Furthermore, the topological constraint was applied to the reconstructed trajectory of the $\XicPlus$ baryon by fitting it to the primary vertex of the collision.

%
\subsection{Background rejection of $\boldsymbol{\Dplus}$, $\boldsymbol{\Ds}$, and $\boldsymbol{\XicPlus}$ hadrons with Boosted Decision Trees}
To further reduce the background contribution for the reconstruction of the $\Ds$-, $\Dplus$-, and $\XicPlus$-hadron signals, machine-learning approaches based on Boosted Decision Trees (BDT) with the gradient boosting algorithm XGBoost~\cite{10.1145/2939672.2939785, barioglio_luca_2022_7014886} were adopted. The algorithm was provided with signal examples of $\Dplus$, $\Ds$, and $\XicPlus$ hadrons from simulations based on the PYTHIA 8 event generator as described in Section~\ref{sec:apparatus}, while the background samples were obtained from data using candidates in regions of the invariant mass distribution far from the signal peak. 
For the $\Dplus$ mesons, these regions were defined by requiring at least $4\sigma$ difference from the expected $\Dplus$-meson mass in data, where $\sigma$ indicates the width of the signal in MC simulations. For the $\Ds$ mesons the intervals $1.72 < M({\rm KK\pi}) < 1.83$~GeV$/c^2$ and $2.01 < M({\rm KK\pi}) < 2.12$~GeV$/c^2$ were considered. These intervals were specifically chosen for the $\Ds$ mesons to exclude contributions from $\Dplus$ mesons decaying through the same decay channel. For the training sample of $\XicPlus$-baryon background the invariant mass intervals $2.168<M({\rm \Xi\pi\pi})<2.411$~GeV$/c^2$ and $2.525<M({\rm \Xi\pi\pi})<2.768$~GeV$/c^2$ were considered. The regions correspond to a separation of at least $8\sigma$ from the signal in data.

Signals decaying either directly to the expected final state or via resonances were weighted in the training according to their natural abundances~\cite{10.1093/ptep/ptac097}.
Loose selection criteria were applied to the D-meson and $\XicPlus$-baryon candidates before the training, following the same procedures as described in Refs.~\cite{ALICE:2019nxm, ALICE:2021mgk, Acharya:2021vjp}. For $\Ds$-meson candidates, the reconstructed invariant mass of the $\mathrm{K^{+}K^{-}}$ pair was further required to match the world-average $\phi$-meson mass within $\pm15$~MeV$/c^{2}$.
The main information provided to the BDT to classify signal candidates from background ones relates to the decay-vertex topology and PID. 
In addition, 
the $\chi^2$ of the $\XicPlus$ baryons to the primary vertex, and the distance of closest approach between the daughters of the $\X$ baryon as well as between the daughters of the $\XicPlus$ baryons were considered to discriminate the $\XicPlus$-baryon signal. Independent BDTs were trained in the different $\pt$ intervals of all the analyses, and samples with signal and background similar to those used for the training were employed to test and validate the trained models. Subsequently, they were applied to the real data sample presented in Section~\ref{sec:apparatus}, in which the type of candidate is unknown. The output of the BDT is related to the candidate probability to be a charm hadron or combinatorial background. Selections on the BDT outputs were optimised to obtain a  large statistical significance for the signal along with a high fraction of prompt charm hadrons. 

%
%
\subsection{$\boldsymbol{\Dzero}$-meson and $\boldsymbol{\Lambdac}$-baryon reconstruction down to $\boldsymbol{\pt}=0$}
For tracks with $\pt\lesssim 500$ MeV$/c$, the spatial resolution of the impact parameter measurement is poor ($\gtrsim 100$ $\upmu$m), weakening the effectiveness of selections based on displaced decay-vertex topology. Furthermore, such selections at low $\pt$ would favour the reconstruction of hadrons originating from beauty-hadron decays (non-prompt), which are by construction more displaced than the prompt ones given the larger proper lifetime of beauty hadrons. For these reasons, the reconstruction of prompt $\Dzero$ mesons and $\Lambdac$ baryons down to $\pt=0$ did not apply any topological selections, and instead was performed by simply building pairs and triplets of identified decay tracks.

The measurement of the $\Dzero$-meson signals exploited the reconstruction of its secondary vertex, as discussed above, only for candidates with $\pt>1.5$~GeV$/c$. In the interval $\pt<1.5$~GeV$/c$, $\Dzero$-meson candidates were reconstructed by pairing opposite-charged tracks compatible with the kaon and pion hypotheses without selection criteria on the displaced decay topology. A similar procedure was also used for the prompt $\Dzero$-meson measurement in \pp collisions at $\s=5.02$~TeV~\cite{ALICE:2019nxm}, where the $\pt$ cut-off value between the two techniques was 1~GeV$/c$. In the present work, the $\Dzero$-meson signal was measured without reconstructing the secondary vertex up to $\pt=1.5$~GeV$/c$ to reduce the total uncertainty. In fact, this reconstruction technique grants a higher efficiency. In addition, the statistical uncertainties of the measured signal at low $\pt$ were reduced compared to the same measurement in \pp collisions at $\s=5.02$~TeV, given that the analysed data sample is larger by about a factor 1.8. 

The reconstruction of the $\Lambdac$ baryons was performed as reported in Ref.~\cite{Acharya:2021vpo} and references therein. Given that $c\tau(\Lambdac)\approx61$ $\upmu$m, which is below the spatial pointing resolution at the primary vertex for $\pt<1$~GeV$/c$, the candidate reconstruction for $\pt<1$~GeV$/c$ does not exploit the reconstruction of the $\Lambdac$-baryon secondary vertex. In this analysis, the $\LambdacTopKpi$ signal was reconstructed for the first time in the interval $0<\pt<1$~GeV$/c$ by combining triplets of tracks identified as a charged pion, kaon, or proton using the Bayesian PID approach based on the \enquote{maximum probability criterion}~\cite{ALICE:2016zzl}.
The $\Lambdac\to\pKzeros$ signal measurement in the same $\pt$ interval was performed as in Ref.~\cite{Acharya:2021vpo}. To select $\Lambdac\to\pKzeros$ signals, a machine-learning approach based on BDTs with AdaBoost was adopted~\cite{Hocker:2007ht}. The BDTs were trained to disentangle signal from background candidates according to the kinematics of the $\kzero$ decay. Some variables utilised in the training were the reconstructed $c\tau$ and the $\pip\pim$ invariant mass, the impact parameter to the primary vertex of the $\kzero$ candidate and its decay daughters, and the PID information of the bachelor track.
More details can be found in Ref.~\cite{ALICE:2020wla}.

\subsection{Raw-yield measurement from an invariant mass analysis}
%
%
\begin{figure}[t!]
    \centering
    \includegraphics[width=0.49\textwidth]{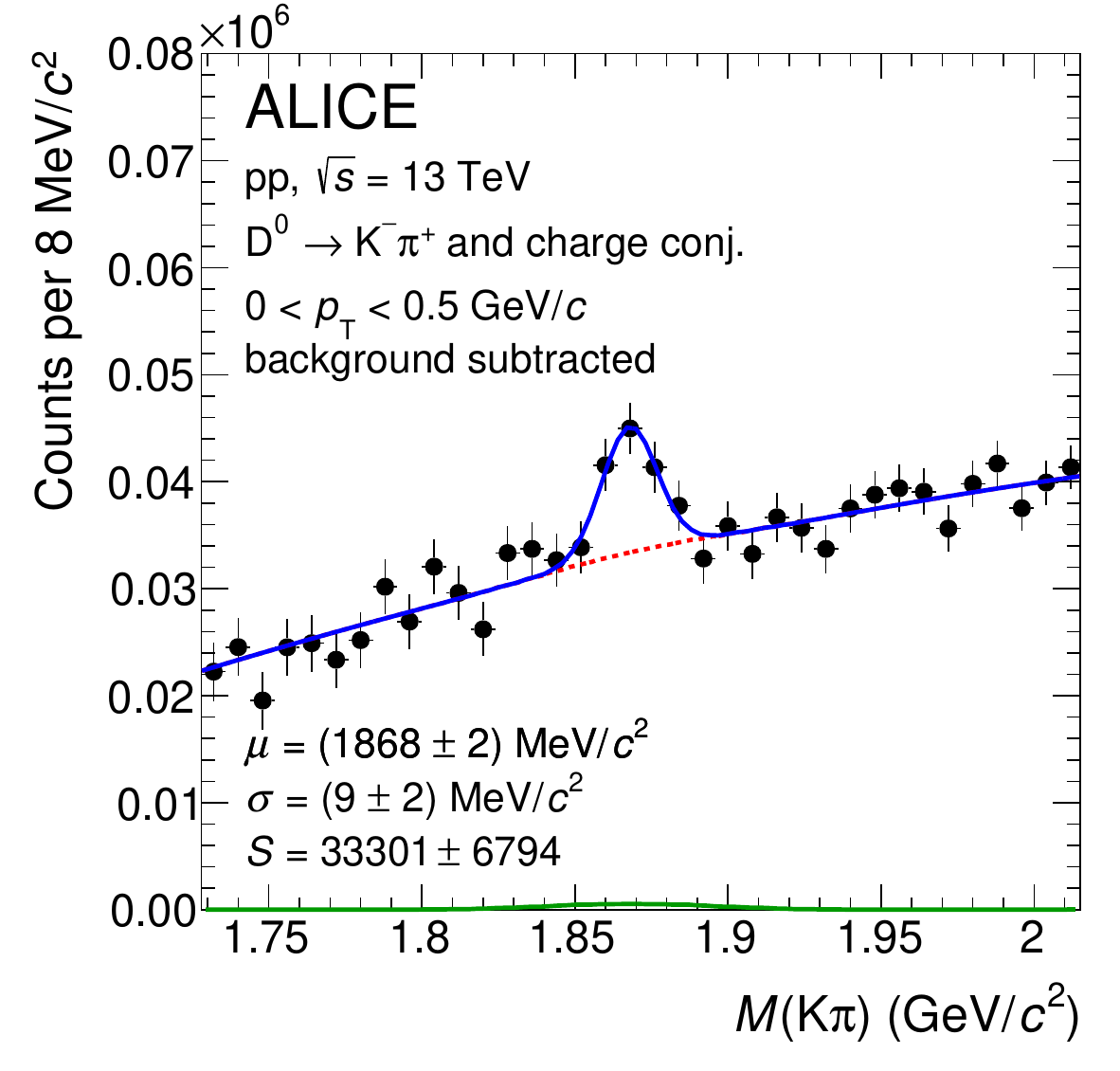}
    \includegraphics[width=0.49\textwidth]{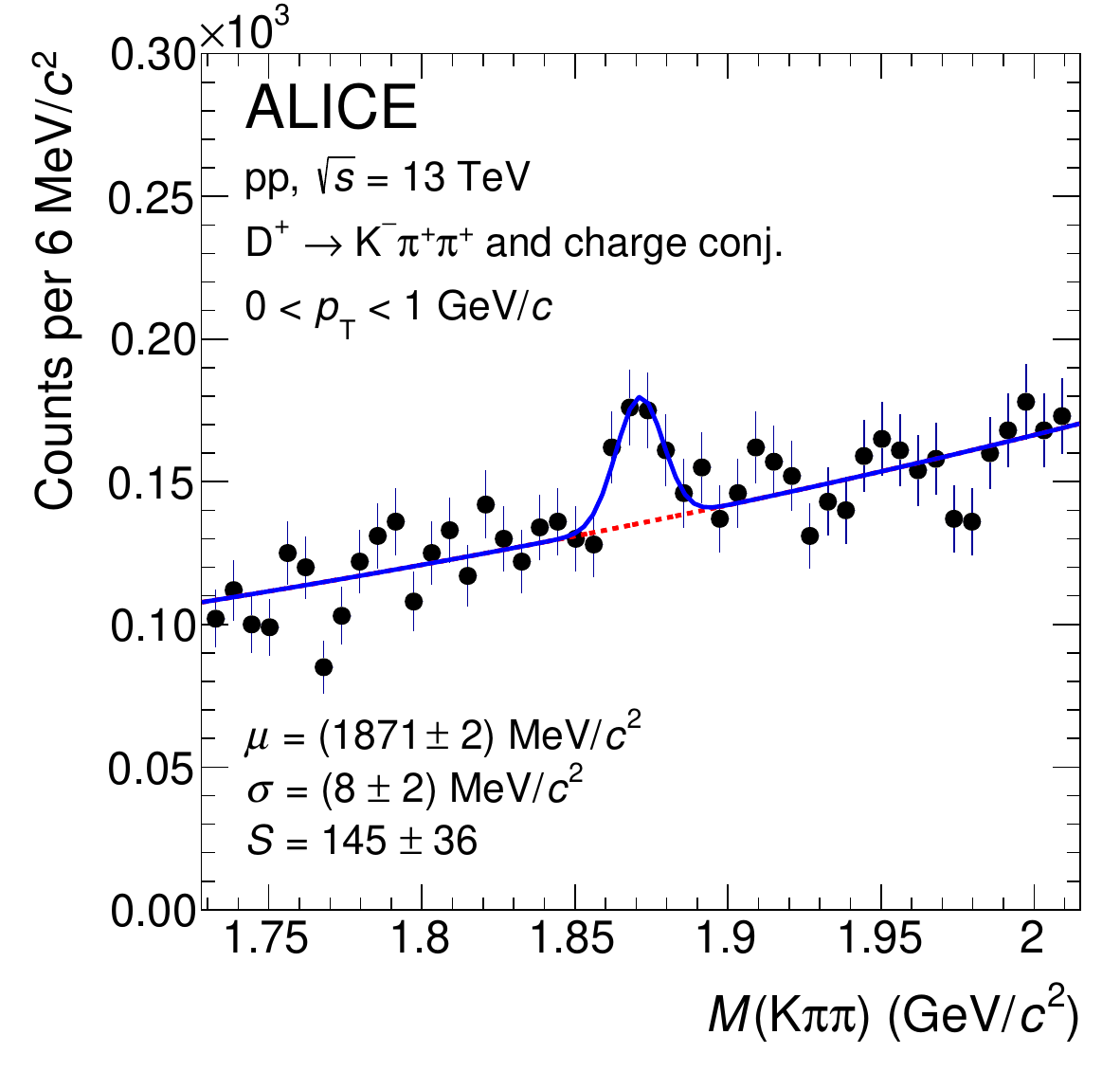}
    \includegraphics[width=0.49\textwidth]{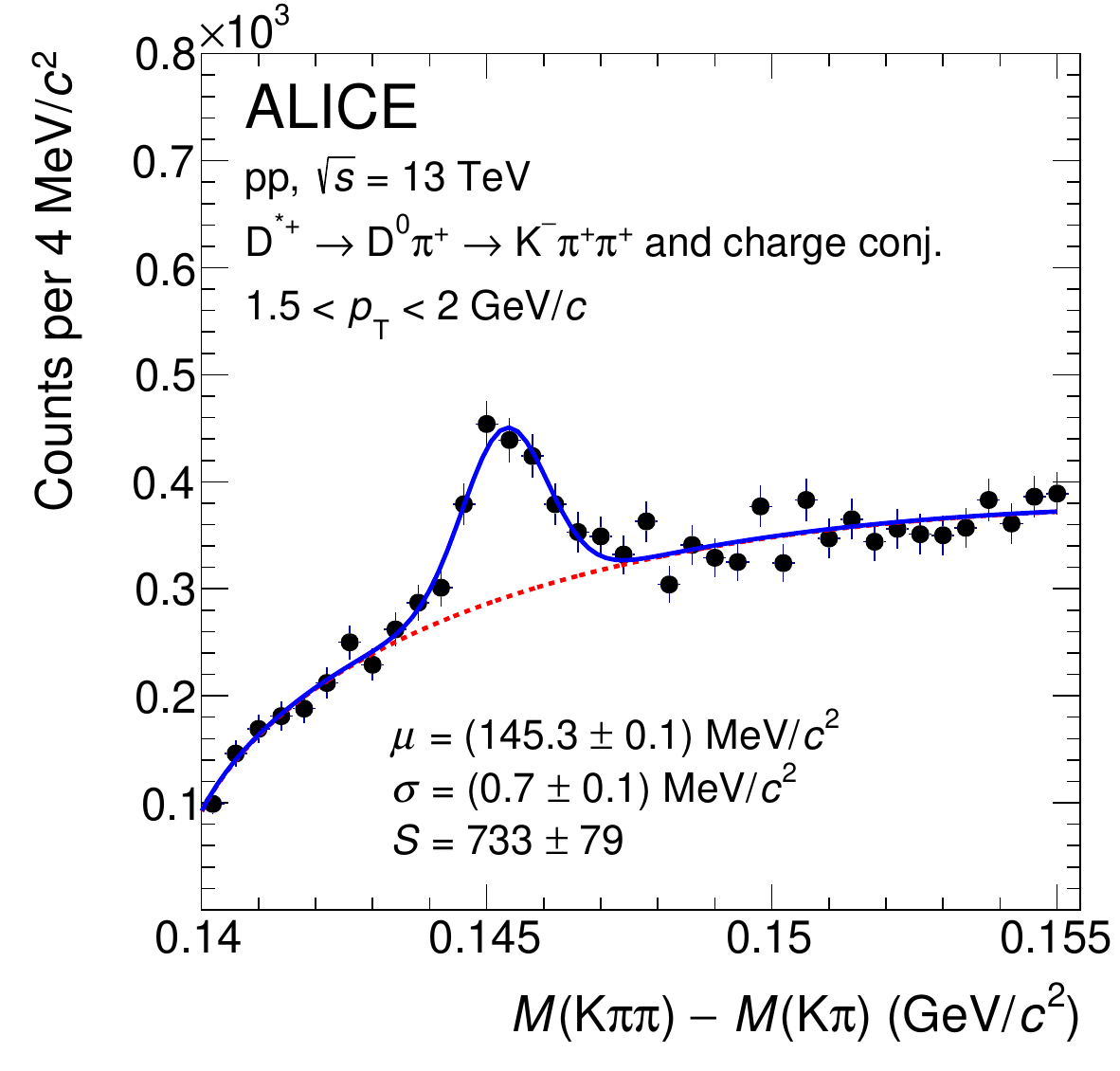}
    \includegraphics[width=0.49\textwidth]{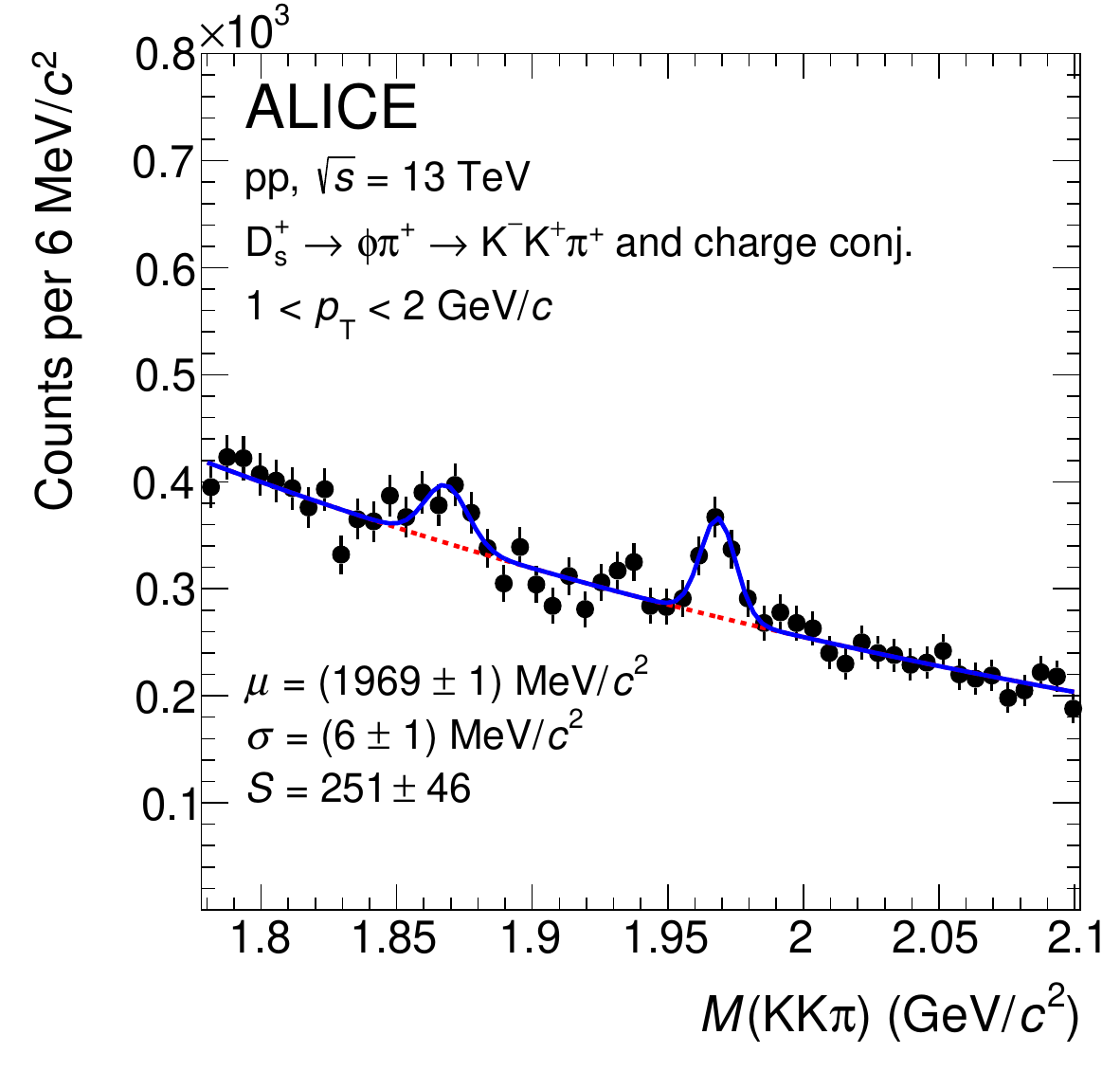}
    \caption{Invariant mass or mass-difference distributions of $\Dzero$-meson candidates (top-left), $\Dplus$-meson candidates (top-right), $\Dstar$-meson candidates (bottom-left), $\Ds$-meson candidates (bottom-right), and charge conjugates in $0< \pt < 0.5$~GeV$/c$, $0< \pt < 1$~GeV$/c$, $1.5< \pt < 2.0$~GeV$/c$, and $1< \pt < 2$~GeV$/c$, respectively. The blue solid lines show the total fit functions as described in the text and the red dashed lines represent the background. In the case of $\Dzero$, the combinatorial background estimated with the track-rotation technique was subtracted, the red-dashed line is the residual background, and the green line represents the contribution of the reflections. In the case of $\Ds$, the peak at lower invariant mass represents the reconstructed $\Dplus\rightarrow\kap\kam\pip$ signal. The values of the mean ($\mu$) and the peak width ($\sigma$) of the signal peak are reported together with the signal counts (\textit{S}). The reported uncertainties are only the statistical uncertainties from the fit.}
    \label{fig:minv_mesons}
\end{figure}
\begin{figure}[t!]
    \centering
    \includegraphics[width=0.49\textwidth]{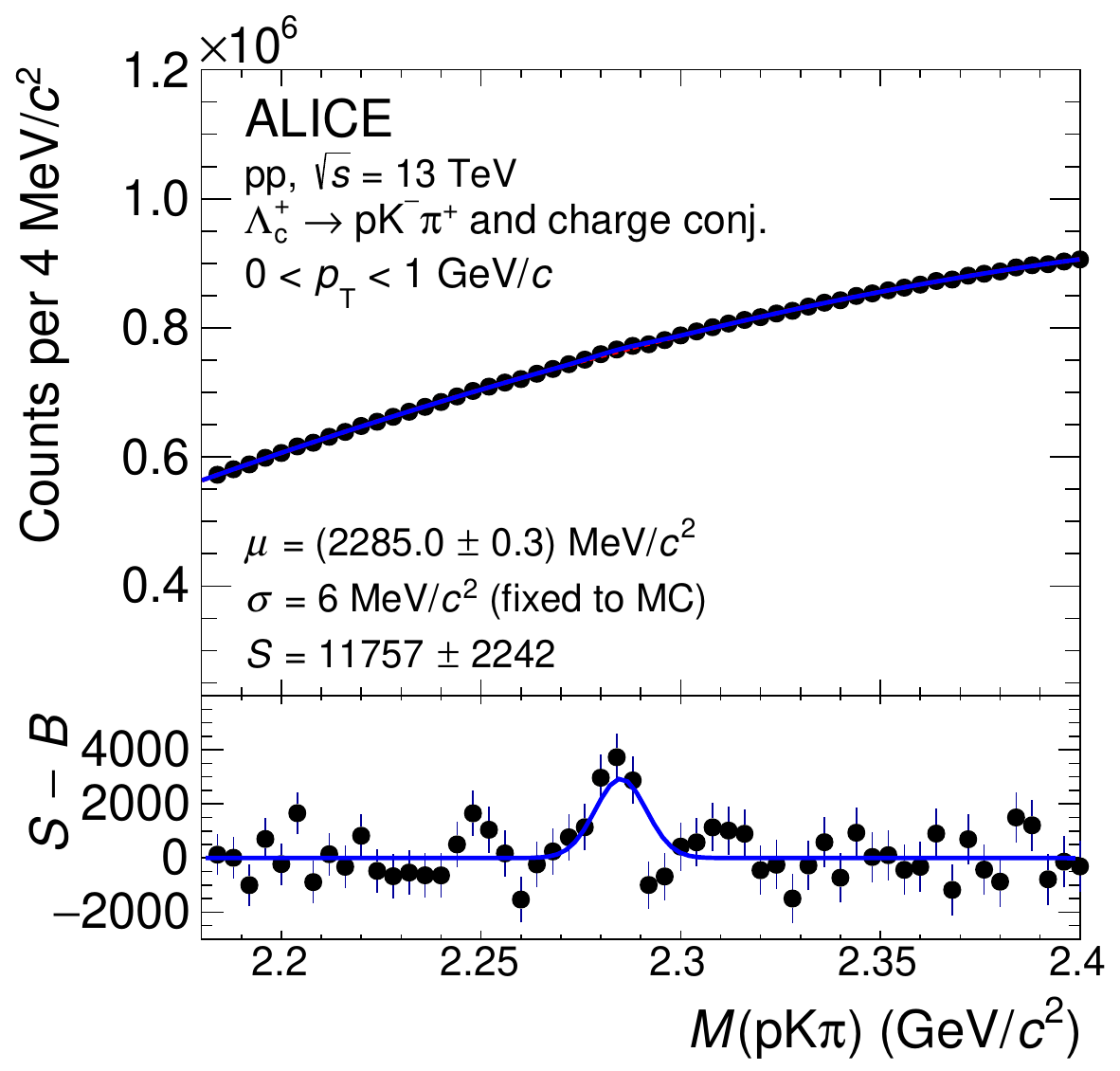}
    \includegraphics[width=0.49\textwidth]{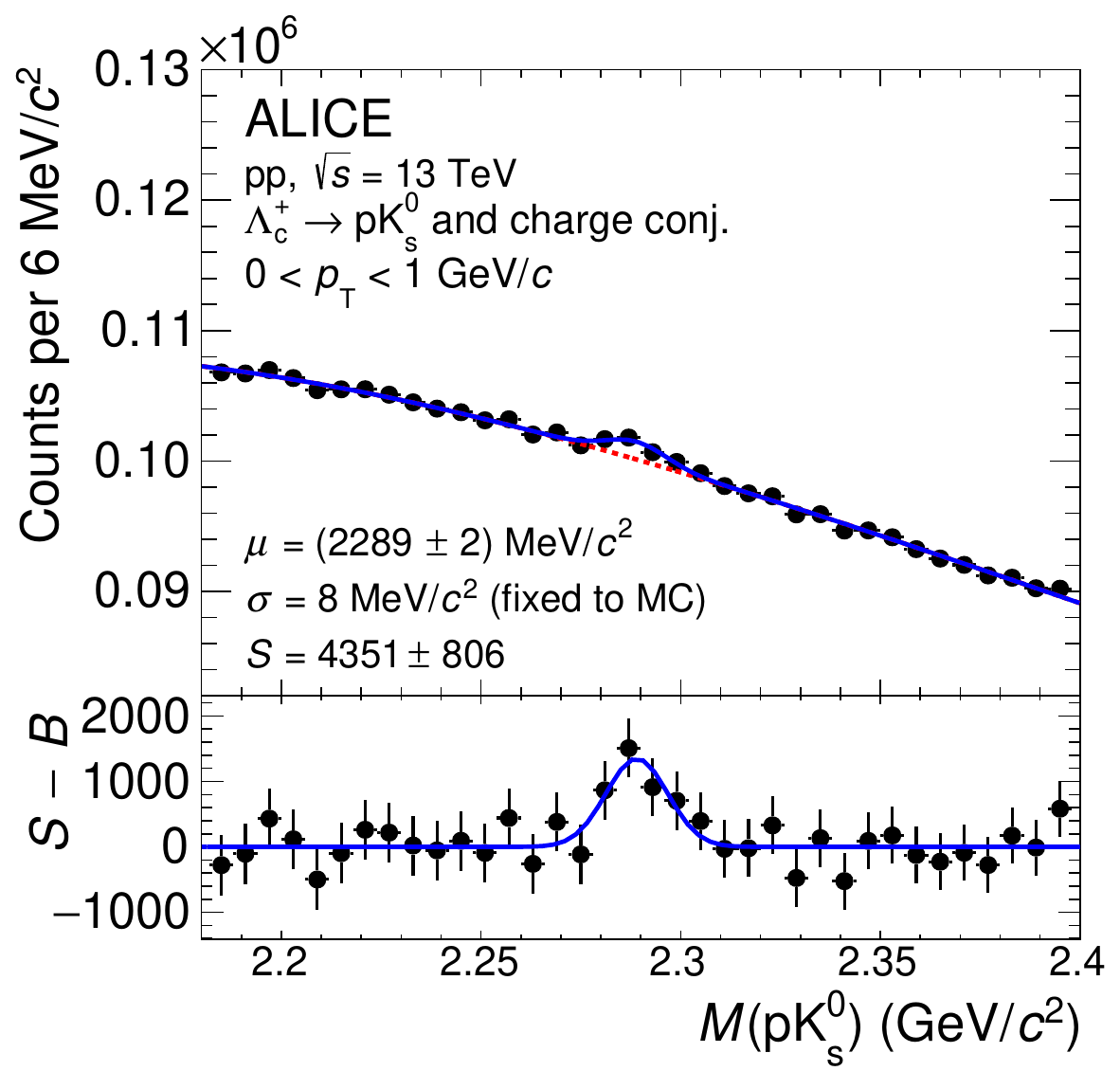}
    \includegraphics[width=0.49\textwidth]{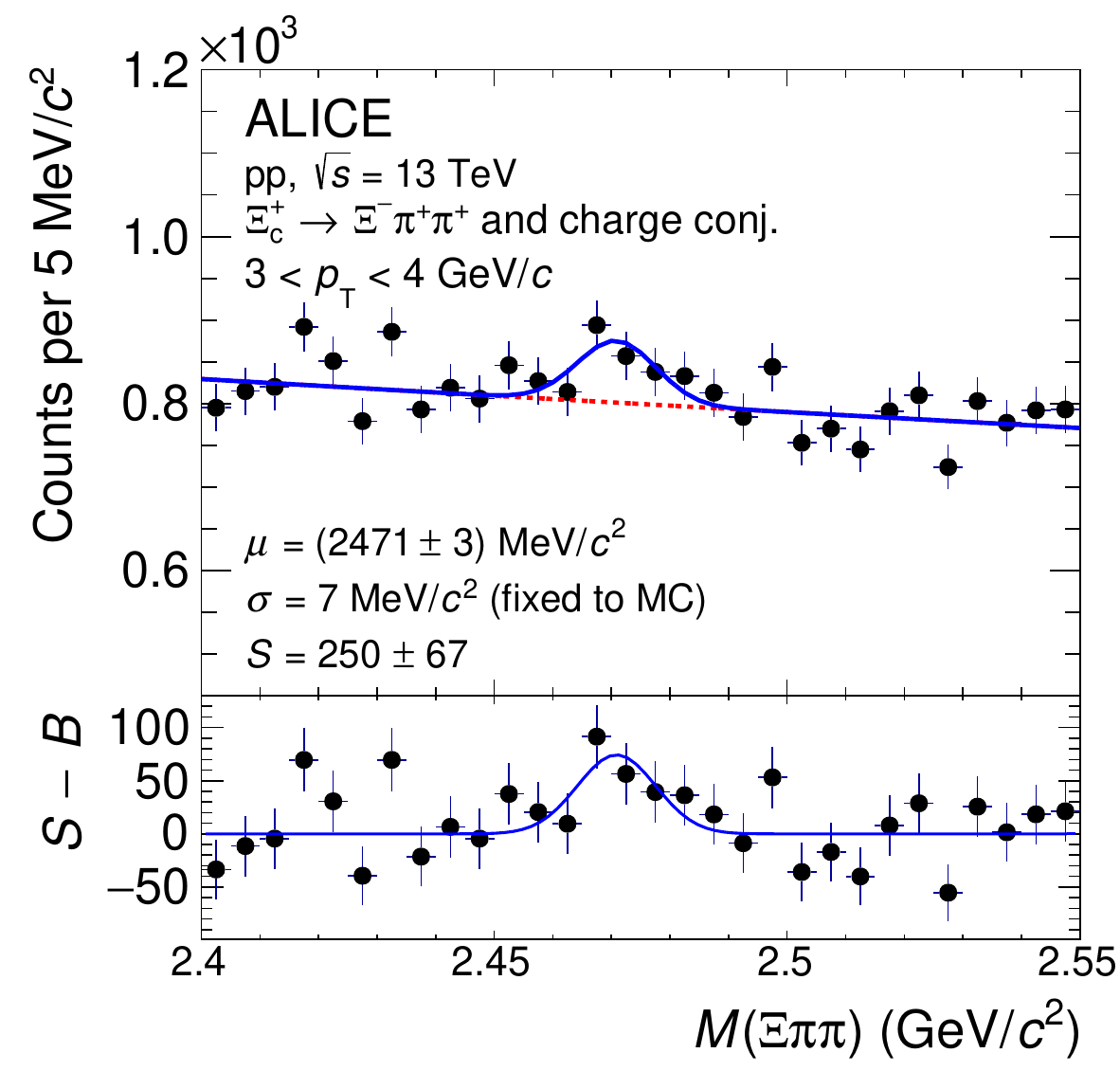}
    \caption{Invariant mass distributions of $\Lambdac$- and $\XicPlus$-baryon candidates and charge conjugates at low transverse momentum in \pp collisions at $\s=13$~TeV. The blue solid lines show the total fit functions as described in the text and the red dashed lines represent the background.
    The values of the mean ($\mu$) and the peak width ($\sigma$) of the signal peak are reported together with the signal counts (\textit{S}). The bottom panels report the charm-baryon invariant mass distribution after the subtraction of the background candidates parametrised as detailed in the main text ($S-B$). The Gaussian $\sigma$ are fixed to values from MC simulations. The reported uncertainties are only the statistical uncertainties from the fit.}
    \label{fig:minv_baryons}
\end{figure}
The $\pt$-differential charm-hadron raw yield was measured in the range $0 < \pt < 50$~GeV$/c$ for $\Dzero$ and $\Dplus$ mesons, $1 < \pt < 50$~GeV$/c$ for $\Dstar$ mesons, $1<\pt<36$~GeV$/c$ for $\Ds$ mesons, $0<\pt<1$~GeV$/c$ for $\Lambdac$ baryons, and $3<\pt<4$~GeV$/c$ for $\XicPlus$ baryons. A binned maximum-likelihood fit of the invariant mass distributions of candidates surviving the selection criteria mentioned above was performed. The signal peak was parameterised with a Gaussian function, whose variance was left free for the D mesons and fixed to the value observed in the simulations for the baryons. For the $\Dzero$-meson signal extraction, the background was described with a polynomial of second order for $\pt<2$~GeV$/c$ and an exponential function in all other $\pt$ intervals. The combinatorial background contribution to the invariant mass of K$\pi$ pairs reconstructed in $\pt<1.5$~GeV$/c$ was described with the track-rotation technique discussed in Ref.~\cite{ALICE:2019nxm}. Following this method, for each candidate K$\pi$ pair, 19 background candidates were built by rotating the kaon-track momentum in the transverse plane from $\pi/10$ to $19\pi/10$ radians. The resulting combinatorial background was finally subtracted before fitting the invariant mass distribution of the reconstructed K$\pi$ pairs. In all the considered $\pt$ intervals, the contribution of $\Dzero$ candidates in the invariant mass distribution reconstructed with the wrong decay particle mass assignment (reflections) was included in the fit. This contribution corresponds to the invariant mass distributions of the reflected signal in MC simulation, as discussed in Ref.~\cite{ALICE:2019nxm}.

For the $\Dplus$- and $\Ds$-meson signal measurement, the background was described with an exponential function. To grant a better stability in the measurement of the $\Ds$-meson signal, an independent Gaussian function was used to fit the peak related to the $\Dplus\to \kap\kam\pim$ reconstructed signal. The $\Dstar$-meson raw yield was measured considering a threshold function multiplied by an exponential for the background ($a \sqrt{\Delta M - m_\pi} \times {\rm e}^{b(\Delta M - m_\pi)}$), where $\Delta M$ is the mass-difference $M(\rm K\pi\pi) - M(K\pi)$, $m_\pi$ is the pion mass, and $a$, $b$ are free parameters.

For the $\Lambdac$ measurement in $0<\pt<1$ GeV$/c$, the combinatorial background in the invariant mass distributions of $\pKpi$ triplets and $\pKzeros$ pairs was described with a polynomial of third and second order, respectively. The combinatorial background for the $\XicPlus\to\X\pip\pip$ raw yield measurement was fitted with an exponential.

Figure~\ref{fig:minv_mesons} shows examples of fits to the invariant mass distributions for D-meson candidates in different $\pt$ intervals. This is also shown for $\Lambdac$ and $\XicPlus$ baryons in Fig.~\ref{fig:minv_baryons}, where the invariant mass distributions after the subtraction of the parametrised background are shown in the sub-panels together with the Gaussian function describing the signal.

\subsection{Cross sections}

\begin{figure}[t!]
    \centering
    \includegraphics[width=0.49\textwidth]{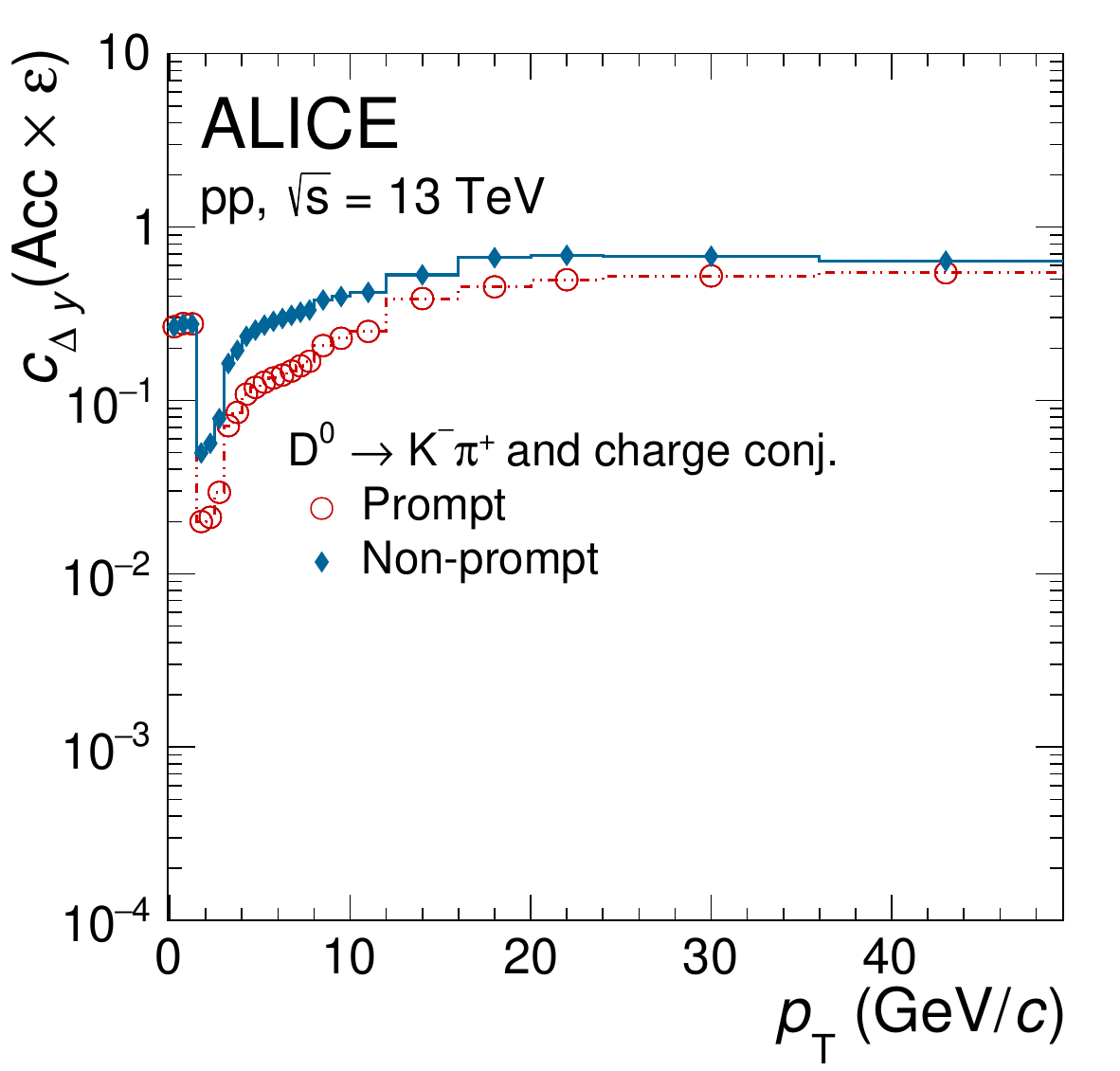}
    \includegraphics[width=0.49\textwidth]{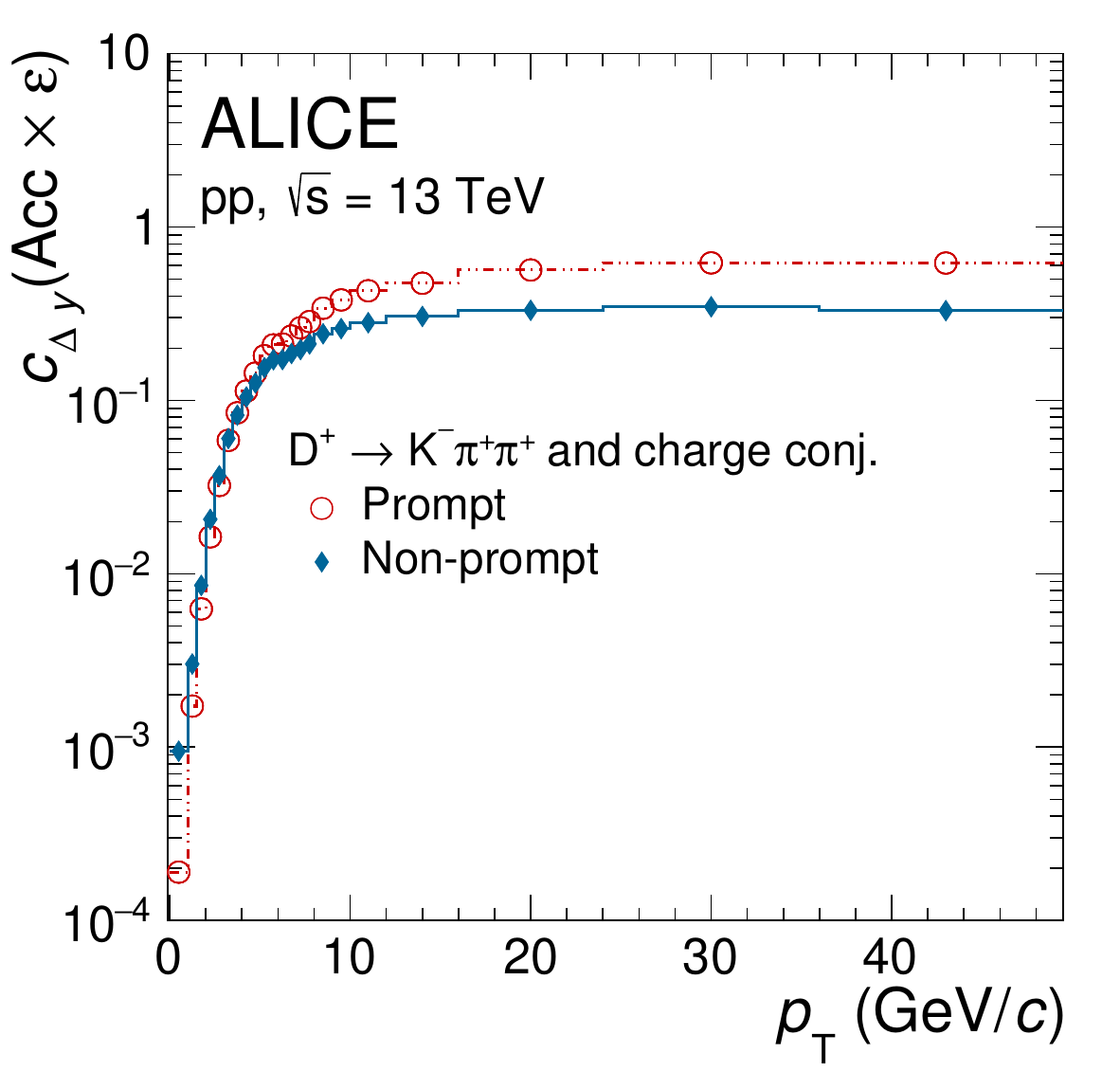}
    \includegraphics[width=0.49\textwidth]{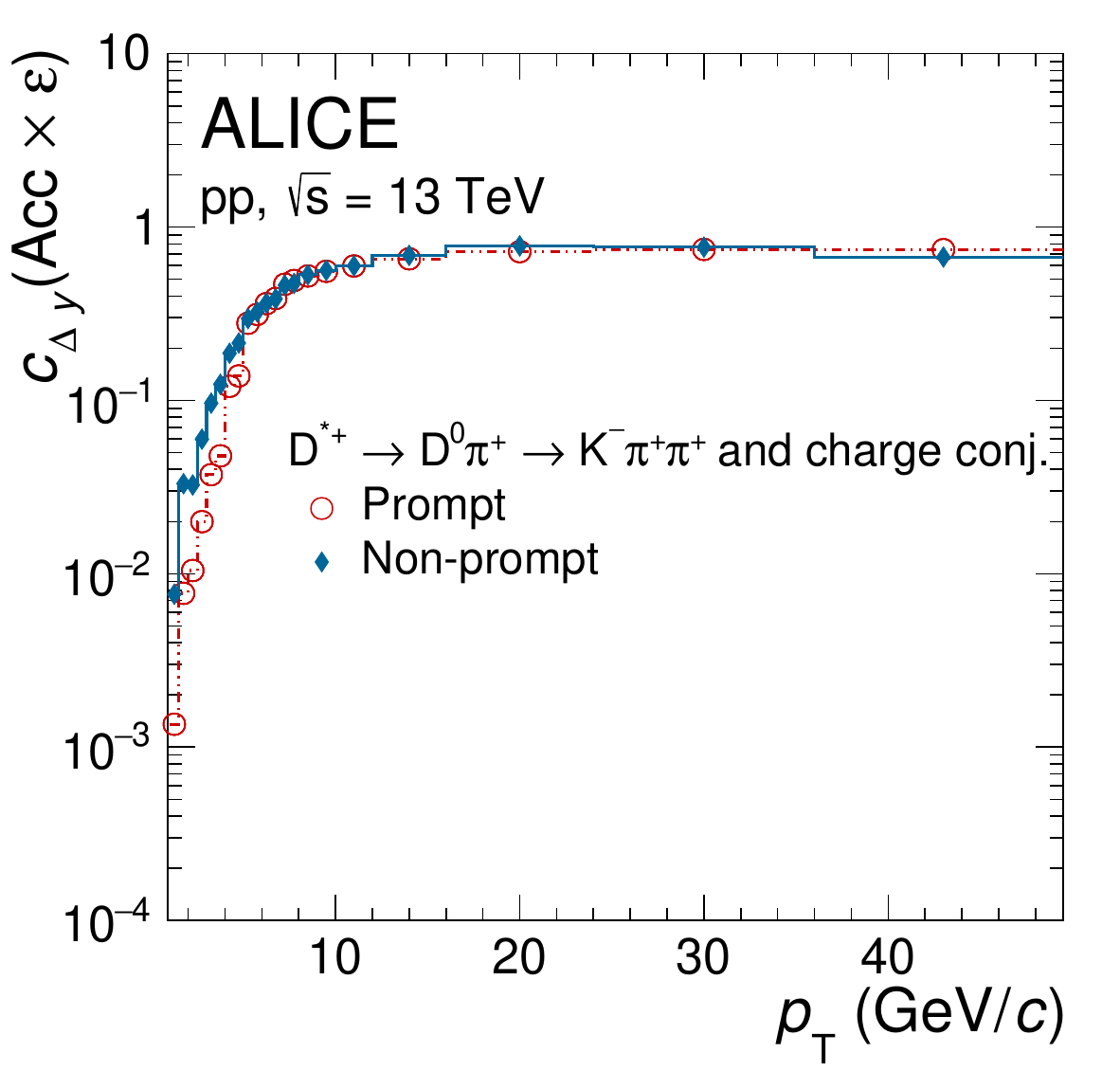}
    \includegraphics[width=0.49\textwidth]{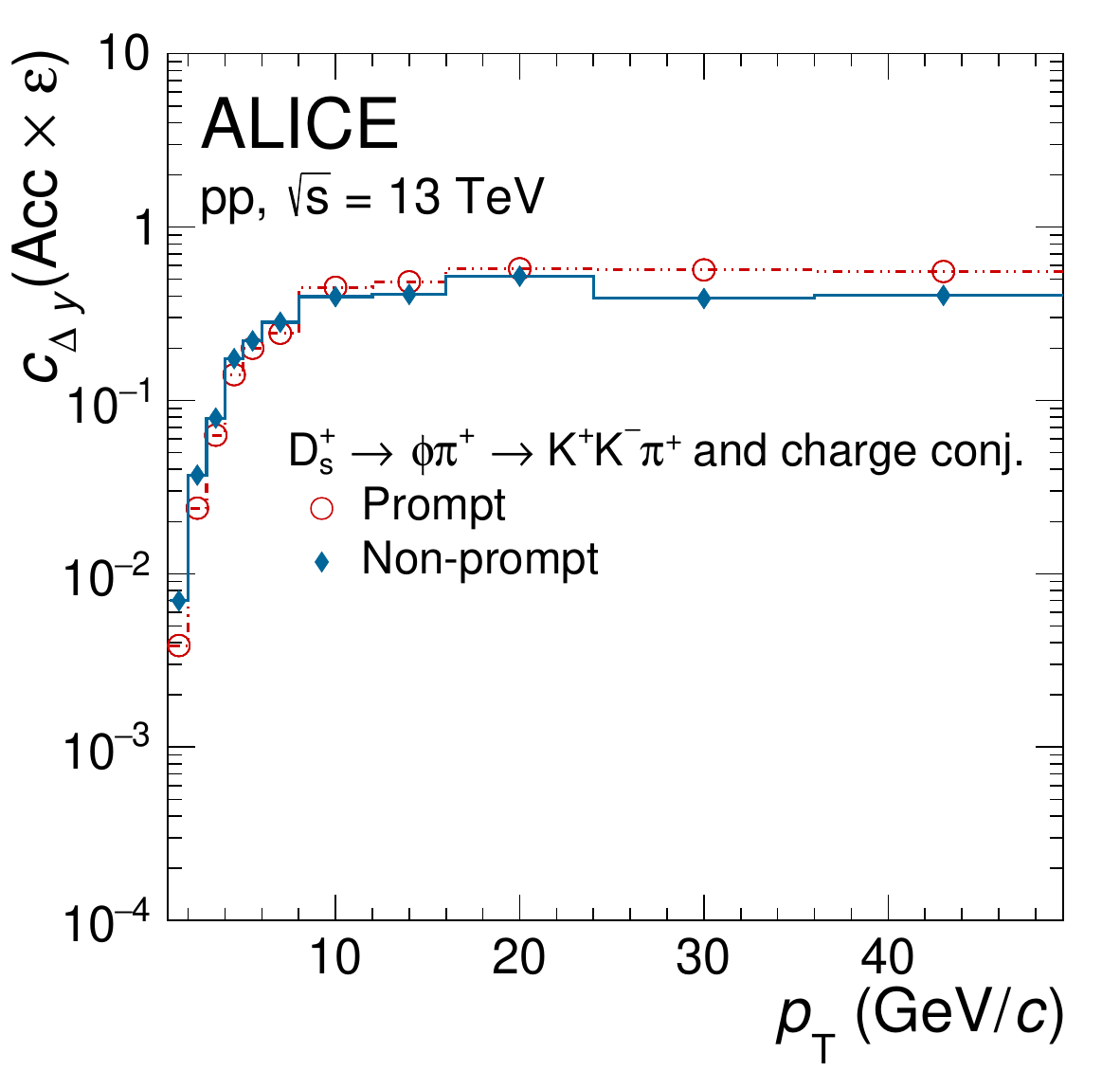}
    \caption{Correction factors $c_{\Delta y}({\rm Acc}\times \varepsilon)$ for the prompt and non-prompt $\Dzero$-meson (top-left), the $\Dplus$-meson (top-right), the $\Dstar$-meson (bottom-left), and the $\Ds$-meson (bottom-right) as a function of $\pt$.}
    \label{fig:accxeff_mesons}
\end{figure}

The $\pt$-differential cross section of prompt charm hadrons was measured as follows:
\begin{equation}
    \left. \frac{{\rm d}\sigma^{\rm H}}{{\rm d}\pt} \right|_{|y|<0.5} = \frac12\frac{1}{\Delta\pt}\times\frac{f_{\rm prompt}\times N^{\rm H+\overline{H}}_{|y|<y_{\rm fid.}}}{c_{\Delta y}({\rm Acc}\times \varepsilon)_{\rm prompt}}\times\frac{1}{\rm BR}\times\frac{1}{\mathcal{L}_{\rm int}}\;.
    \label{eq:cross_section_SigmaC}
\end{equation}
The term $N^{\rm H+\overline{H}}_{|y|<y_{\rm fid.}}$ refers to the raw yield, the sum of reconstructed particles and antiparticles obtained from the invariant mass fits shown in Figs.~\ref{fig:minv_mesons} and~\ref{fig:minv_baryons}. This quantity was divided by 2 to obtain the averaged yields between particles and antiparticles, and it was scaled by the prompt fraction $f_{\rm prompt}$ to correct for the charm-hadron signal originating from beauty-hadron decays. The raw yield was corrected by the $c_{\Delta y}({\rm Acc}\times \varepsilon)_{\rm prompt}$ term, which accounts for the rapidity coverage, the detector acceptance, and the reconstruction and selection efficiency of the prompt charm hadron signal~\cite{ALICE:2019nxm}.
The production cross section in each $\pt$ interval was obtained by further scaling the raw yield by the $\pt$-interval width ($\Delta\pt$), the branching ratio of the decay channel chosen to reconstruct the signal (BR), and the integrated luminosity ($\mathcal{L}_{\rm int}$). 

Figure~\ref{fig:accxeff_mesons} shows $c_{\Delta y}({\rm Acc}\times \varepsilon)$ as a function of $\pt$ for prompt and non-prompt $\Dzero$, $\Dplus$, $\Dstar$, and $\Ds$ mesons, 
where the non-prompt D mesons are those produced in beauty-hadron decays.
Given the average larger displacement from the primary vertex of beauty-hadron decay vertices due to their
long lifetime, the selection criteria applied on the decay length in general enhance the $c_{\Delta y}({\rm Acc}\times \varepsilon)$ factor of non-prompt D mesons with respect to that of prompt ones, especially at low $\pt$.
The $c_{\Delta y}({\rm Acc}\times \varepsilon)$ values of non-prompt $\Dzero$ mesons for $\pt>1.5$~GeV$/c$ are higher than those of prompt ones. On the other hand, the values for prompt and non-prompt $\Dzero$ mesons are compatible for $\pt<1.5$~GeV$/c$, because in this interval no selection criteria based on the intrinsic displacement of the decay are applied. This explains the significant drop of the efficiency at $\pt=1.5$~GeV$/c$, as shown in the top-left panel of Fig.~\ref{fig:accxeff_mesons}.
The $c_{\Delta y}({\rm Acc}\times \varepsilon)$ factor of prompt and non-prompt $\Dstar$ mesons are compatible above $\pt=5$~GeV$/c$, while the $c_{\Delta y}({\rm Acc}\times \varepsilon)$ factor for prompt $\Dplus$ and $\Ds$ mesons is larger than that of non-prompt mesons for $\pt$ higher than 5~GeV$/c$. Conversely, in the low-$\pt$ region, the trend is reversed.

The $c_{\Delta y}({\rm Acc}\times \varepsilon)$ term for the prompt and non-prompt $\Lambdac\to\pKpi$ signal reconstruction without selections on topological variables is about 3\% in the interval $0<\pt<1$~GeV$/c$, while that for the $\LambdacTopKzeroS$ in the same $\pt$ interval with selections based on the BDT classification is about 11\% for the prompt signal and about 10\% for the non-prompt one. The same term for the prompt and non-prompt $\XicPlus$-baryon reconstruction in the interval $3 < \pt < 4$~GeV$/c$ is around $0.7\%$. In the case of the $\XicPlus$ baryon, the efficiency was calculated after weighting the simulated $\pt$ distributions to match the $\XicZero$-baryon $\pt$-differential cross section measured in \pp collisions at $\s=13$~TeV following the procedure described in Ref.~\cite{Acharya:2021vjp}.
The $c_{\Delta y}({\rm Acc}\times \varepsilon)$ correction factors of all mesons and baryons were computed taking into account all the possible resonant channels which produce the final states chosen for the reconstruction, as described in Refs.~\cite{ALICE:2020wla, Acharya:2021vjp, ALICE:2017thy}.

The measured raw yield was further corrected by the fraction of prompt reconstructed hadrons ($\rm\textit{f}_{prompt}$). This fraction was calculated similarly to previous measurements (see e.g.~Refs.~\cite{ALICE:2019nxm, ALICE:2017olh}) adopting the beauty-hadron production cross sections from FONLL calculations, the beauty-hadron decay kinematics modelled with PYTHIA 8~\cite{SJOSTRAND2015159}, and the efficiencies reported in Fig.~\ref{fig:accxeff_mesons} for the non-prompt D mesons. 
In the beauty-hadron production cross section calculations from FONLL, the beauty-quark fragmentation fractions were taken from LHCb measurements~\cite{LHCb:2019fns} for $\rm b\to\Lambda_b^0$ and the averaged results from LEP for B mesons~\cite{Gladilin:2014tba}. A prediction for the non-prompt $\XicPlus$-baryon cross section was made by scaling the prediction for the non-prompt $\Lambdac$-baryon cross section with the ratio of the sums of the beauty hadron fragmentation fractions times branching ratios, $\rm\sum_{h_b} b\rightarrow h_{b}\rightarrow \rm\XicPlus$ and $\rm\sum_{h_b} b\rightarrow h_{b}\rightarrow\Lambdac$. Using the fragmentation fractions from LHCb measurements ~\cite{LHCb:2019fns} and the BR from PYTHIA 8~\cite{SJOSTRAND2015159} simulations the ratio of the beauty-hadron sums can be approximated as $(\rm b\rightarrow\Xi_{b}^{0,-}\rightarrow\XicPlus) / (\rm b\rightarrow\Lambda_{b}^0\rightarrow\Lambdac)$.
Furthermore, the $\rm b\to\Xi_b$ fragmentation fraction was assumed to be equal to $\rm c\to\Xi_c$ and the non-prompt $\rm\XicPlus/\Lambdac$ ratio was taken to be equal to the prompt ratio. Under these assumptions the scaling factor for the non-prompt $\Lambdac$-baryon cross section can be approximated as the prompt cross section ratio of $\Lambdac$- and $\rm\Xi_{c}$-baryons, which was taken from Ref.~\cite{Acharya:2021vjp}.

For $\Dzero$, $\Dplus$, $\Dstar$, and  $\Ds$ mesons, the values of $\rm\textit{f}_{prompt}$ range between 81\% and 96\% depending on the D-meson species and $\pt$ interval. The $\rm\textit{f}_{prompt}$ fractions for $\Lambdac$ baryons reconstructed in $0<\pt<1$~GeV$/c$ in the $\pKpi$ and $\pKzeroS$ decay channels are about 98\% and 97\%, respectively. The same quantity for $\XicPlus$ baryons was found to be 97\% in the $\pt$ interval $3<\pt<4$~GeV$/c$.

\section{Systematic uncertainties}
\label{sec:systematics}

\begin{table}[t!]
    \centering
    \caption{Relative systematic uncertainties of the measured cross section of prompt $\Dzero$, $\Dplus$, $\Dstar$, and $\Ds$ mesons in \pp collisions at $\s=13$~TeV. The values reported in this table are those from the lowest and highest $\pt$ intervals considered in the analyses.  Uncertainties found to be negligible are indicated as such (negl.).
    }
    \begin{tabular}{|l|c c|c c|c c|c c|}
         \hline
         & \multicolumn{2}{c|}{$\Dzero$} & \multicolumn{2}{c|}{$\Dplus$} & \multicolumn{2}{c|}{$\Dstar$} & \multicolumn{2}{c|}{$\Ds$} \tabularnewline

         $\pt$ (GeV$/c$) &
         0$-$0.5& 36$-$50 & 
         0$-$1 & 36$-$50 & 
         1$-$1.5 & 36$-$50 & 
         1$-$2   & 24$-$36 \tabularnewline 

         \hline
         Signal extraction &
         9\%& 5\%& 
         10\% & 5\% & 
         10\% & 2\% & 
         8\% & 5\% \tabularnewline 

         Tracking efficiency &
         4\% & 6\%& 
         5.5\% & 8\% & 
         4.5\% & 6.5\% & 
         4\% & 8\% \tabularnewline 

         Selection efficiency &
         negl. & 3\%& 
         7\% & 2\% & 
         10\% & 2\% & 
         10\% & 3\% \tabularnewline 

         PID efficiency &
         negl. & negl. & 
         negl. & negl. & 
         negl. & negl. & 
         negl. & negl. \tabularnewline 

         $\pt$ shape in MC &
         1\% & negl. & 
         5\% & negl. & 
         3\% & negl. & 
         1\% & negl. \tabularnewline 

         Material budget in MC &
         negl. & negl. & 
         negl. & negl. & 
         negl. & negl. & 
         negl. & negl. \tabularnewline 

         Prompt fraction & 
         $^{+2.4\%}_{-2.3\%}$ & $^{+1.8\%}_{-0.8\%}$ & 
         $^{+6.1\%}_{-6.0}$ & $^{+0.9\%}_{-0.8\%}$ & 
         $^{+8.6\%}_{-8.6\%}$ & $^{+1.2\%}_{-0.8\%}$ & 
         $^{+6\%}_{-6\%}$ & $^{+2\%}_{-2\%}$ \tabularnewline 
         \hline

         Branching ratio &
         \multicolumn{2}{c|}{0.8\%} & 
         \multicolumn{2}{c|}{1.7\%} & 
         \multicolumn{2}{c|}{1.1\%} & 
         \multicolumn{2}{c|}{2.7\%} \tabularnewline 
         \hline

         Luminosity &
         \multicolumn{8}{c|}{1.6\%} \tabularnewline
         \hline

         Total uncertainty &
         $10\%$ & $9\%$ & 
         $16\%$ & $9\%$ & 
         $18\%$ & $7\%$ & 
         $15\%$ & $11\%$ 
         \tabularnewline
         \hline
    \end{tabular}
    \label{tab:syst_Dmesons}
\end{table}

\begin{table}[t!]
    \centering
    \caption{Relative systematic uncertainties of the measured cross section of prompt $\Lambdac$ and $\XicPlus$ baryons in \pp collisions at $\s=13$~TeV. The systematic uncertainties evaluated for the two $\Lambdac$-baryon decay channels are reported separately. Uncertainties found to be negligible are indicated as such (negl.).
    }
    \begin{tabular}{|l|c|c|c|}
          \hline
          &  $\Lambdac\to\pKpi$ & $\Lambdac\to\pKzeros$ & $\XicPlus$ \tabularnewline
         $\pt$ (GeV$/c$) & 0$-$1 & 0$-$1 & 3$-$4 \tabularnewline

         \hline
         Signal extraction &
         12\% & 
         11\% & 
         9\% \tabularnewline 

         Tracking efficiency &
         5\% & 
         5\% & 
         6\% \tabularnewline 

         Selection efficiency &
         10\% & 
         5\% & 
         3\% \tabularnewline 

         PID efficiency &
         5\% & 
         1\% & 
         negl. \tabularnewline 

         $\pt$ shape in MC &
         1\% & 
         negl. & 
         3\% \tabularnewline 

         Material budget in MC &
         negl. & 
         4\% & 
         4\% \tabularnewline 

         Prompt fraction & 
         $^{+0.9\%}_{-1.3\%}$ & 
         $^{+1.6\%}_{-2.1\%}$ & 
         $^{+2.9\%}_{-3.3\%}$ \tabularnewline 

         \hline
         Branching ratio &
         5.1\% & 
         5.0\% & 
         44.4\% \tabularnewline 

         \hline
         Luminosity &
         \multicolumn{3}{c|}{1.6\%} \tabularnewline
         \hline

         Total uncertainty &
         18\% & 
         15\% & 
         46\% \tabularnewline 
         \hline
    \end{tabular}
    \label{tab:syst_baryons}
\end{table}

The systematic uncertainties of the prompt charm-hadron production cross sections were estimated taking into account the following sources: (i) the stability of the signal extraction from the fits to the invariant mass distribution described in Section~\ref{sec:analysis}; (ii) the track reconstruction efficiency; (iii) the selection efficiency of charm hadrons; (iv) the PID selection efficiency; (v) the shape of the generated $\pt$ distribution for charm hadrons in MC simulations; (vi) the detector material budget description in MC simulations; (vii) the estimation of the fraction of prompt hadrons; (viii) the branching ratios of the decay channels used in the analyses; (ix) the collected luminosity. 
The values of the estimated uncertainties in some representative $\pt$ intervals are reported in Tables~\ref{tab:syst_Dmesons} and~\ref{tab:syst_baryons} for the prompt D mesons and charm baryons, respectively.
The total uncertainties in the analyses of each cross section measurement were calculated as the quadratic sum of these contributions. The uncertainty sources were assumed to be uncorrelated among the charm-hadron species, with the exception of (ii), (vi), (vii) and (ix). 
In the following, the strategies used to estimate the values of each source of uncertainty are briefly described. More details on the methodologies used to estimate the systematic uncertainties can be found in previous publications (see Refs.~\cite{ALICE:2021mgk,Acharya:2021vpo,Acharya:2021dsq,ALICE:2017olh,ALICE:2012inj,ALICE:2012gkr,ALICE:2019nxm, Acharya:2021vjp,ALICE:2022cop}).

The systematic uncertainty of the charm-hadron raw yield extraction was estimated in each \pt interval by repeating the fits several hundred times by varying the fit configurations. Such variations included, for example, the change of the lower and the upper limits of the fit range and of the background fit function. The same approach was considered using a method based on bin counting after subtracting the background estimated from a fit of the sidebands to test the description of the line shape of the signal. The uncertainty was assigned by calculating the RMS of the distribution of the signal yields obtained from all these variations. The uncertainty for the D-meson measurements ranges from about 2\% to 10\% depending on the $\pt$ interval and the particle species. The uncertainty for the $\XicPlus$ baryon is about $9\%$ in $3<\pt<4$ GeV$/c$. For the $\Lambdac$-baryon signal in the interval $0<\pt<1$ GeV$/c$, additional trials to further test the stability of the background parametrisation were included, similarly to what was done for the $\SigmacZeroPlusPlus$-baryon signal measurement in Ref.~\cite{Acharya:2021vpo}. The combinatorial background was described with a template distribution multiplied by a parabola. The template was obtained by recalculating the invariant mass after rotating a daughter track similarly to what was done for the $\Dzero$-meson signal extraction at low $\pt$ (see Section~\ref{sec:analysis}). The uncertainty for the $\Lambdac$-baryon raw yield measurement in $0<\pt<1$ GeV$/c$ was estimated to be about 12\% and 11\% in the analyses of the $\pKpi$ and $\pKzeroS$ channels, respectively. 

The systematic uncertainty of the track reconstruction efficiency accounts for possible discrepancies between data and MC in the TPC-ITS track prolongation efficiency and in the selection efficiency due to track-quality criteria. The per-track systematic uncertainty was evaluated by varying the track-quality selection criteria in the TPC detector and by comparing the prolongation probability of the TPC tracks to the ITS hits in data and MC simulations. They are subsequently propagated to the charm-hadron candidates via their decay kinematics evaluated with MC simulations. For the D mesons, this source introduces an uncertainty that grows with increasing $\pt$ from about 4\% at low transverse momenta up to about 7\% in the highest $\pt$ intervals. For the $\Lambdac$- and $\XicPlus$-baryon reconstruction, the value of this uncertainty is about 5\% and 6\%, respectively.

The systematic uncertainty related to the selection efficiency was studied by repeating the full analyses with varied selection criteria compared to the reference ones, resulting in a noticeable modification of the efficiencies, raw yield, and background values. The magnitude for this source of uncertainty was then assigned considering the dispersion and the shift of the production cross sections with respect to the reference one. The value of this uncertainty generally decreases with increasing $\pt$. For the D mesons, it ranges from roughly 10\% at low $\pt$ to about 2\% at high $\pt$. In the case of the $\Dzero$-meson analysis at $\pt<1.5$ GeV$/c$, where no selections on the displaced decay vertex topologies are applied, the stability was tested against variations of the single-track $\pt$ selection, and no systematic effect was observed. With a similar procedure, a 10\% uncertainty was assigned in the $\Lambdac\to\pKpi$ analysis. The uncertainties for the $\LambdacTopKzeroS$ and the $\XicPlus$-baryon selection amount to 5\% and 3\%, respectively.

The systematic uncertainty of the PID selection efficiency for the D-meson species was estimated by comparing the PID selection efficiencies in the data and in the simulation for pions and kaons. A pure sample of pions was isolated considering decays of strange hadrons. To evaluate the systematic uncertainty of the kaon identification using the TPC information, a pure sample of kaons was isolated applying a strict selection on the TOF information, and vice versa. The resulting per-track uncertainty was then propagated to the D mesons using their decay kinematics. As an additional test, the analyses were repeated without any PID requirement on the candidate D-meson daughter tracks. The systematic uncertainty of the PID selection efficiency was found to be negligible in the analysed $\pt$ intervals. A similar strategy was adopted to estimate an uncertainty related to the PID selections adopted for the $\LambdacTopKzeroS$ candidate reconstruction before the BDT application, and a 1\% uncertainty was assigned in this case. For the $\LambdacTopKpi$ reconstruction, this uncertainty was estimated as done in Refs.~\cite{ALICE:2020wla,Acharya:2021vpo, ALICE:2021npz}, namely repeating the cross section measurement using the threshold probability criterion for the application of the Bayesian PID approach. The assigned uncertainty in this case is around 5\%.

Possible differences between the generated hadron $\pt$ distributions from simulation and the spectra observed in data influence the calculation of the $c_{\Delta y}({\rm Acc}\times \varepsilon)$ factor and introduce an additional source of systematic uncertainty. To estimate this effect, the simulated $\pt$ distributions were weighted to match the charm-hadron $\pt$ spectra from different model calculations. The $c_{\Delta y}({\rm Acc}\times \varepsilon)$ correction factor was recomputed using the weighted spectra and an uncertainty was assigned based on the difference between the charm-hadron production cross sections obtained with the default and the weighted $c_{\Delta y}({\rm Acc}\times \varepsilon)$.
For D mesons, this uncertainty was estimated using FONLL as an alternative with respect to PYTHIA 8 to simulate the D-meson $\pt$ distributions. It was found to range between 1\% and 5\% for $\pt < 2$ GeV$/c$ depending on the D-meson species, and to be negligible at higher $\pt$.
For the $\Lambdac$-baryon analyses, the reconstruction efficiency was recalculated by considering the shape from the PYTHIA 8 Monash tune, which is the default adopted in the MC simulations, and that of PYTHIA 8 CR-BLC Mode 2. The assigned uncertainty is about 1\%, corresponding to the maximum variation observed in the efficiency-times-acceptance quantity. In a similar way, spectra from Catania, SHM+RQM, QCM, and PYTHIA 8 CR-BLC Mode 0, 2, 3 were considered for recalculating the reconstruction efficiency for the $\XicPlus$-baryon signal and an uncertainty of 3\% was assigned. 

To provide an unbiased efficiency correction, the real detector geometry must be accurately implemented in the MC simulations. Recent investigations based on the reconstruction of photons in the material of the ALICE apparatus highlighted a difference of 24\% and 7\% in the current MC simulations for the material budget description of the silicon pixels in the ITS and the thermal shield in the TPC inner containment vessel~\cite{ALICE:2023kzv}. Such discrepancies may introduce biases in the description of multiple scattering and absorption probability in the material. This may affect in particular the heaviest particles (i.e.~protons), for which such effects are expected to be more significant. The $c_{\Delta y}({\rm Acc}\times \varepsilon)$ term was recomputed in dedicated MC productions covering the material budget variations mentioned above. A systematic uncertainty was assigned according to the difference to the $c_{\Delta y}({\rm Acc}\times \varepsilon)$ correction factor calculated with the default MC simulations used to correct the data. A value of 4\% corresponding to the maximum observed variation was assigned to the measurement of the $\Lambdac\to\pKzeros$ and 
$\XicPlus$ signals. No significant effects were observed for the prompt $\Lambdac\to\pKpi$ and $\Dzero$-meson signal reconstruction, given the looser selections employed in the reconstruction, as well as in the other D-meson analyses. Therefore, no systematic uncertainty was assigned in these cases.

The systematic uncertainty related to the correction for the fraction of charm hadrons originating from beauty-hadron decays accounts for the uncertainties accompanying the several ingredients adopted in the calculation of $f_{\rm prompt}$, as described in Section~\ref{sec:analysis}. The uncertainties for the D mesons were estimated by varying the FONLL parameters, namely the b-quark mass, the renormalisation ($\mu_{\mathrm{R}}$), and factorisation scales ($\mu_{\mathrm{F}}$), as discussed in Ref.~\cite{ALICE:2019nxm}. 
The assigned values range from 1\% to 9\% depending on the $\pt$ interval and the D-meson species. Similarly, those assigned to the $\Lambdac$-baryon analyses range from 1\% to 2\%.
For the estimation of the fraction of prompt $\XicPlus$ baryons, the uncertainties of the non-prompt $\Lambda_\mathrm{c}^+$-baryon cross section were taken into account. In addition, in order to account for possible differences between the $\rm \Xi_\mathrm{c}^0/\Lambda_\mathrm{c}^+$ and $\rm \Xi_\mathrm{b}^-/\Lambda_\mathrm{b}^0$ ratios, the latter cross section ratio was scaled up by a conservative factor of $\rm2$. The lower uncertainty of the ratio was obtained by scaling it down by a factor of $\rm0.05$ to capture the $\rm \Xi_{b}^-/\Lambda_{b}^0$ value measured at forward rapidity by the LHCb Collaboration~\cite{LHCb:2019}. The upper and lower limits of the uncertainty band of $f_\mathrm{prompt}$ were taken as the quadratic sum of the two described contributions. The systematic uncertainty assigned to the $\XicPlus$-baryon measurement is around 2\%.

\section{Results}
\label{sec:results}

\subsection{Prompt D-meson $\boldsymbol{\pt}$-differential cross sections}
\label{sec:DmesonCrossSection}

\begin{figure}[t!]
    \centering
    \includegraphics[width=0.7\textwidth]{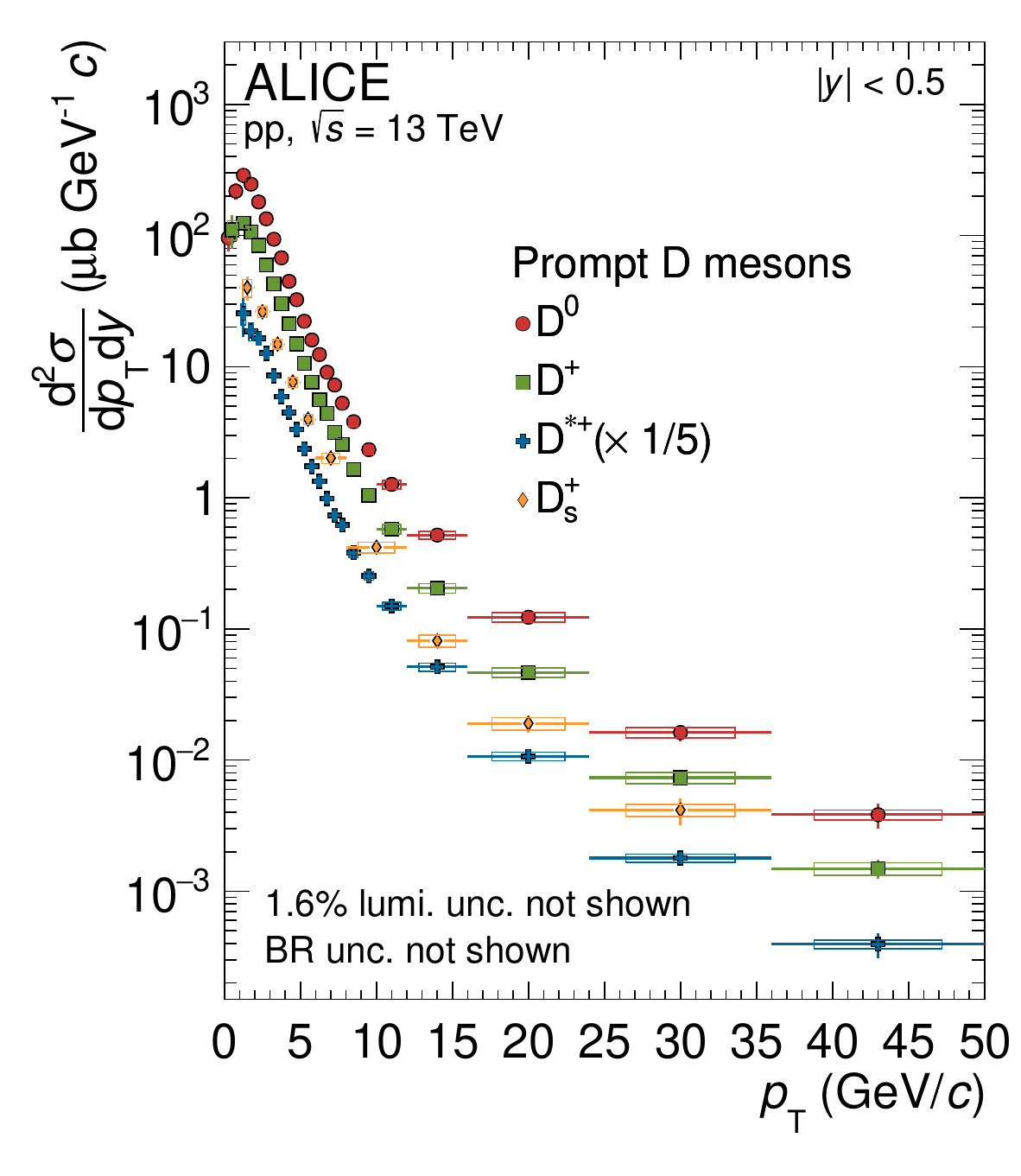}
    \caption{$\pt$-differential production cross sections of prompt $\Dzero$, $\Dplus$, $\Dstar$, and $\Ds$ mesons at midrapidity ($|y|<0.5$) in pp collisions at $\s=13$~TeV.
    The
    vertical bars and boxes report the statistical and systematic uncertainties, respectively. The total systematic uncertainties reported in the plots do not include the contributions of the luminosity and the branching ratio, which are reported separately. 
The measured prompt $\Dstar$-meson production cross section was scaled by a factor of $1/5$ for better visibility reasons.}
    \label{fig:mesonCrossSection}
\end{figure}

The $\pt$-differential production cross sections of prompt $\Dzero$, $\Dplus$, $\Dstar$, and $\Ds$ mesons at midrapidity ($|y|<0.5$) in \pp collisions at $\s=13$~TeV are shown in Fig.~\ref{fig:mesonCrossSection}. Statistical and systematic uncertainties are depicted as vertical lines and empty boxes, respectively. The $\pt$-differential production cross sections of prompt $\Dzero$ and $\Ds$ mesons are compatible with previously published results~\cite{Acharya:2021vpo, ALICE:2021npz}, providing an extended $\pt$ coverage and a finer $\pt$ binning.

\begin{figure}[]
    \centering
    \includegraphics[width=0.46\textwidth]{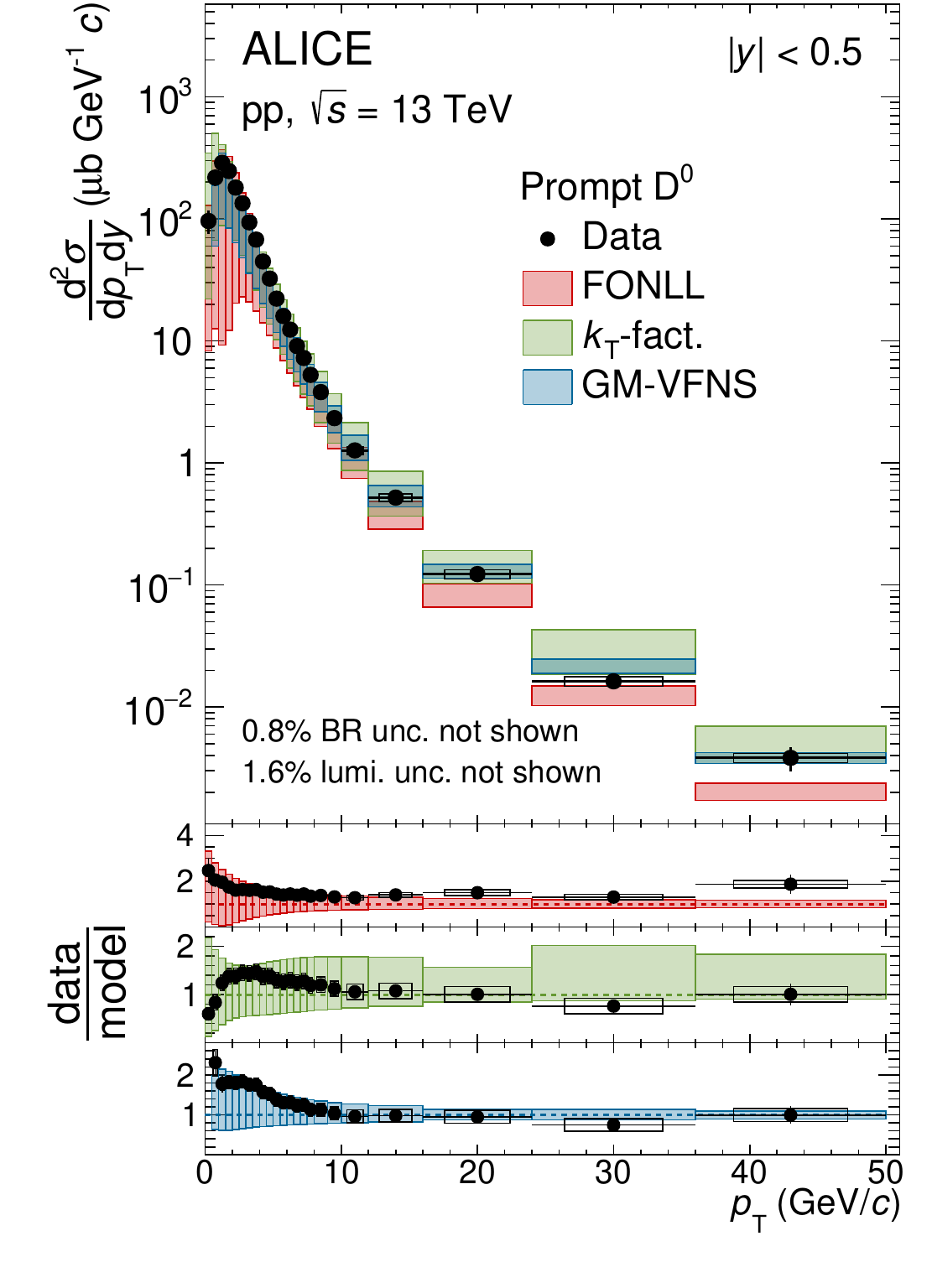}
    \includegraphics[width=0.46\textwidth]{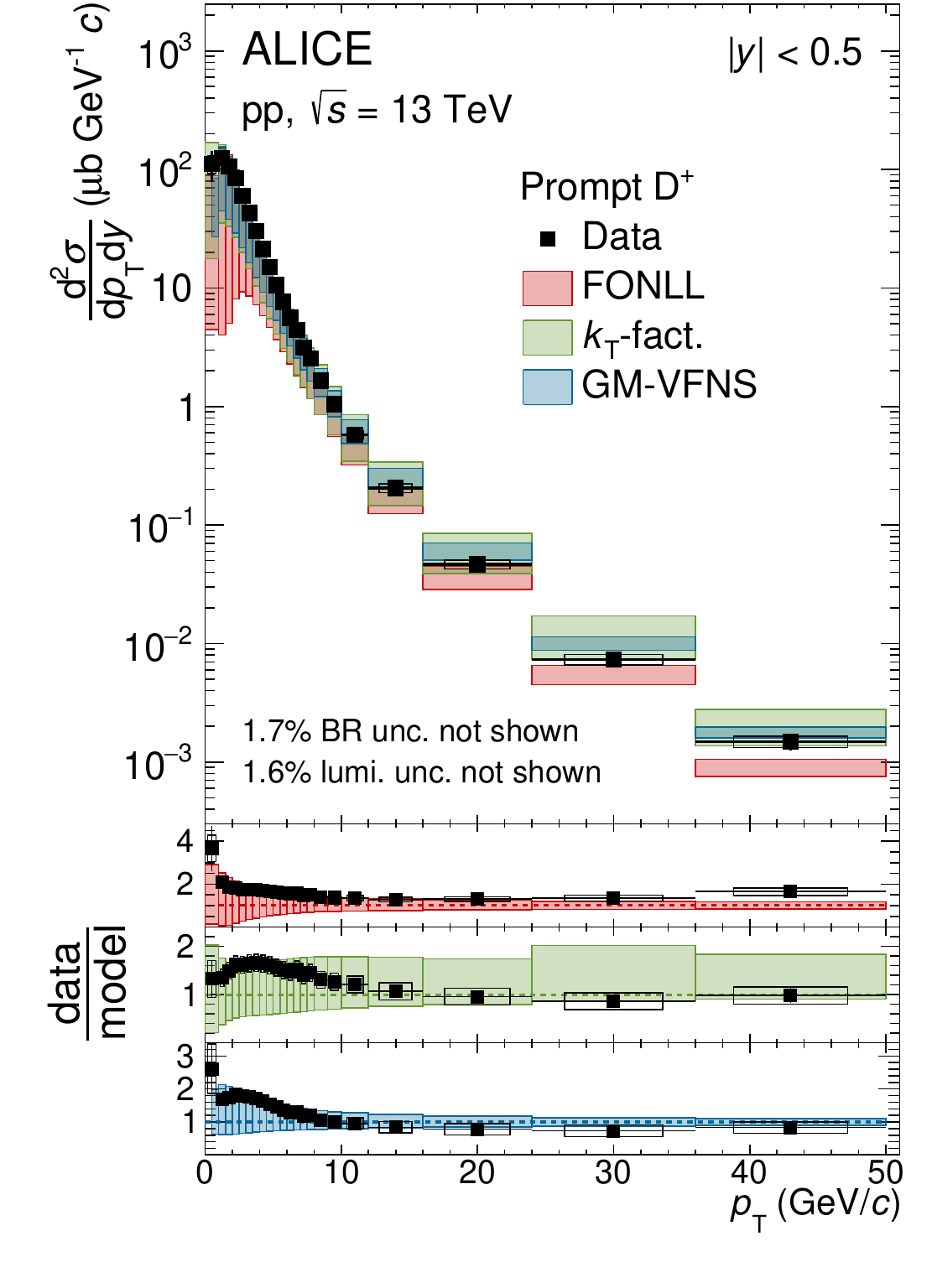}
    \includegraphics[width=0.46\textwidth]{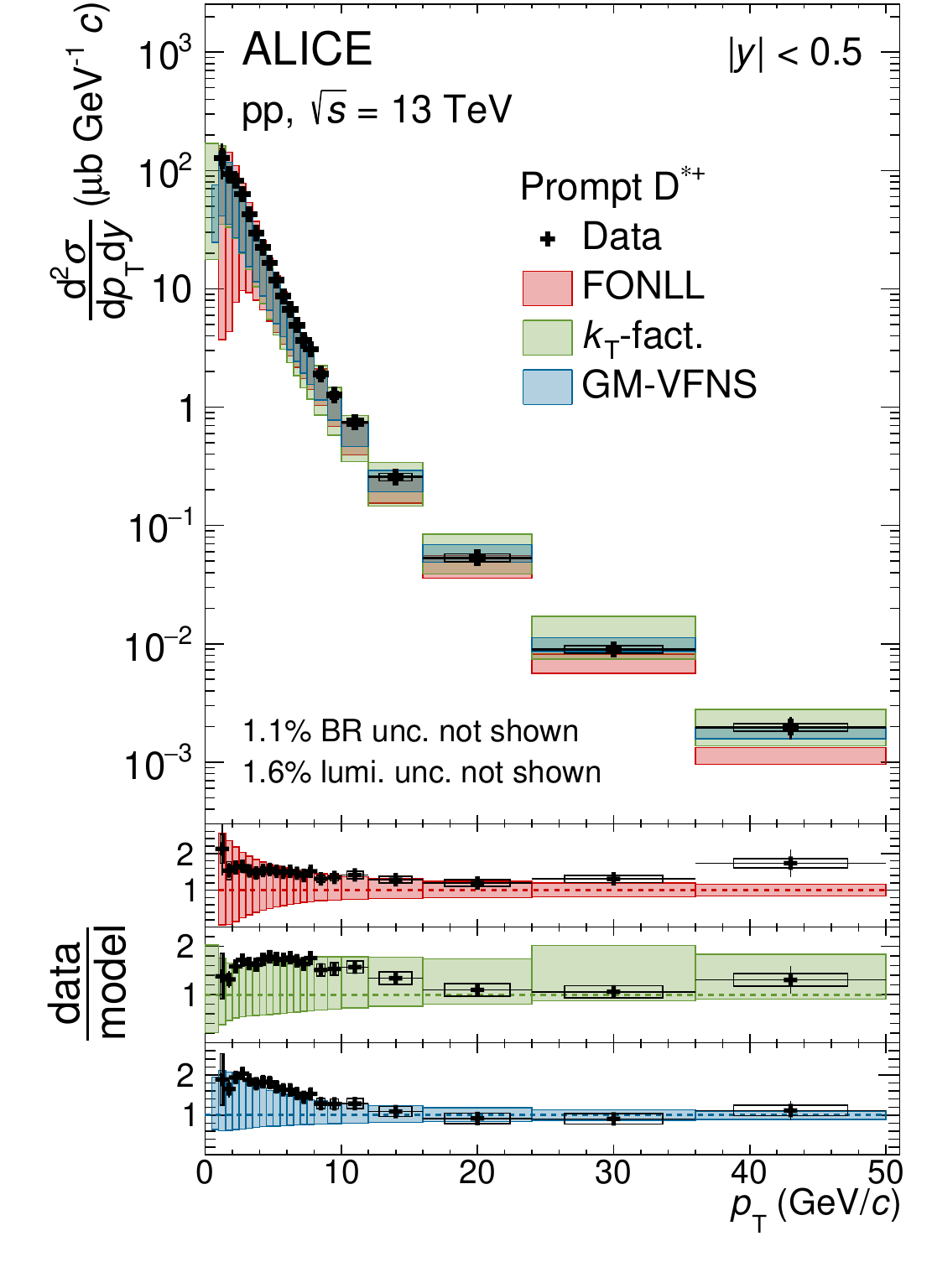}
    \includegraphics[width=0.46\textwidth]{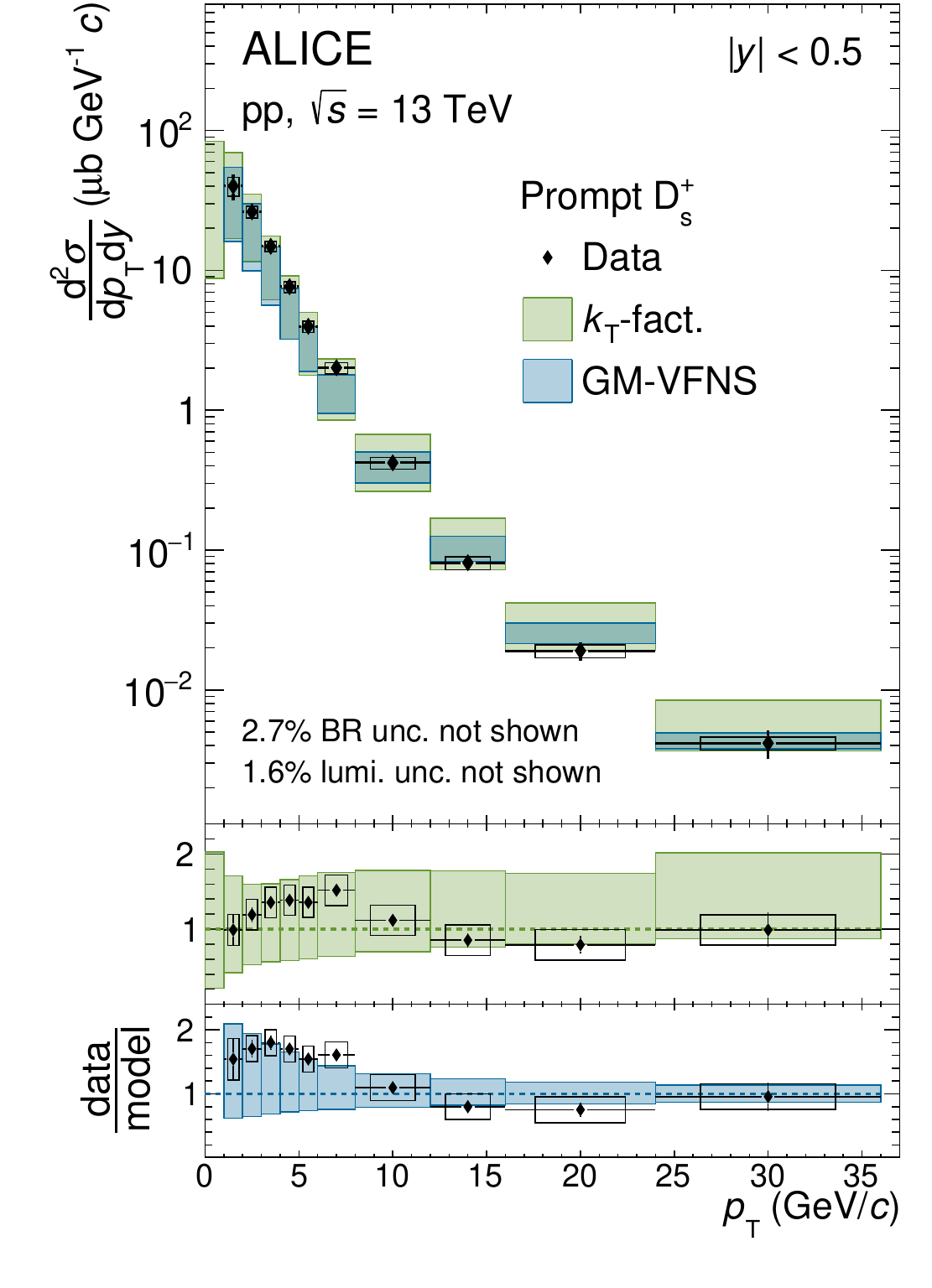}
    \caption{$\pt$-differential production cross sections for prompt D mesons in comparison with pQCD calculations: FONLL~\cite{Cacciari:1998it, Cacciari:2012ny}, GM-VFNS~\cite{Benzke:2017yjn, Kramer:2017gct}, and $k_{\mathrm{T}}$-factorisation~\cite{Guiot:2021vnp}. The uncertainty on the predictions is depicted as coloured boxes.
    The ratios of the data to the theoretical predictions are shown in the lower part of each panel. The statistical (systematic) uncertainties are depicted as vertical bars (boxes).} 
    \label{fig:mesonCrossSection_vs_models}
\end{figure}

The measured prompt D-meson $\pt$-differential cross sections are compared in Fig.~\ref{fig:mesonCrossSection_vs_models} with results from pQCD calculations performed with different schemes: FONLL~\cite{Cacciari:1998it, Cacciari:2012ny} (not available for the $\Ds$ mesons), GM-VFNS framework~\cite{Benzke:2017yjn, Kramer:2017gct}, and a calculation based on the $k_{\mathrm{T}}$-factorisation approach~\cite{Guiot:2021vnp}. The theoretical uncertainties of the predictions based on these calculations are depicted as boxes.

In the case of FONLL, the theoretical uncertainty includes the variation of the factorisation and renormalisation scales, the variation of the charm-quark mass value, and the uncertainties on the PDFs employed on the calculations from the reference case (CTEQ6.6~\cite{Pumplin:2002vw}), as discussed in Refs.~\cite{Cacciari:2015fta, Cacciari:2012ny}. The D-meson fragmentation fractions, $f(\mathrm{c}\rightarrow\mathrm{D})$ adopted in the FONLL calculations were taken from Ref.~\cite{Gladilin:2014tba}.

The configuration of the GM-VFNS calculations was the same as the one employed in Ref.~\cite{ALICE:2021mgk}. The CTEQ6.6 PDFs~\cite{Pumplin:2002vw} are used as default. The factorisation and renormalisation scales $\mu_{\mathrm{F}}$ and $\mu_{\mathrm{R}}$ used for the central values correspond to those adopted in Ref.~\cite{ALICE:2019nxm}. The calculations based on GM-VFNS were performed in the same $\pt$ intervals as the measurements, except for the first interval of the $\Dzero$- and $\Dplus$-meson results that start from 0.5~GeV$/c$. According to the authors, calculations down to $\pt=0$ at such values of $\s$ were compromised by numerical instabilities.

A variable-flavour-number scheme was adopted in the case of QCD calculations within the $k_{\mathrm{T}}$-factorisation framework. The $k_{\mathrm{T}}$-factorisation calculations overcame the factorisation scheme employed in the predictions reported in Refs.~\cite{ALICE:2021mgk, ALICE:2017olh} for the estimation of the unintegrated PDFs and the final production cross section.
As discussed in Ref.~\cite{Guiot:2021vnp}, the variable-flavour-number scheme was demonstrated to be more efficient, resumming to all orders some large
logarithms $\ln(\pt^2/ m_{\rm Q}^2)$ thanks to the heavy-quark distribution function.
The authors also demonstrated the importance of excitation quantum processes to correctly describe the production of charm quarks. The theoretical uncertainties were estimated by varying the $\mu_{\mathrm{F}}$ scale.

In analogy to what was observed at $\s = 5.02$~TeV~\cite{ALICE:2021mgk, ALICE:2019nxm} and $\s = 7$~TeV~\cite{ALICE:2017olh}, the pQCD calculations implementing the factorisation approach describe within uncertainties the measured D-meson production cross sections in pp collisions at LHC energies. In the case of FONLL calculations, the data systematically lie on the upper edge of the theoretical uncertainty up to $\pt=36$ GeV$/c$. In the interval $36<\pt<50$~GeV$/c$, the data are in agreement with the upper edge of the calculations within about 2$\sigma$. The central values of the calculation in the GM-VFNS framework slightly underestimate (overestimate) the data in the low-(high-)$\pt$ region.
The $k_{\mathrm{T}}$-factorisation calculation is in good agreement with the data in the low-$\pt$ and intermediate-$\pt$ region. However, the ratio between the measurement and the calculation is not flat as a function of $\pt$, and the model prediction tends to overestimate the data at high $\pt$.

\subsection{Prompt charm-baryon measurements down to low $\boldsymbol{\pt}$}
\label{sec:baryonCrossSection}
\begin{figure}[t!]
    \centering
    \includegraphics[width=0.8\textwidth]{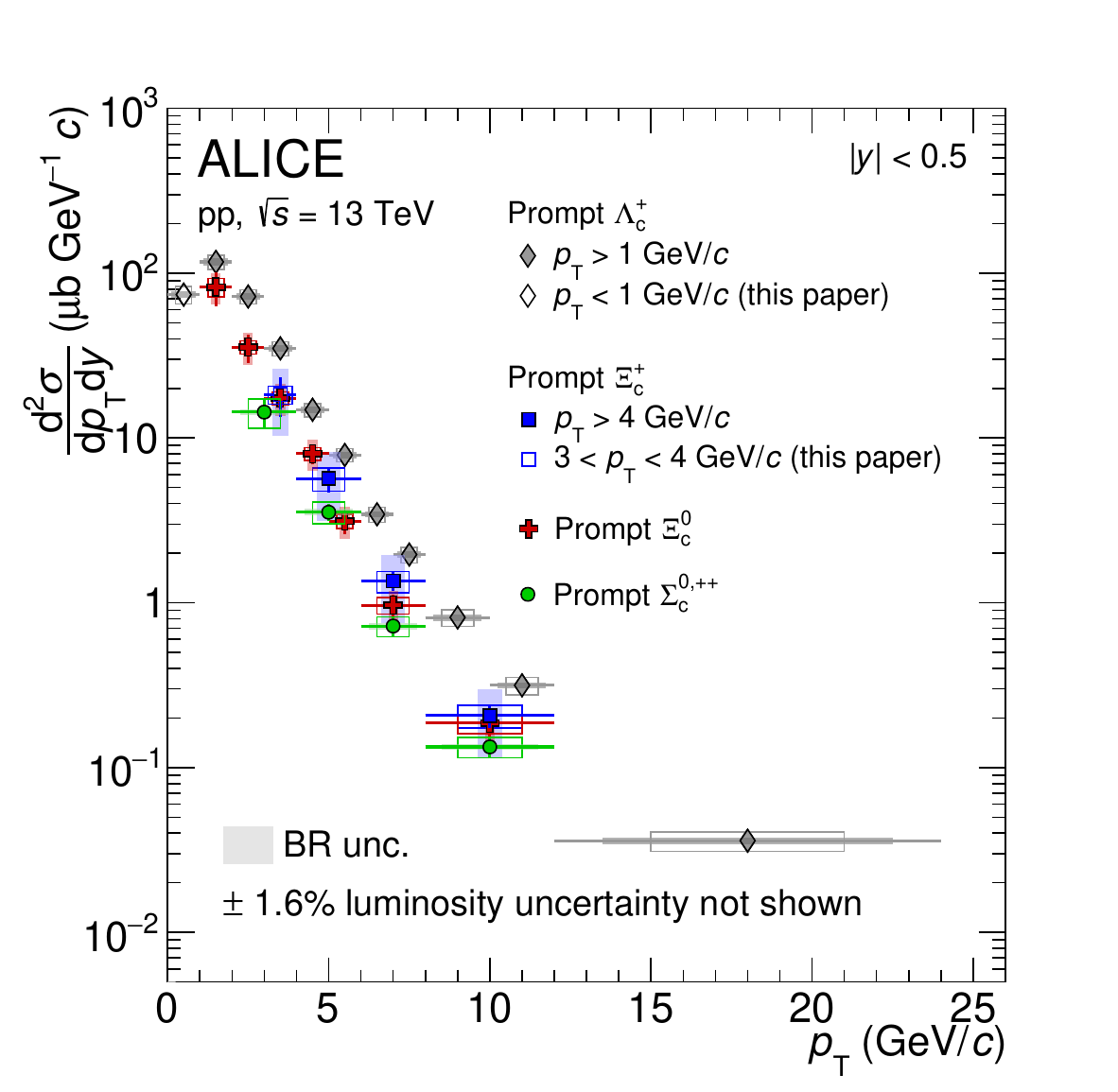}
    \caption{$\pt$-differential production cross sections of prompt $\Lambdac$, $\XicZero$, $\XicPlus$, and $\SigmacZeroPlusPlus$ baryons at midrapidity ($|y|<0.5$) in pp collisions at $\s=13$~TeV~\cite{Acharya:2021vpo, Acharya:2021vjp}. The statistical (systematic) uncertainties are shown as vertical bars (boxes). The shaded boxes report the BR uncertainty.}
    \label{fig:baryonCrossSection}
\end{figure}

A compilation of the $\pt$-differential production cross sections of prompt charm baryons measured at midrapidity ($|y|<0.5$) in pp collisions at $\s=13$~TeV is shown in Fig.~\ref{fig:baryonCrossSection}.
The $\pt$-differential cross section of prompt $\Lambdac$ baryons for $\pt>1$~GeV$/c$ and that of $\SigmacZeroPlusPlus$ baryons are those published in Ref.~\cite{Acharya:2021vpo}.

The new measurement of the prompt $\Lambdac$-baryon production cross section in the interval $0<\pt<1$~GeV$/c$ was obtained by averaging the $\Lambdac$-baryon cross sections measured via both the $\Lambdac\to\pKpi$ and $\Lambdac\to\pKzeros$ decay channels, which are compatible within about $ 2\sigma$. The weights adopted in the average were calculated using the relative uncertainties of the sources assumed as uncorrelated between the two decay channels, and accounting for the partial correlation between their branching ratios. The strategy was the same as the one followed in Ref.~\cite{Acharya:2021vpo}.

The production cross section of the prompt $\XicPlus$ baryons in the interval $3<\pt<4$~GeV$/c$ is the first measurement of the prompt $\XicPlus$-baryon cross section down to $\pt=3$~GeV$/c$, and it extends the $\XicPlus$-baryon results already published in Ref.~\cite{Acharya:2021vjp}. 
Within the current uncertainties, especially those related to the branching ratio of the $\XicPlus$ baryons, the $\pt$-differential cross sections of the $\XicPlus$ and $\XicZero$ baryons do not show any significant charge dependence in the common range of the measurements. 

\subsection{Charm-hadron cross section ratios}
\label{sec:ratios}

\subsubsection{Meson-to-meson ratios}
\begin{figure}[t!]
    \centering
    \includegraphics[width=0.875\textwidth]{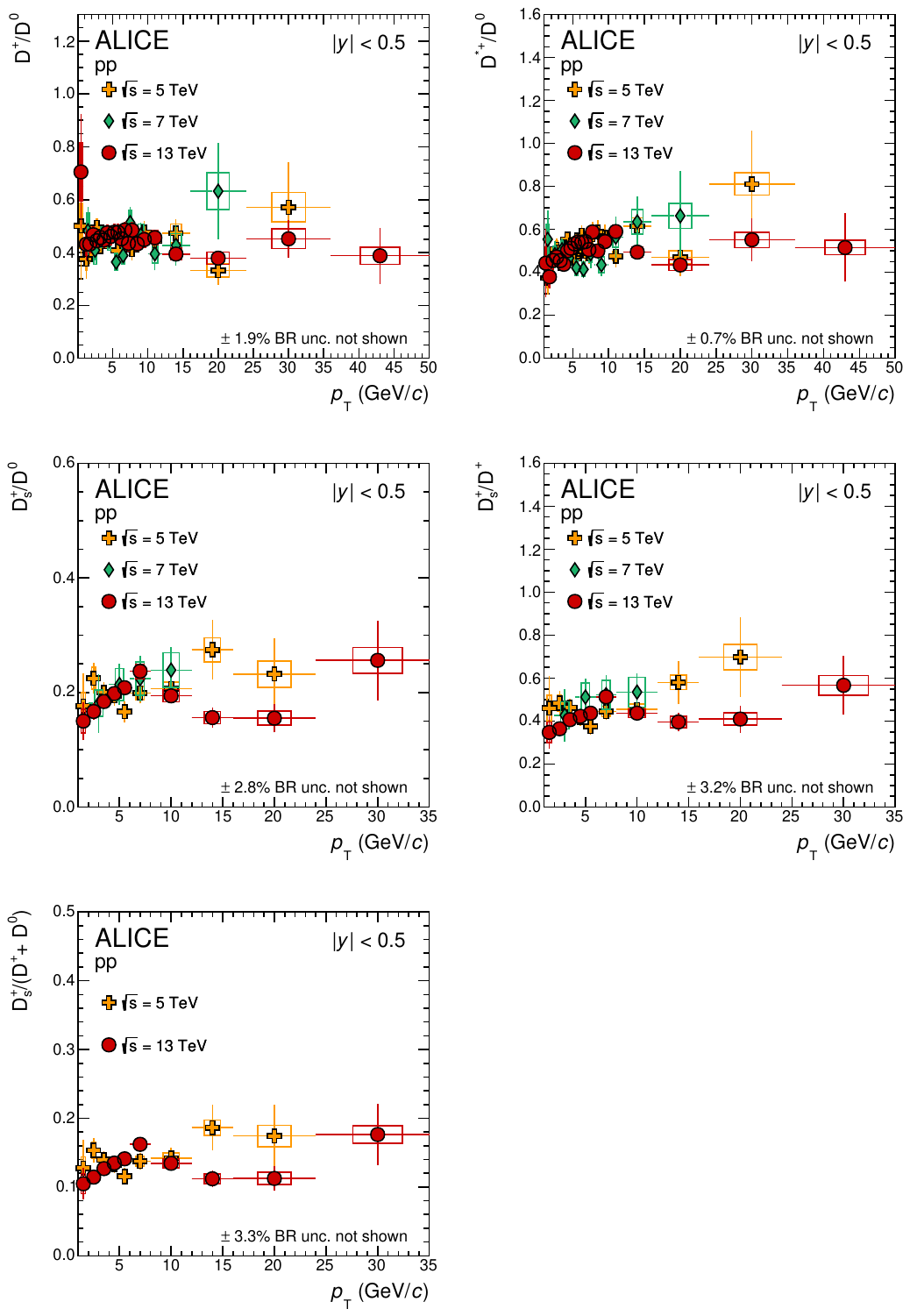}
    \caption{Ratios of production cross sections as a function of $\pt$ of prompt $\Dplus/\Dzero$, $\Dstar/\Dzero$, $\Ds/\Dzero$, $\Ds/\Dplus$, and $\Ds/(\Dzero+\Dplus)$ mesons in pp collisions at $\s = 5.02$~TeV~\cite{ALICE:2021mgk, ALICE:2019nxm}, $\s = 7$~TeV~\cite{ALICE:2017olh}, and $\s = 13$~TeV. Vertical bars (boxes) report the statistical (systematic) uncertainties.}
    \label{fig:MesonToMesonRatio-5-7-13}
\end{figure}

The ratios of the $\pt$-differential cross sections of prompt $\Dplus$, $\Dstar$, and $\Ds$ mesons to prompt $\Dzero$ mesons as well as prompt $\Ds$ to prompt $\Dplus$ meson in pp collisions at $\s = 13$~TeV are reported in Fig.~\ref{fig:MesonToMesonRatio-5-7-13}. In the evaluation of the systematic uncertainties, the contributions of the yield extraction and selection efficiency were considered as uncorrelated, while those of the prompt-fraction correction, the tracking efficiency and the luminosity were treated as fully correlated among the different D-meson species.

The results in \pp collisions at $\s=13$~TeV are compared with the ones obtained at $\s = 5.02$~TeV~\cite{ALICE:2021mgk, ALICE:2019nxm} and $\s = 7$~TeV~\cite{ALICE:2017olh}. A hint of increase with $\pt$ is visible for the $\Ds/\Dzero$, $\Ds/\Dplus$, and $\Ds/(\Dzero + \Dplus)$ ratios in the interval $\pt<8$ GeV/$c$. In the other cases, the $\pt$-differential cross section ratios do not show any significant dependence on the $\pt$ of the D mesons considering the uncertainties.
No appreciable dependence on the collision energy is observed within the current experimental uncertainties. This suggests common fragmentation functions of charm quarks to pseudoscalar and vector mesons and to mesons with and without strange quark content at different LHC energies.

\subsubsection{Baryon-to-meson ratios}
\begin{figure}[t!]
    \centering
    \includegraphics[width=0.49\textwidth]{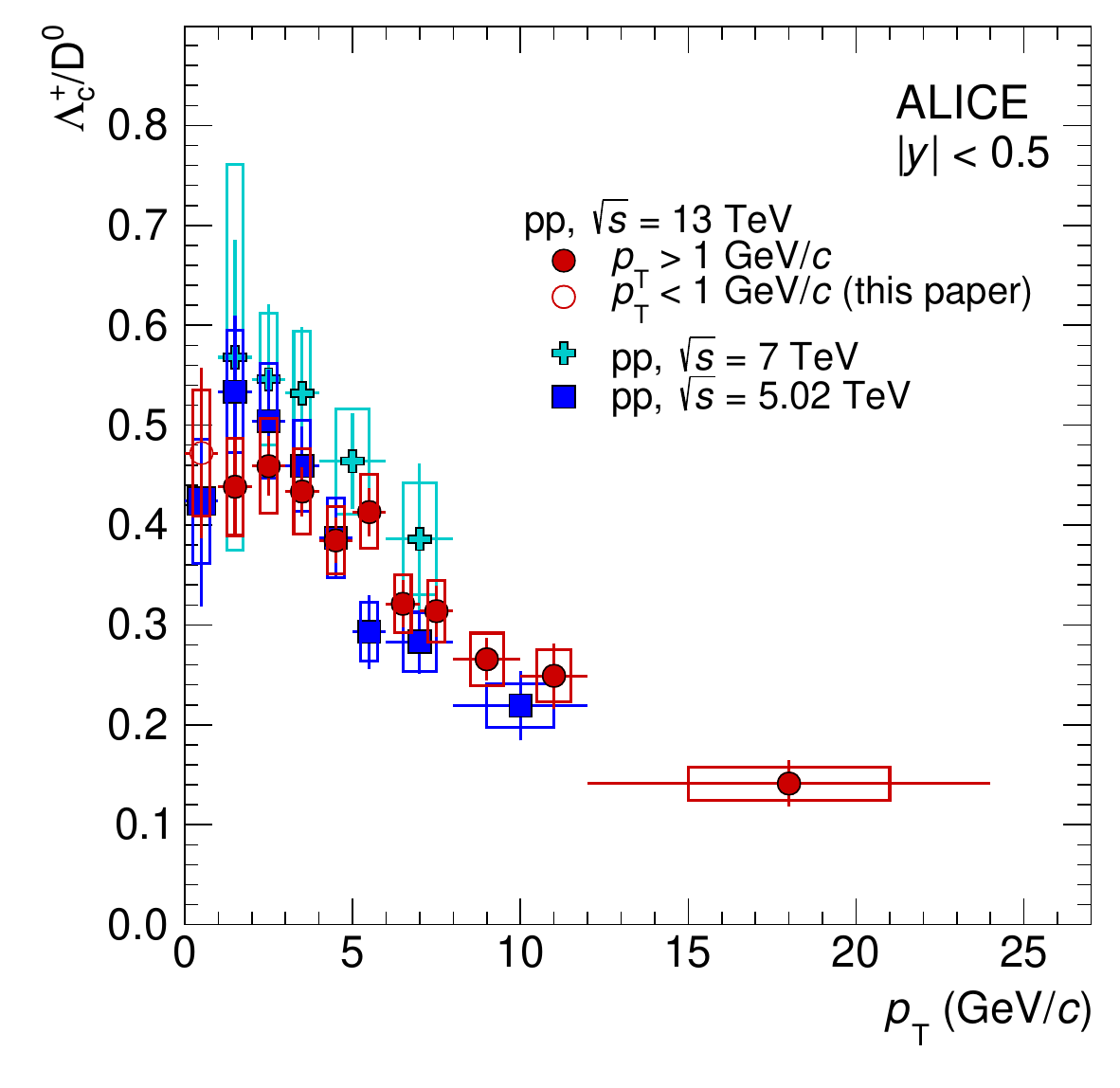}
    \includegraphics[width=0.49\textwidth]{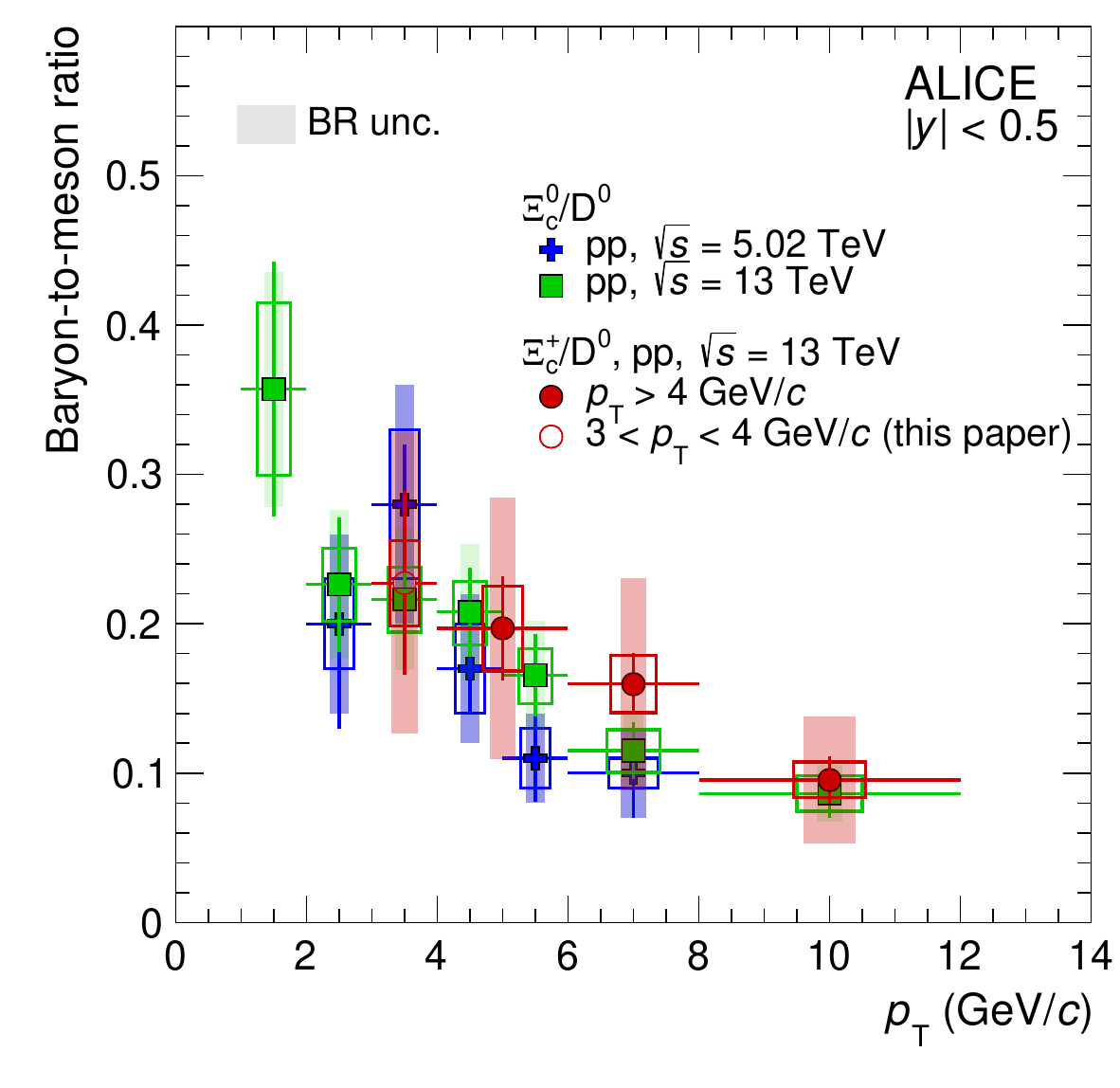}
    \includegraphics[width=0.49\textwidth]{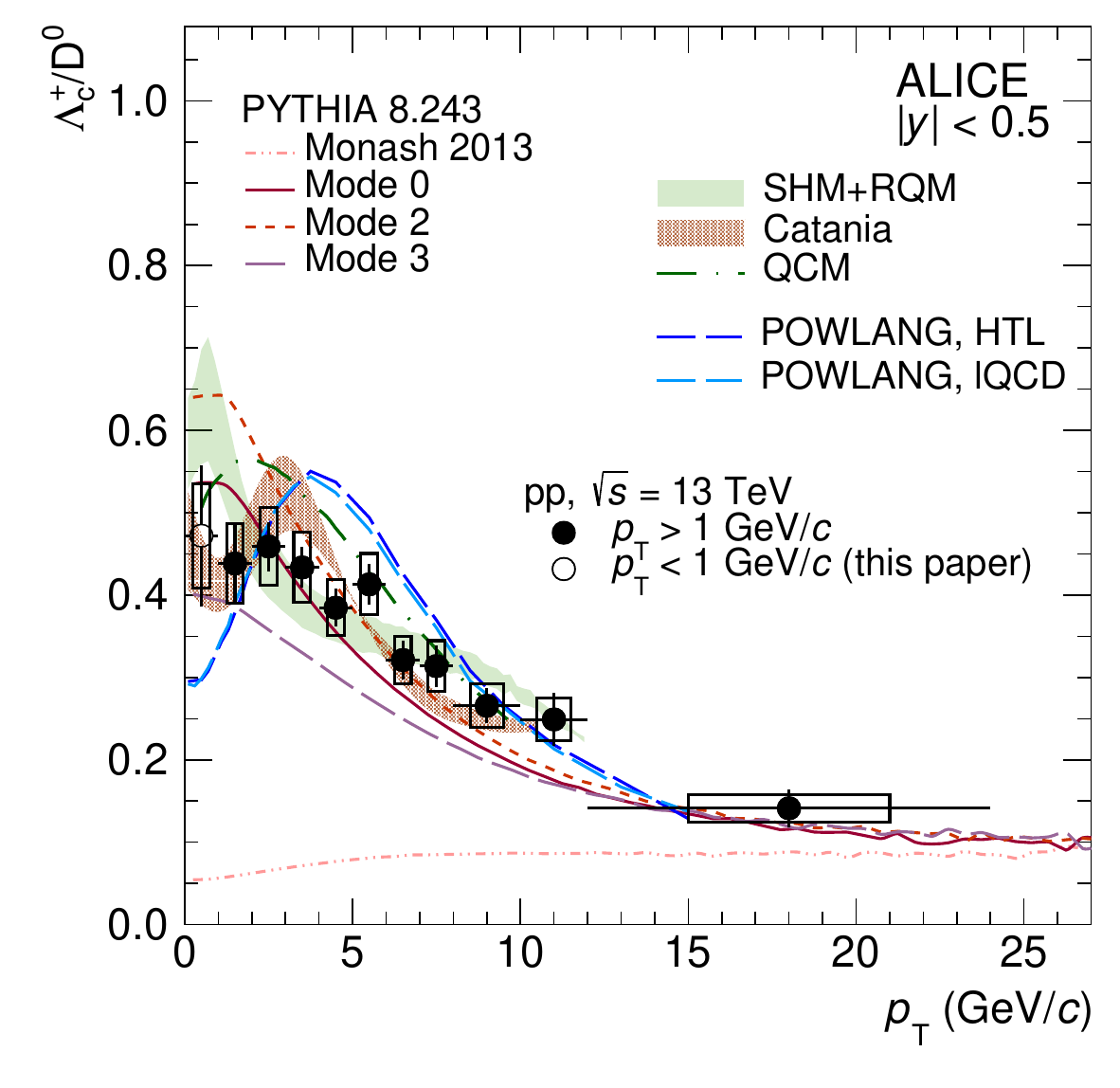}
    \includegraphics[width=0.49\textwidth]{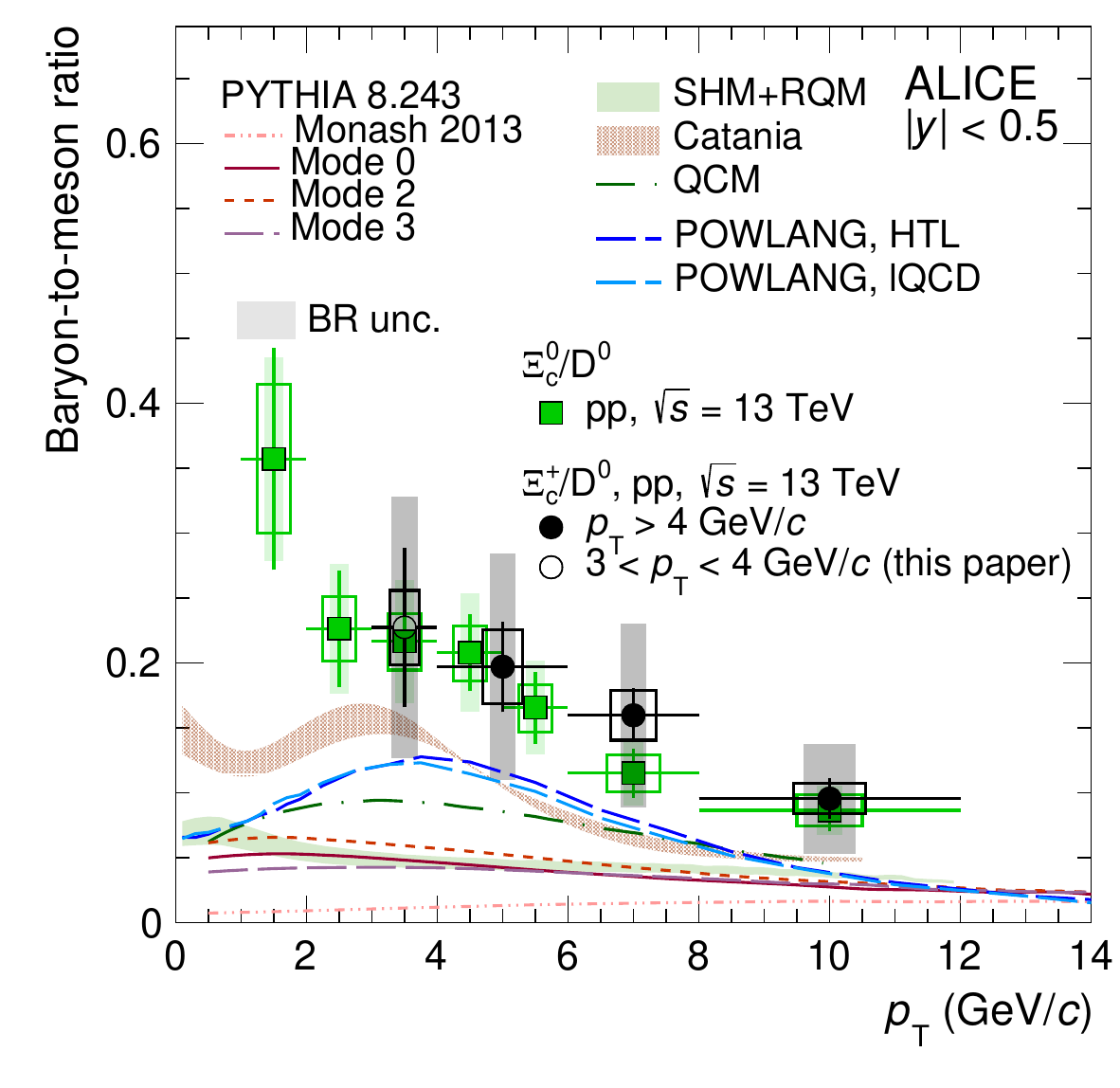}
    \caption{
    Top left: ratio between the $\pt$-differential cross sections at midrapidity ($|y|<0.5$) of prompt $\Lambdac$ baryons and $\Dzero$ mesons in pp collisions at $\s=5.02$ TeV~\cite{ALICE:2020wfu,ALICE:2020wla,ALICE:2022ych}, 7 TeV~\cite{ALICE:2017thy} and $13$~TeV~\cite{Acharya:2021vpo}.
    The measurement of $\Lambdac/\Dzero$ ratio in \pp collisions at $\s=13$ TeV for $\pt>1$ GeV$/c$ uses the $\Lambdac$-baryon cross section published in Ref.~\cite{Acharya:2021vpo}.
    Bottom left: ratio between the $\pt$-differential cross sections at midrapidity ($|y|<0.5$) of prompt $\Lambdac$ baryons and $\Dzero$ mesons in pp collisions at $\s=13$~TeV compared with the predictions from PYTHIA Monash tune~\cite{Skands:2014pea}, PYTHIA CR-BLC Mode 0, 2 and 3~\cite{Christiansen:2015yqa}, SHM+RQM~\cite{HE2019117}, Catania~\cite{Minissale:2020bif}, QCM~\cite{Song:2018tpv}, and POWLANG~\cite{Beraudo:2023nlq} models in pp collisions at $\s=13$~TeV.
    Top right: $\pt$-differential $\XicPlus/\Dzero$ ratio in pp collisions at $\s=13$~TeV and $\XicZero/\Dzero$ ratio in pp collisions at $\s=5.02$~TeV~\cite{Acharya:2021dsq} and $\s=13$~TeV~\cite{Acharya:2021vjp}. The $\XicPlus/\Dzero$ ratio in pp collisions at $\s=13$~TeV for $\pt>4$ GeV$/c$ uses the $\XicPlus$ published in Ref.~\cite{Acharya:2021vjp}. Statistical (systematic) uncertainties are reported as vertical bars (open boxes). The shaded boxes show the BR uncertainty.
    Bottom right: $\pt$-differential $\XicZero/\Dzero$ and $\XicPlus/\Dzero$ ratio in pp collisions at $\s=13$~TeV compared with the predictions from the models reported above.}
    \label{fig:baryonToMesonRatio}
\end{figure}
The ratio of the $\pt$-differential cross sections of the prompt $\Lambdac$ baryons to $\Dzero$ mesons at midrapidity ($|y|<0.5$) in pp collisions at $\s=13$~TeV is shown in the left panels of Fig.~\ref{fig:baryonToMesonRatio}.
The measurement at $\pt>1$~GeV$/c$ was performed with the prompt $\Lambdac$-baryon cross section published in Ref.~\cite{Acharya:2021vpo}, which was extended down to $\pt=0$ with the measurement in $0<\pt<1$~GeV$/c$ from this paper. The ratio was then obtained by using the prompt $\Dzero$-meson cross section reported in Section ~\ref{sec:DmesonCrossSection} as denominator.
In the ratio, the systematic uncertainties related to the tracking efficiency, the prompt fraction correction, and the luminosity were propagated as correlated, while those from other sources were treated as fully uncorrelated. 

The results were compared with several model calculations, namely different tunes of PYTHIA 8, the Catania and QCM models implementing quark recombination, and the SHM+RQM model based on statistical hadronisation with additional excited charm-baryon states. The uncertainty band  assigned to the predictions from Catania model is related to the variations on the width of the Wigner function used to calculate the probability of baryon formation. The uncertainty band assigned to the predictions from the SHM+RQM model accounts for the uncertainty on the branching ratios of resonance decays to ground-state charm hadrons. 
As discussed in Ref.~\cite{Acharya:2021vpo}, the measured baryon-to-meson ratio is underestimated  in the interval $\pt<5$~GeV$/c$ by a factor 4$-$5 by the prediction from PYTHIA 8 with the Monash tune, which is not able to describe the $\pt$ dependence of the measurement. On the other hand, the predictions from the PYTHIA 8 CR-BLC Mode 0, 2, 3, SHM+RQM, and Catania models describe the measurement within uncertainties in the full range ($\pt>0$), but the current precision and granularity of the measurement does not allow one model to be favoured over the others. The prediction from the QCM model is compatible with the measured $\Lambdac/\Dzero$ ratio within 2$\sigma$, tending to overestimate it in the interval $3<\pt<8$ GeV$/c$.

In the bottom-left panel of Fig.~\ref{fig:baryonToMesonRatio}, the prompt $\Lambdac/\Dzero$ baryon-to-meson ratios in pp collisions at $\s=13$~TeV are compared also with the predictions from the POWLANG model~\cite{Beraudo:2023nlq}. In these calculations, the formation of a small, deconfined, and expanding fireball even in pp collisions is assumed, where the same in-medium hadronization mechanism developed for heavy-ion collisions is employed. In this model, the formation of charm baryons is promoted by the recombination of charm quarks with light diquark excitations in the hot medium. The first predictions from this model were provided in pp collisions at $\s=5.02$~TeV for the prompt $\Lambdac/\Dzero$ ratio at midrapidity in Ref.~\cite{Beraudo:2023nlq}, employing transport coefficients calculated by weak-coupling (Hard-Thermal-Loop, HTL) and the most recent lattice-QCD calculations~\cite{Altenkort:2023oms}. 
The model predictions were found to qualitatively describe the measurement, tending to overestimate the magnitude of the ratio in the interval $3<\pt<8$ GeV$/c$. In the bottom-left panel of Fig.~\ref{fig:baryonToMesonRatio}, the same comparison for the prompt $\Lambdac/\Dzero$ ratio in pp collisions at $\s=13$~TeV, extended down to $\pt=0$, is shown. In particular, in the interval $0<\pt<1$ GeV$/c$ the prediction tends to underestimate the measurement. Overall, the model does not describe the $\pt$ dependence of the prompt $\Lambdac/\Dzero$ ratio in pp collisions at $\s=13$~TeV.

The $\Lambdac/\Dzero$ ratio in pp collisions at $\s=13$~TeV is also compared to the same results obtained in pp collisions at $\s=5.02$~TeV~\cite{ALICE:2020wfu,ALICE:2020wla,ALICE:2022ych} and $\s=7$~TeV~\cite{ALICE:2017thy}. Within the current uncertainties, no significant energy dependence is observed in pp collisions at midrapidity at the LHC.

In the top-right panel of Fig.~\ref{fig:baryonToMesonRatio}, the $\pt$-differential ratio of prompt $\XicPlus/\Dzero$ at midrapidity in pp collisions at $\s=13$~TeV is reported. This result was obtained from the $\XicPlus$-baryon measurement in $\pt>4$~GeV$/c$ published in Ref.~\cite{Acharya:2021vjp} and it was extended with the $\XicPlus$-baryon measurement in $3<\pt<4$~GeV$/c$ reported in this paper. The same strategy as for the $\Lambdac/\Dzero$ measurement was used for the error propagation.
The results are also compared with the $\XicZero/\Dzero$ ratio at midrapidity in pp collisions at $\s=5.02$~TeV~\cite{Acharya:2021dsq} and $\s=13$~TeV. The $\XicPlus/\Dzero$ and $\XicZero/\Dzero$ ratios in pp collisions at $\s=13$~TeV are found to be compatible within uncertainties (top-right panel of Fig.~\ref{fig:baryonToMesonRatio}). Also in this case no significant energy dependence is observed for the $\pt$-differential baryon-to-meson ratio.

As visible in the bottom-right panel of Fig.~\ref{fig:baryonToMesonRatio}, only the Catania and POWLANG models are compatible with the measured $\XicPlus/\Dzero$ ratio within about 1$\sigma$. The other model predictions are compatible with the measurement within about 2$\sigma$, even if they systematically underestimate the $\XicPlus/\Dzero$ ratio for all $\pt$. None of these models is significantly disfavoured by the measurement. 
Given the better precision of the measured $\XicZero/\Dzero$ ratio, only the predictions from Catania, POWLANG and QCM coalescence models are in agreement within at most 3$\sigma$ with the measured $\XicZero/\Dzero$ ratio. On the other hand, a larger tension with SHM+RQM prediction and PYTHIA predictions is observed in all the $\pt$ intervals. As discussed in Ref.~\cite{ALICE:2022cop}, only coalescence models including contributions from strong decays of additional excited charm baryons can describe the measured productions of strange-charm baryons. This comparison suggests that the coalescence models provide the best description of the data. Meanwhile, the tension between the measured $\XicZero/\Dzero$ ratio and the SHM+RQM suggests that additional excited charm baryons predicted by the RQM are not enough to account for the $\XicZero$- and $\XicPlus$-baryon abundance, despite the good description of $\Lambdac/\Dzero$ ratio.

\subsubsection{Ratios of charm-hadron cross sections at different collision energies}

\begin{figure}[t!]
    \centering
    \includegraphics[width=\textwidth]{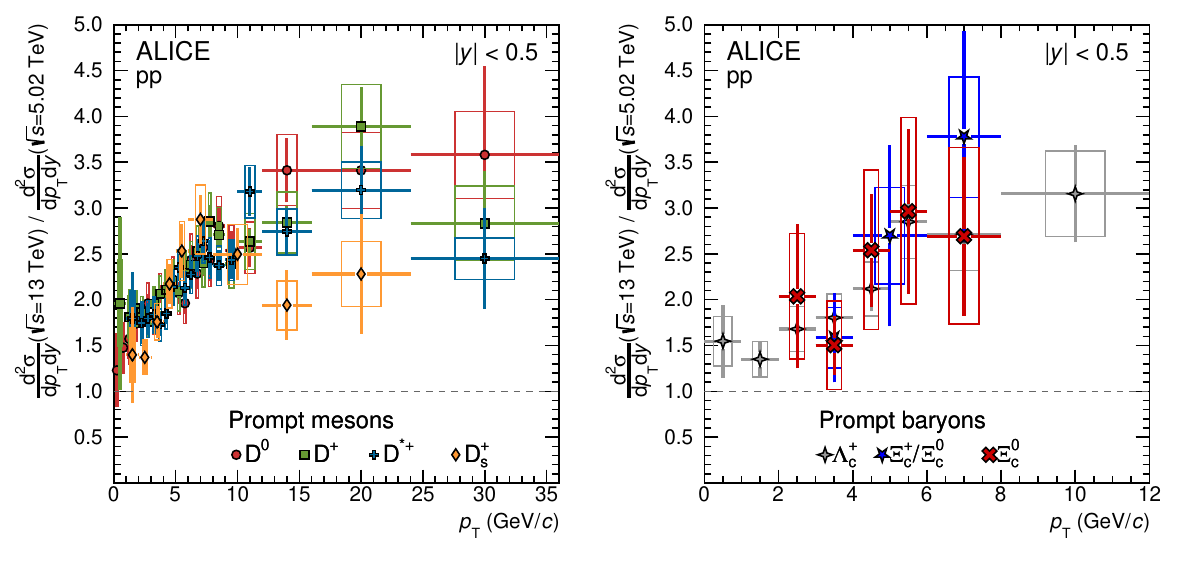}
    \caption{Left: ratios between  prompt $\Dzero$, $\Dplus$, $\Dstar$, and $\Ds$ mesons production cross sections in \pp collisions at $\s = 13$~TeV and those in \pp collisions at $\s = 5.02$~TeV~\cite{ALICE:2019nxm, ALICE:2021mgk} as a function of $\pt$. Right: ratios between the prompt $\Lambdac$-, $\XicZero$-, and $\XicPlus$-baryon production cross sections in \pp collisions at $\s = 13$~TeV and those in \pp collisions at $\s = 5.02$~TeV~\cite{Acharya:2021dsq, ALICE:2020wfu, ALICE:2020wla, ALICE:2022ych} as a function of $\pt$. Vertical bars (boxes) report the statistical (systematic) uncertainties.}
    \label{fig:MesonToMesonRatio-13to5}
\end{figure}

To further investigate the dependence of the D-meson production on the \pp collision centre-of-mass energy, the ratio of the $\pt$-differential D-meson cross sections at $\s = 13$~TeV to the ones at $\s~=~5.02$~TeV~\cite{ALICE:2019nxm, ALICE:2021mgk} was computed for the different D-meson species. The results are shown in the left panel of Fig.~\ref{fig:MesonToMesonRatio-13to5}. 
The systematic uncertainties were propagated as fully uncorrelated between the results at the two energies, with the exception of those related to the prompt-fraction correction, and the branching ratio which were propagated as correlated. The latter uncertainties cancel out in the ratios.
The ratios for the different D-meson species are compatible within the uncertainties and show a common increase with increasing $\pt$.
This effect is similar to that seen for the D-meson production cross sections ratios between $\s = 7$~TeV and $\s = 5.02$~TeV~\cite{ALICE:2019nxm}. Furthermore, as discussed in Ref.~\cite{ALICE:2019nxm}, these results are in agreement with pQCD calculations, which benefit from the cancellation of a portion of the uncertainties in the ratio. This cancellation enables a precise description of the observed trend in the data.

An analogous study was performed in the baryon sector by measuring the ratios of the $\Lambdac$-, $\XicZero$-, and $\XicPlus$-baryon $\pt$-differential cross sections at $\s = 13$~TeV to the ones at $\s = 5.02$~TeV~\cite{Acharya:2021dsq, ALICE:2020wfu, ALICE:2020wla, ALICE:2022ych}. For the $\XicPlus$ baryon, the ratio was computed with respect to the measurement of the $\XicZero$ baryon, since due to isospin symmetry the two baryons are expected to be produced with equal yields. This was 
supported by the fact that their $\pt$-differential cross sections were found to be fully compatible in the measured $\pt$ range at $\s = 13$~TeV~\cite{Acharya:2021vjp}. The ratios for the baryon sector are reported in the right panel of Fig.~\ref{fig:MesonToMesonRatio-13to5}, and they share a similar increasing trend as a function of $\pt$. This result suggests a compatible $\pt$-spectrum hardening between mesons and baryons from $\s = 5.02$~TeV to $\s = 13$~TeV.

\subsubsection{Ratios of $\boldsymbol{\Dzero}$-meson cross section at different rapidities and collision energies}
\begin{figure}[t!]
    \centering
    \includegraphics[width=.95\textwidth]{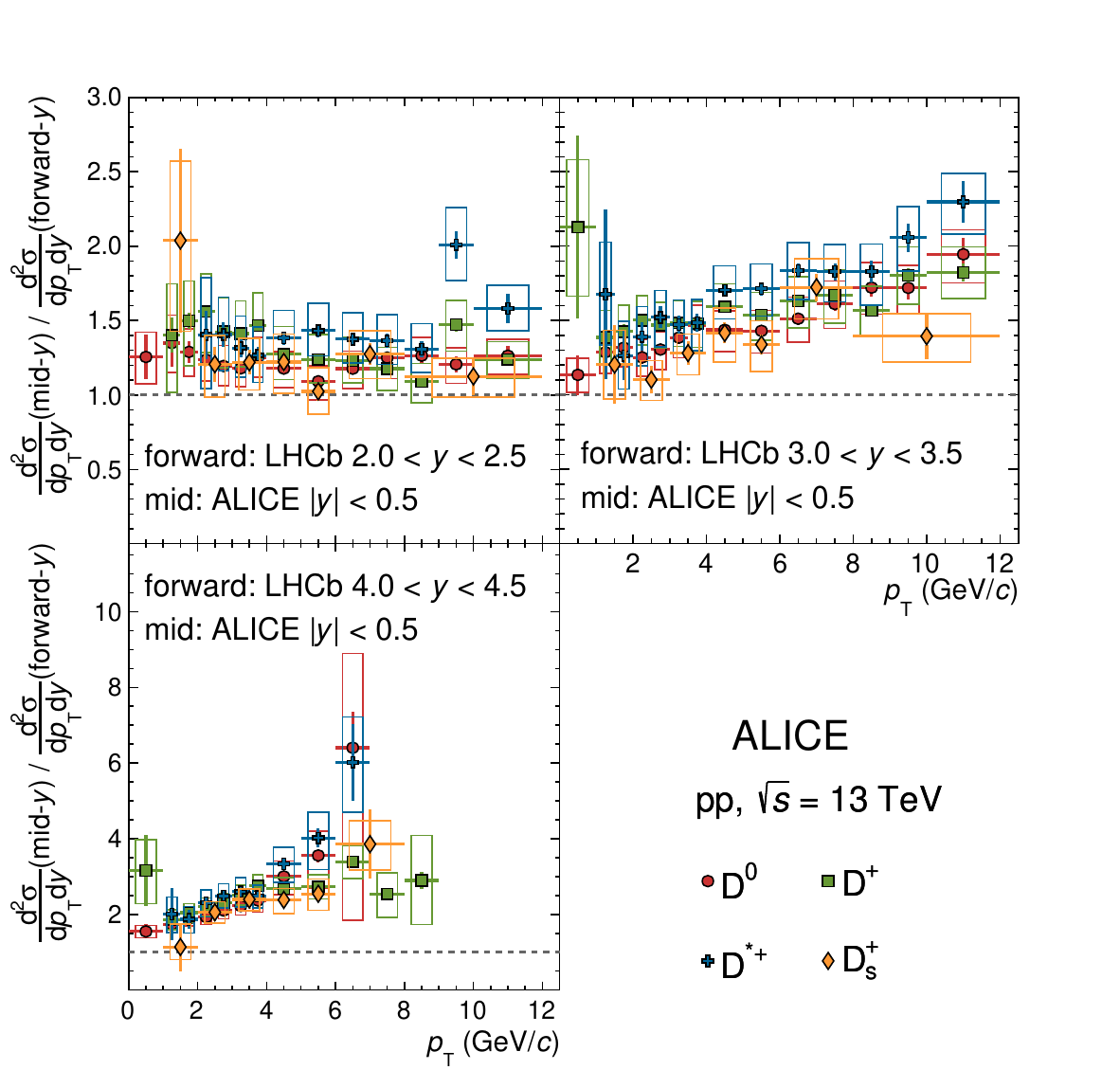}
    \caption{Ratios of D-meson production cross sections per unit of rapidity at midrapidity ($|y| < 0.5$) to those measured by LHCb~\cite{LHCb:2015swx} in three rapidity intervals: $2 < y < 2.5$ (top-left panel), $3 < y < 3.5$ (top-right panel), and $4 < y < 4.5$ (bottom-left panel), as a function of $\pt$. Statistical (systematic) uncertainties are reported as vertical bars (boxes).}
    \label{fig:MesonToMesonRatio-ALICEvsLHCb}
\end{figure}
The rapidity dependence of the D-meson production in pp collisions at $\s = 13$~TeV was studied by computing the ratio between the presented measurements at midrapidity and the results from the LHCb Collaboration at forward rapidity at the same collision energy~\cite{LHCb:2015swx}. Figure~\ref{fig:MesonToMesonRatio-ALICEvsLHCb} shows the ratios between the D-meson production cross sections measured by ALICE at midrapidity and by LHCb in three intervals at forward rapidity (top-left: $2 < y < 2.5$, top-right: $3 < y < 3.5$, bottom-left: $4 < y < 4.5$) in \pp collisions at $\s = 13$~TeV. The uncertainties of the measurement involved in the ratios were propagated as fully uncorrelated. Within current uncertainties, a common trend and magnitude are observed among the different D-meson species in all three rapidity intervals. The significant increase at high $\pt$ of such ratios when going to more forward regions suggests softer $\pt$-spectra at forward rapidity. Such behaviour is reproduced by FONLL calculations, as discussed in Refs.~\cite{ALICE:2019nxm, ALICE:2017olh}.

\begin{figure}[t!]
    \centering
    \includegraphics[width=1\textwidth]{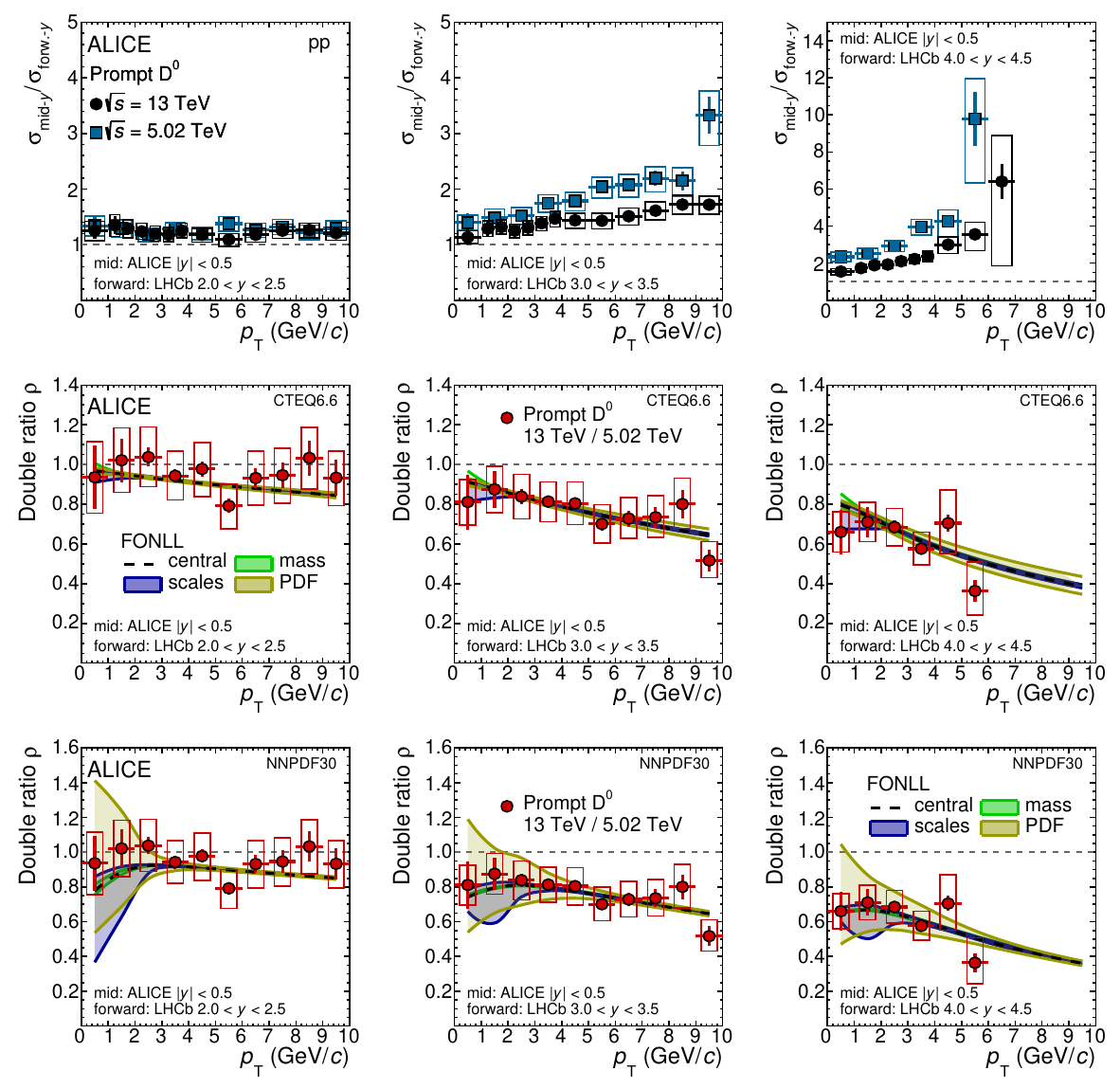}
    \caption{Ratio between $\pt$-differential cross sections of prompt $\Dzero$ mesons at midrapidity ($|y|<0.5$) in pp collisions at $\s=5.02$~TeV~\cite{ALICE:2019nxm} and $13$~TeV and those at forward rapidity in three rapidity intervals (left: $2<y<2.5$, middle: $3<y<3.5$, right: $4<y<4.5$) measured by the LHCb Collaboration~\cite{LHCb:2015swx}. The double ratios $\rho$, defined in Eq.~\ref{eq:double_ratio}, are shown in the middle and bottom rows together with FONLL calculations employing the CTEQ6.6~\cite{Pumplin:2002vw} (middle row) and NNPDF30~\cite{NNPDF:2014otw} (bottom row) PDF sets. The statistical (systematic) uncertainties are shown as vertical bars (boxes).}
    \label{fig:ratioALICEoverLHCb_13over5_D0}
\end{figure}

As discussed in Ref.~\cite{Cacciari:2015fta}, the uncertainty on the PDFs in FONLL calculation can be severely constrained at small values of Bjorken-$x$ ($10^{-4}-10^{-5}$)~\cite{Cacciari:2015fta} by performing precise measurements of ratios at different centre-of-mass energies of D-meson production cross sections in different rapidity intervals. The computation of such ratios between measurements at different energies and rapidity intervals was performed considering the $\pt$-differential cross section of the prompt $\Dzero$ mesons measured at midrapidity ($|y|<0.5$) and those measured at forward rapidity by the LHCb Collaboration~\cite{LHCb:2016ikn, LHCb:2015swx}. The choice of using the prompt $\Dzero$ mesons for this study is motivated by the observation that the $\pt$ dependence of the cross sections of the different D-meson species are compatible among each other (as can be seen in Fig.~\ref{fig:MesonToMesonRatio-ALICEvsLHCb}) and that the measurements of prompt $\Dzero$ mesons are the most precise down to $\pt = 0$. The ratios are shown in Fig.~\ref{fig:ratioALICEoverLHCb_13over5_D0} and compared with FONLL predictions. In the top row, the $\pt$-differential ratios between the prompt $\Dzero$-meson production cross section at midrapidity and that at forward rapidity (left: $2<y<2.5$, middle: $3<y<3.5$, right: $4<y<4.5$) measured by the LHCb Collaboration are shown in \pp collisions at $\s=5.02$~TeV and 13~TeV. The results at the two energies are fully compatible within uncertainties when the forward rapidity interval $2<y<2.5$ is considered. However, when moving to more forward rapidities the ratio in \pp collisions at $\s=5.02$~TeV gets systematically higher than the one at $\s=13$~TeV, with a hint of a harder $\pt$ shape for the ratio at lower energy. This behaviour reflects the different Bjorken-$x$ values, which depend on $\s$ for the same hadron rapidity and $\pt$, that are probed by measuring a charm hadron in several rapidity intervals at different centre-of-mass energies. Such values at low $\pt$ go from $x\sim 10^{-4}$ at midrapidity to $x\sim 10^{-6}$ at $y=4.5$.

The results obtained at $\s=13$~TeV can be further divided by those at $\s=5.02$~TeV for each rapidity interval, providing the \enquote{double ratios} that are useful to constrain the PDF uncertainties, as discussed in Ref.~\cite{Cacciari:2015fta}. The resulting double ratios
\begin{equation}
    \rho = (\sigma_{{\rm mid}\text{-}y}^{\rm 13\text{ }TeV}/\sigma_{{\rm forward}\text{-}y}^{\rm 13\text{ }TeV}) / (\sigma_{{\rm mid}\text{-}y}^{\rm 5.02\text{ }TeV}/\sigma_{{\rm forward}\text{-}y}^{\rm 5.02\text{ }TeV})\,,
    \label{eq:double_ratio}
\end{equation}
where the symbol $\sigma$ indicates the $\pt$-differential production cross section of prompt $\Dzero$ mesons, are shown in the middle and bottom rows of Fig.~\ref{fig:ratioALICEoverLHCb_13over5_D0}. Both the statistical and systematic uncertainties associated to the measurements were propagated as uncorrelated. The double ratios are compared with FONLL calculations performed employing the CTEQ6.6~\cite{Pumplin:2002vw} and NNPDF30~\cite{NNPDF:2014otw} PDF sets respectively. The CTEQ6.6 PDF set is the same used for the FONLL calculations shown in Fig.~\ref{fig:mesonCrossSection_vs_models}, as well as in comparison with the measured ratio of the prompt $\Dzero$-meson production cross section at midrapidity with that at forward rapidity in \pp collisions at $\s=5.02$~TeV and $\s=7$~TeV~\cite{ALICE:2012mhc, ALICE:2019nxm}. The NNPDF30 PDF set is the one considered in Ref.~\cite{Cacciari:2015fta}. It was obtained with a more robust and efficient fitting code to include also LHC measurements, 
as discussed in Ref.~\cite{NNPDF:2014otw}.

The different sources of uncertainty in the FONLL calculations are shown separately as coloured bands. They correspond to: (i) the value assumed for the charm-quark mass, by default $m_{\rm c}=1.5$~GeV$/c^2$ and varied to 1.3 and 1.7~GeV$/c^2$; (ii) the values assumed for the factorisation and renormalisation scales, as discussed in Ref.~\cite{Averbeck:2011ga};
(iii) the uncertainty on the PDFs. The calculation of the theoretical uncertainties related to the charm-quark mass and scale variations was performed by computing the double-ratio considering the same variation for all the four cross sections involved, and the uncertainty band of each source corresponds to the envelope of the calculated double ratios. The bands associated to the PDF uncertainty were provided by the authors of Ref.~\cite{Cacciari:2015fta}, employing the recipes prescribed for the CTEQ6.6 and NNPDF30 PDF sets in Ref.~\cite{Pumplin:2002vw}
 and Ref.~\cite{NNPDF:2014otw} respectively. The theoretical predictions reproduce both the magnitude and the $\pt$-dependence of the measured double ratios computed in all the three forward rapidity ranges. The uncertainties of the FONLL calculations employing the CTEQ6.6 PDF sets (middle row in Fig.~\ref{fig:ratioALICEoverLHCb_13over5_D0}) are significantly lower than the experimental uncertainties for $\pt>2$~GeV$/c$, while at lower $\pt$ their magnitude is of the same order. In this case, the uncertainties are dominated by the variations of the charm-quark mass and the $\mu_{\rm F,R}$ scale values. The uncertainties of the PDFs become dominant for $\pt<2$~GeV$/c$ when employing the NNPDF30 PDF sets in the FONLL calculations (bottom row in Fig.~\ref{fig:ratioALICEoverLHCb_13over5_D0}), as discussed in Ref.~\cite{Cacciari:2015fta}. In this $\pt$ range, the measured double ratios are within the PDF uncertainty band of the calculations and the measured uncertainties are about a factor 2$-$3 smaller than the PDF ones.

The comparisons of $\rho$ between data and FONLL indicate that the measured double ratios of prompt $\Dzero$ mesons shown in Fig.~\ref{fig:ratioALICEoverLHCb_13over5_D0} are precise enough in the range $\pt<2$~GeV$/c$ to put quantitative constraints on the gluon PDFs employed in the NNPDF30 set. For this PDF set, the PDF uncertainties become equivalent to the scale ones at high $\pt$. This happens at about $\pt=2$~GeV$/c$ for the double ratios considering the forward-$y$ range $2<y<2.5$. This threshold increases up to $\pt=6$~GeV$/c$ going to more forward rapidities, where the PDF uncertainty remains dominant in a wider $\pt$ interval. At high $\pt$, the overall uncertainties become negligible compared to the experimental ones and the calculations reproduce the decreasing trend of the measurement, which becomes steeper going to $4<y<4.5$. As discussed in Ref.~\cite{Cacciari:2015fta}, (double-)ratios at such forward rapidity and at high $\pt$ ($\pt\gtrsim 20$~GeV$/c$) would probe gluon densities in the range of Bjorken-$x\sim0.2$, which has not been well constrained by the experiments so far. However, given the lack of measurements at forward rapidity for $\pt\gtrsim 15$~GeV$/c$, this regime cannot be tested experimentally with the current measurements.

\subsection{Charm-quark production in pp collisions at $\boldsymbol{\s=13}$~TeV}
\label{sec:FFccbar}

\subsubsection{Charm-hadron $\boldsymbol{\pt}$-integrated cross sections and extrapolation down to $\boldsymbol{\pt=0}$}
\label{sec:XicSigmacExtrap}
For the measurement of the charm-quark fragmentation fractions and the $\ccbar$ production cross section per unit of rapidity at midrapidity in pp collisions at $\s=13$~TeV (Sections~\ref{sec:FFVsE} and~\ref{sec:ccbarVsE}), the total production cross section of each hadron species must be considered. In the case of $\Dzero$ and $\Dplus$ mesons and $\Lambdac$ baryons, for which the $\pt$-differential cross sections were measured down to $\pt = 0$, the total cross section was calculated by integrating the $\pt$-differential results over the $\pt$ interval of the measurement, considering that the contribution to the cross section in the unmeasured interval ($\pt > 24$ GeV$/c$ for the $\Lambdac$ baryon, $\pt > 36$ GeV$/c$ for the $\Ds$ meson, and $\pt > 50$ GeV$/c$ for the $\Dzero$ meson) is negligible.
All the systematic uncertainties were propagated assuming them as correlated among the $\pt$ intervals, with the exception of those related to the signal extraction, which were assumed to be uncorrelated. For the other charm-hadron species, for which the analysis down to $\pt = 0$ was not possible, the measured $\pt$-differential cross sections were extrapolated as described below.

The measurement of the $\Dstar$-meson cross section at midrapidity ($|y|<0.5$) was integrated over the measured momentum interval ($\pt > 1$~GeV$/c$) to obtain $\sigma_\text{visible}$ and was then extrapolated down to $\pt~=~0$~GeV$/c$ with the strategy described in Ref.~\cite{ALICE:2019nxm}. The $\pt$-integrated cross section in the full $\pt$ range, $\sigma_\text{full}$, was measured by scaling $\sigma_\text{visible}$ by the extrapolation factor $C^\text{FONLL}\equiv\sigma_\text{full}^\text{FONLL}/\sigma_\text{visible}^\text{FONLL}$ calculated from the FONLL predictions for the $\Dstar$ meson at midrapidity. The systematic uncertainties of the extrapolation factor accounted for the uncertainties of the PDF, the variation of the charm-quark mass, and the values of the renormalisation and factorisation scales for the FONLL calculations.

The extrapolation of the $\Ds$-meson cross section was performed as described in Refs.~\cite{ALICE:2019nxm, ALICE:2021mgk}. Due to the lack of FONLL predictions for the $\Ds$-meson production, the extrapolation factor was calculated using the $\pt$-differential cross section of charm quarks provided by FONLL calculations, the fragmentation fractions $f({\rm c}\to\Ds)$ and $f({\rm c}\to{\rm D_s^{*+}})$ from ALEPH measurements~\cite{ALEPH:1999syy}, and the charm-quark fragmentation functions from Ref.~\cite{Braaten:1994bz}.

The extrapolation of the $\SigmacZeroPlusPlusPlus$-baryon cross section was performed by computing a scaling factor based on the prediction from PYTHIA CR-BLC Mode 0, which among the CR-BLC modes was observed to better describe both the magnitude and the $\pt$ dependence of the measured cross section. The other modes were used to estimate the extrapolation systematic uncertainty, together with the SHM-RQM, QCM, and the Catania models, which provided predictions only for the $\SigmacZeroPlusPlusPlus/\Dzero$ ratio. In this case, the model predictions were used to fit the measured ratio from Ref.~\cite{Acharya:2021vpo}, leaving only the normalisation as a free parameter. The fit function was scaled by the measured $\Dzero$-meson cross section in the range $0<\pt<2$ GeV$/c$, where no $\SigmacZeroPlusPlusPlus$ measurement is currently available. 
The extrapolation systematic uncertainty was calculated as the envelope of the values of the $\pt$-integrated cross sections obtained with all the considered variations. To a minimum and maximum $\SigmacZeroPlusPlusPlus$-baryon cross section extrapolated from these ratios, the measured $\Dzero$-meson cross section in $0<\pt<2$~GeV$/c$ was shifted up and down by one standard deviation, defined by the quadrature sum of the statistical and systematic uncertainties.

The extrapolation of the $\XicPlus$-baryon cross section was performed following the same strategy as for the $\SigmacZeroPlusPlusPlus$ baryon. The Catania model was used as the central value prediction as it was observed to best describe the $\pt$-dependence and magnitude of the measured cross section. A Tsallis fit to the measurement and predictions from the PYTHIA CR-BLC Modes 0, 2 and 3, the SHM+RQM model, and the QCM model were used for the estimation of the systematic uncertainty.

\begin{table}[t!]
    \renewcommand*{\arraystretch}{1.4}
    \centering
    \caption{${\rm d}\sigma/{\rm d}y|_{|y|<0.5}$ of all measured charm-hadron species in \pp collisions at $\s=13$~TeV. These results are obtained by integrating the measured $\pt$-differential cross section at midrapidity and extrapolating down to $\pt=0$ if necessary.}
    \begin{tabular}{|ll|}
        \hline
          & \multicolumn{1}{c|}{${\rm d}\sigma/{\rm d}y|_{|y|<0.5}$ ($\mu$b), $\pt>0$} \\
        \hline
        $\Dzero$ & 749 $\pm$ 27 (stat.) $^{+48}_{-50}$ (syst.) $\pm$ 12 (lumi.) $\pm$ 6 (BR) \\
        $\Dplus$ & 375 $\pm$ 32 (stat.) $^{+35}_{-35}$ (syst.) $\pm$ 6 (lumi.) $\pm$ 6 (BR) \\
        $\Ds$ & 120 $\pm$ 11 (stat.) $^{+12}_{-13}$ (syst.) $^{+25}_{-10}$ (extrap.) $\pm$ 2 (lumi.) $\pm$ 3 (BR) \\
        $\Lambdac$ & 329 $\pm$ 15 (stat.) $^{+28}_{-29}$ (syst.) $\pm$5 (lumi.) $\pm$15 (BR) \\
        $\XicZero$~\cite{Acharya:2021vjp} & 194 $\pm$ 27 (stat.) $^{+46}_{-46}$ (syst.) $^{+18}_{-12}$ (extrap.) $\pm$ 3 (lumi.)\\
        $\XicPlus$ & 187 $\pm$ 25 (stat.) $^{+19}_{-19}$ (syst.) $^{+13}_{-59}$ (extrap.) $\pm$ 3 (lumi.) $\pm$ 82 (BR)\\
        $\Jpsi$~\cite{ALICE:2021edd} & 7.29 $\pm$ 0.27 (stat.) $^{+0.52}_{-0.52}$ (syst.) $^{+0.04}_{-0.01}$ (extrap.) \\
        \hline
        $\Dstar$   & 306 $\pm$ 26 (stat.) $^{+33}_{-34}$ (syst.) $^{+48}_{-17}$ (extrap.) $\pm$ 5 (lumi.) $\pm$ 3 (BR)\\
        $\SigmacZeroPlusPlusPlus$ & 142 $\pm$ 22 (stat.) $^{+24}_{-24}$ (syst.) $^{+24}_{-32}$ (extrap.) $\pm$ 2 (lumi.) $\pm$ 6 (BR)\\
        \hline
    \end{tabular}
    \label{tab:crossSecHadrons}
\end{table}

The $\pt$-integrated cross sections of all the charm hadrons at midrapidity $(|y|<0.5)$ in \pp collisions at $\s=13$~TeV are listed in Table~\ref{tab:crossSecHadrons}.


\subsubsection{Strange to-non strange charm-meson production ratio $f_{\rm s}/(f_{\rm u}+f_{\rm d})$}

The relative production of strange to non-strange D-meson production can be studied using the ratio of fragmentation fractions $f_{\rm s}/(f_{\rm u}+f_{\rm d})$, where $f_{\rm x}$ represents the probability for a charm quark to hadronise with another quark of flavour x. In the charm-meson sector, this ratio corresponds experimentally to the prompt cross section ratio $\Ds/(\Dzero + \Dplus)$, as all $\Dstar$ and $\rm D^{*0}$ mesons decay to $\Dzero$ and $\Dplus$ mesons, and all $\rm D^{*+}_s$ mesons decay to $\Ds$ mesons. The contribution of the decays of excited states that change the strange/non-strange D-meson content (e.g. $\rm D_{s1}^+ \to \Dstar\kzero$ or $\rm D_{s2}^{*+} \to \Dplus \kzero$) was neglected in the computation of the ratio $f_{\rm s}/(f_{\rm u}+f_{\rm d})$.

The $\pt$-integrated cross sections reported in Table~\ref{tab:crossSecHadrons} were used to compute the ratios of production yields
among the different D-meson species in \pp collisions at $\s=13$~TeV. The ratios of prompt D mesons $\Dplus/\Dzero$, $\Ds/\Dzero$, $\Ds/\Dplus$, and $\Ds/(\Dzero+\Dplus)$ for $\pt>0$ are reported in Table~\ref{tab:ptIntRatios}. The systematic uncertainties related to the tracking efficiency, luminosity, extrapolation, and to the subtraction of the component from beauty-hadron decays were propagated as correlated among the D-meson species. All the other sources of systematic uncertainties were propagated as uncorrelated. The same ratios measured in \pp collisions at $\s=5.02$~TeV~\cite{ALICE:2021mgk} are also reported in Table~\ref{tab:ptIntRatios}. The results do not show any significant dependence on the collision energy.

\begin{table}[t!]
    \renewcommand*{\arraystretch}{1.4}
    \centering
    \caption{Ratios of the measured production cross sections of prompt D mesons for $\pt>0$ at midrapidity $(|y|<0.5)$ in pp collisions at $\s=5.02$~TeV~\cite{ALICE:2021mgk} and $\s=13$~TeV.}
    \begin{tabular}{|ll|}
    \hline
        \multicolumn{2}{|c|}{\pp, $\s=5.02$~TeV~\cite{ALICE:2021mgk}} \\
        [0.5mm]
        \hline
        $\Dplus/\Dzero$         & $0.442 \pm 0.055\text{ (stat.)} \pm 0.033\text{ (syst.)} \pm 0.008\text{ (BR)}$ \\
        [0.5mm]
        $\Ds/\Dzero$            & $0.186 \pm 0.028\text{ (stat.)} \pm 0.015\text{ (syst.)} ^{+0.051}_{-0.018}\text{ (extrap.)} \pm 0.007\text{ (BR)}$ \\
        [0.5mm]
        $\Ds/\Dplus$            & $0.419 \pm 0.078\text{ (stat.)} \pm 0.041\text{ (syst.)} ^{+0.116}
_{-0.040}\text{ (extrap.)} \pm 0.017\text{ (BR)}$ \\
[0.5mm]
        $\Ds/(\Dzero + \Dplus)$ & $0.128 \pm 0.020\text{ (stat.)} \pm 0.010\text{ (syst.)} ^{+0.035}
_{-0.012}\text{(extrap.)}\pm0.005\text{ (BR)}$ \\
[0.5mm]
        \hline
        \multicolumn{2}{|c|}{\pp, $\s=13$~TeV} \\
        \hline
        $\Dplus/\Dzero$         & $0.500 \pm 0.047\text{ (stat.)} \pm 0.033\text{ (syst.)} \pm 0.009\text{ (BR)}$ \\
        [0.5mm]
        $\Ds/\Dzero$            & $0.160 \pm 0.015\text{ (stat.)} \pm 0.012\text{ (syst.)} ^{+0.034} _{-0.013}\text{ (extrap.)} \pm 0.004\text{ (BR)}$ \\
        [0.5mm]
        $\Ds/\Dplus$            & $0.319 \pm 0.039\text{ (stat.)} \pm 0.030\text{ (syst.)} ^{+0.068} _{-0.024}\text{ (extrap.)} \pm 0.010\text{ (BR)}$ \\
        [0.5mm]
        $\Ds/(\Dzero + \Dplus)$ & $0.106 \pm 0.010\text{ (stat.)} \pm 0.008\text{ (syst.)} ^{+0.023} _{-0.009}\text{ (extrap.)} \pm 0.003\text{ (BR)}$ \\
        [0.5mm]
        \hline
    \end{tabular}
    \label{tab:ptIntRatios}
\end{table}

\begin{table}[t!]
\renewcommand*{\arraystretch}{1.4}
    \centering
    \caption{Production cross sections of prompt D mesons for $\pt>1$ GeV$/c$ at midrapidity $(|y|<0.5)$ in pp collisions at $\s=13$~TeV.}
    \begin{tabular}{|ll|}
    \hline
        \multicolumn{2}{|c|}{${\rm d}\sigma/{\rm d}y|_{|y|<0.5}$ ($\mu$b), $\pt>1$ GeV$/c$} \\
        [0.5mm]
        \hline
        $\Dzero$         & $592 \pm 19 \text{ ({stat.})} ^{+40}_{-42} \text{ ({syst.})} \pm 9 \text{ ({lumi.})} \pm 4 \text{ ({BR.})}$ \\
        [0.5mm]
        $\Dplus$         & $264 \pm 5 \text{ ({stat.})} ^{+20}_{-20} \text{ ({syst.})} \pm 4 \text{ ({lumi.})} \pm 5 \text{ ({BR.})}$ \\
        [0.5mm]
        $\Ds$            & $99 \pm 9 \text{ ({stat.})} ^{+10}_{-11} \text{ ({syst.})} \pm 2 \text{ ({lumi.})} \pm 3 \text{ ({BR.})}$ \\
        \hline
    \end{tabular}
    \label{tab:crossSecHadronsPtGtr1GeV}
\end{table}

\begin{figure}[t!]
    \centering
    \includegraphics[width=0.9\textwidth]{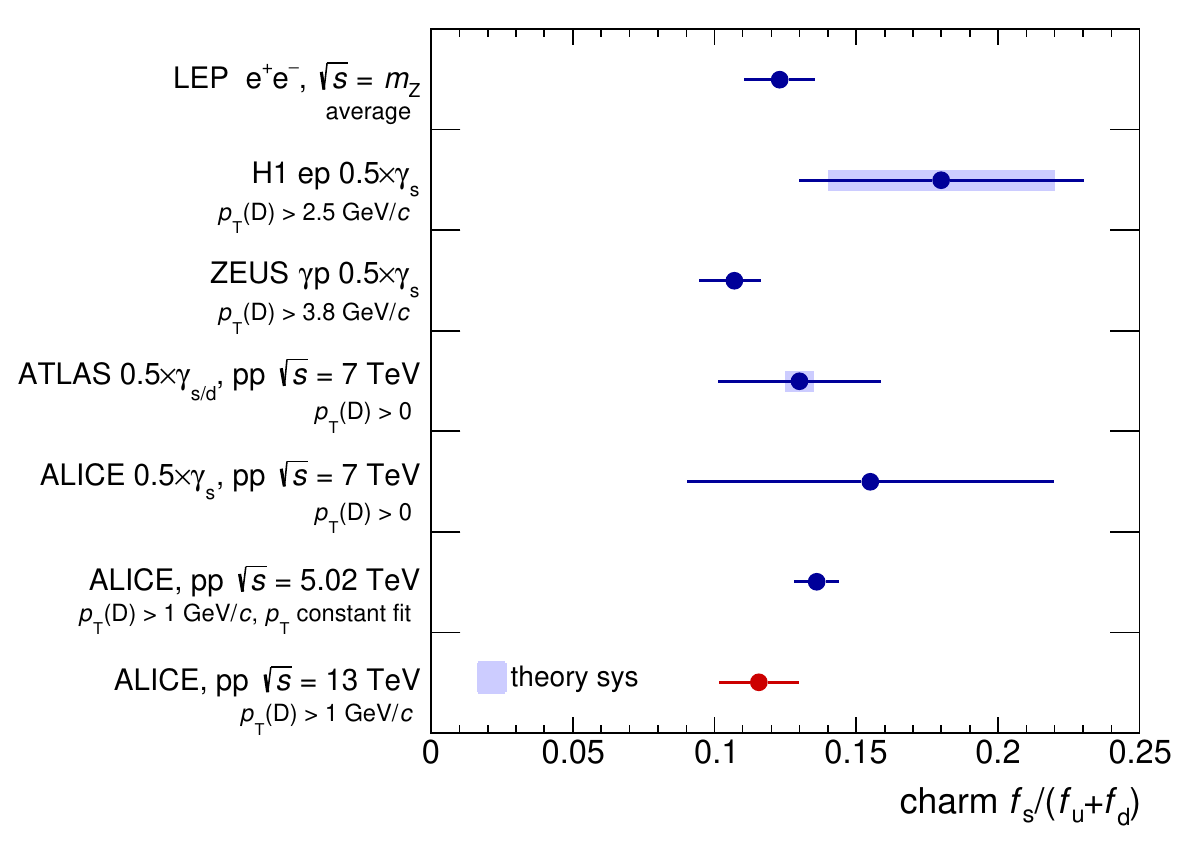}
    \caption{Charm-quark fragmentation-fraction ratio $f_{\rm s}/(f_{\rm u} + f_{\rm d})$ (red) compared with previous measurements performed by the ALICE~\cite{ALICE:2021mgk, ALICE:2012gkr}, H1~\cite{H1:2004bwe}, ZEUS~\cite{ZEUS:2013fws}, and ATLAS~\cite{ATLAS:2015igt} Collaborations, and to the average of LEP measurements~\cite{Gladilin:2014tba}. The total experimental uncertainties (bars) and the theoretical uncertainties (shaded boxes) are shown.}
    \label{fig:gammaS}
\end{figure}

The uncertainty on the $\Ds/(\Dzero+\Dplus)$ ratios reported in Table~\ref{tab:ptIntRatios} are dominated by the limited precision of the measurements in the low-$\pt$ intervals, and by the uncertainty related to the extrapolation of the cross section of prompt $\Ds$ mesons down to $\pt=0$. For the measurement in \pp collisions at $\s=13$~TeV, the total relative uncertainty amounts to about 22\%.
As discussed in Ref.~\cite{ALICE:2021mgk}, the measurement of the ratio $f_{\rm s}/(f_{\rm u}+f_{\rm d})$ in \pp collisions at $\s=5.02$~TeV was performed with a fit of the $\pt$-differential $\Ds/(\Dzero + \Dplus)$ ratios using a constant function, since the ratio was found to be constant within uncertainties. With this strategy, the total uncertainty of the measurement was reduced.
Given the better precision, the $\Ds/(\Dzero+\Dplus)$ ratio in \pp at $\s=13$~TeV suggests a hint of increasing trend with $\pt$ in the interval $\pt<8$ GeV$/c$, as visible in Fig.~\ref{fig:MesonToMesonRatio-5-7-13}. Therefore, the approach based on a fit to a constant is no longer justified and the ratio $f_{\rm s}/(f_{\rm u}+f_{\rm d})$ in \pp collisions at $\s=13$~TeV was measured considering the prompt $\Dzero$-, $\Dplus$-, and $\Ds$-meson cross sections for $\pt>1$ GeV$/c$, corresponding to the interval of the $\Ds$ measurement, reported in Table~\ref{tab:crossSecHadronsPtGtr1GeV}. This led to:

\begin{equation}
    \frac{\gamma_{\rm s}}{2} \equiv \left( \frac{f_{\rm s}}{f_{\rm u} + f_{\rm d}} \right)_{\rm charm} = 0.116 \pm 0.011\,\text{(stat.)} \pm 0.009\,\text{(syst.)} \pm 0.003\,\text{(BR)} \;,
    \label{eq:fs_fufd}
\end{equation}

where $\gamma_{\rm s}$ denotes the strangeness suppression factor, as defined in Ref.~\cite{Becattini:1997ii}.
The systematic uncertainties were propagated as for the measurement for $\pt>0$. The result shown in Eq.~\ref{eq:fs_fufd} agrees with the $\Ds/(\Dzero + \Dplus)$ ratio reported in Table~\ref{tab:ptIntRatios}, and it does not depend on the extrapolation down to $\pt=0$ of the prompt $\Ds$ mesons.

In Fig.~\ref{fig:gammaS}, the ratio $f_{\rm s}/(f_{\rm u} + f_{\rm d})$ (red) is compared with previous measurements of strangeness suppression factor $\gamma_{\rm s}$ or $f_{\rm s}/(f_{\rm u} + f_{\rm d})$ from the ALICE~\cite{ALICE:2021mgk, ALICE:2012gkr}, H1~\cite{H1:2004bwe},
ZEUS~\cite{ZEUS:2013fws}, and ATLAS~\cite{ATLAS:2015igt} Collaborations. In the cases where $\gamma_{\rm s}$ was used, the measurements were scaled by a factor of 0.5, accounting for the different normalisation between the two observables, as shown in Eq.~\ref{eq:fs_fufd}. The total experimental uncertainties are reported as bars, and the theoretical ones as shaded boxes. The theoretical uncertainties in the H1 measurement denote  the branching ratio uncertainty and the model dependencies of the acceptance determination. In the case of the ATLAS measurement, they correspond to the extrapolation uncertainties to the full phase space. The values are compatible within uncertainties, and they are in agreement with the average of measurements at LEP~\cite{Gladilin:2014tba}. This indicates that the production of prompt strange D mesons relative to that of prompt non-strange D mesons ($f_{\rm s}/(f_{\rm u}+f_{\rm d})$) in $\ee$, $\ep$ and \pp collisions does not show any significant dependence on the collision system and energy. Furthermore, the $f_{\rm s}/(f_{\rm u} + f_{\rm d})$ is compatible with the ratio of the $\pt$-integrated cross sections at forward rapidity ($2<y<4.5$) of prompt $\Ds$, $\Dplus$ and $\Dzero$-meson  in the interval $1<\pt<8$ GeV$/c$ measured by the LHCb Collaboration~\cite{LHCb:2015swx}, suggesting that the relative production of strange and non-strange D mesons does not depend on the meson rapidity.

\subsubsection{Charm-quark fragmentation fractions in $\rm{\bold{pp}}$ collisions}
\label{sec:FFVsE}

\begin{figure}[t!]
    \centering
    \includegraphics[width=0.54\textwidth]{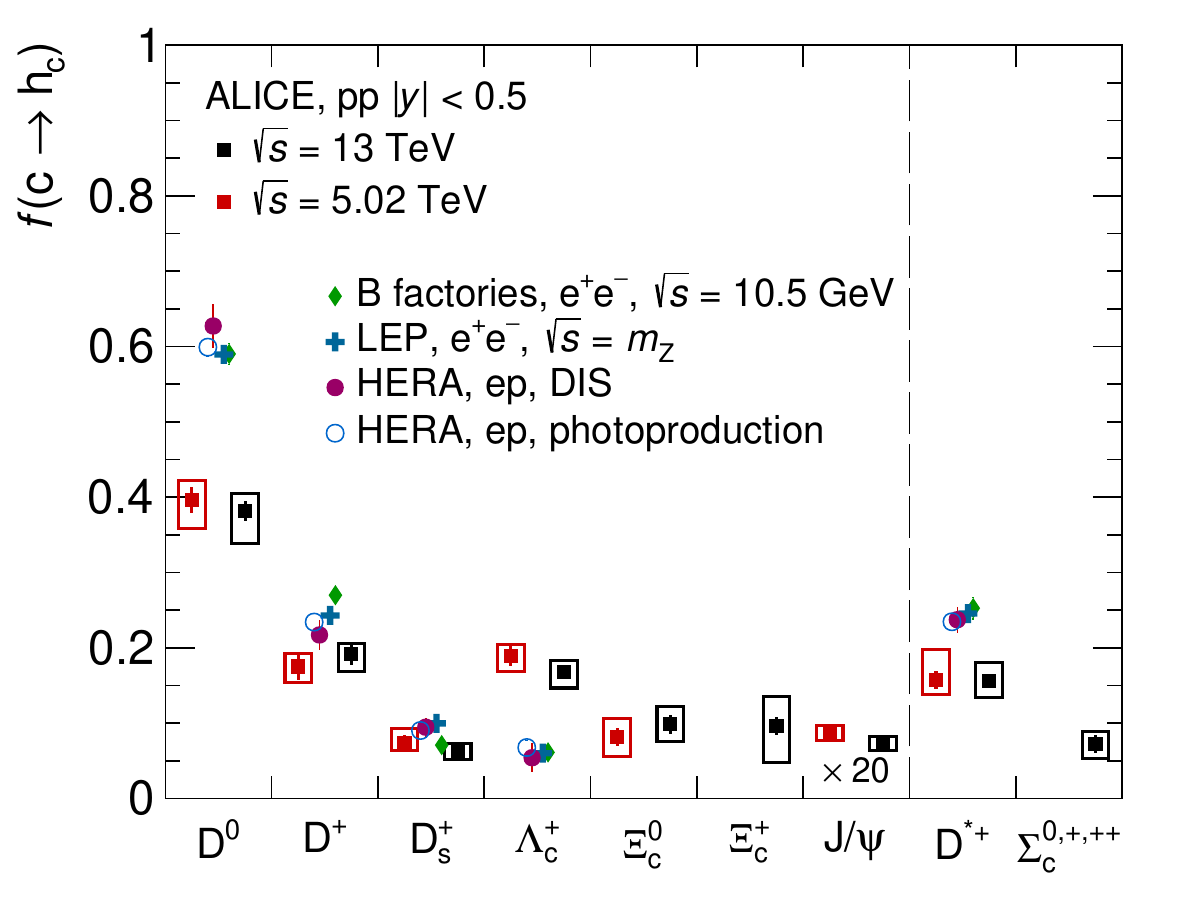} \quad
    \includegraphics[width=0.425\textwidth]{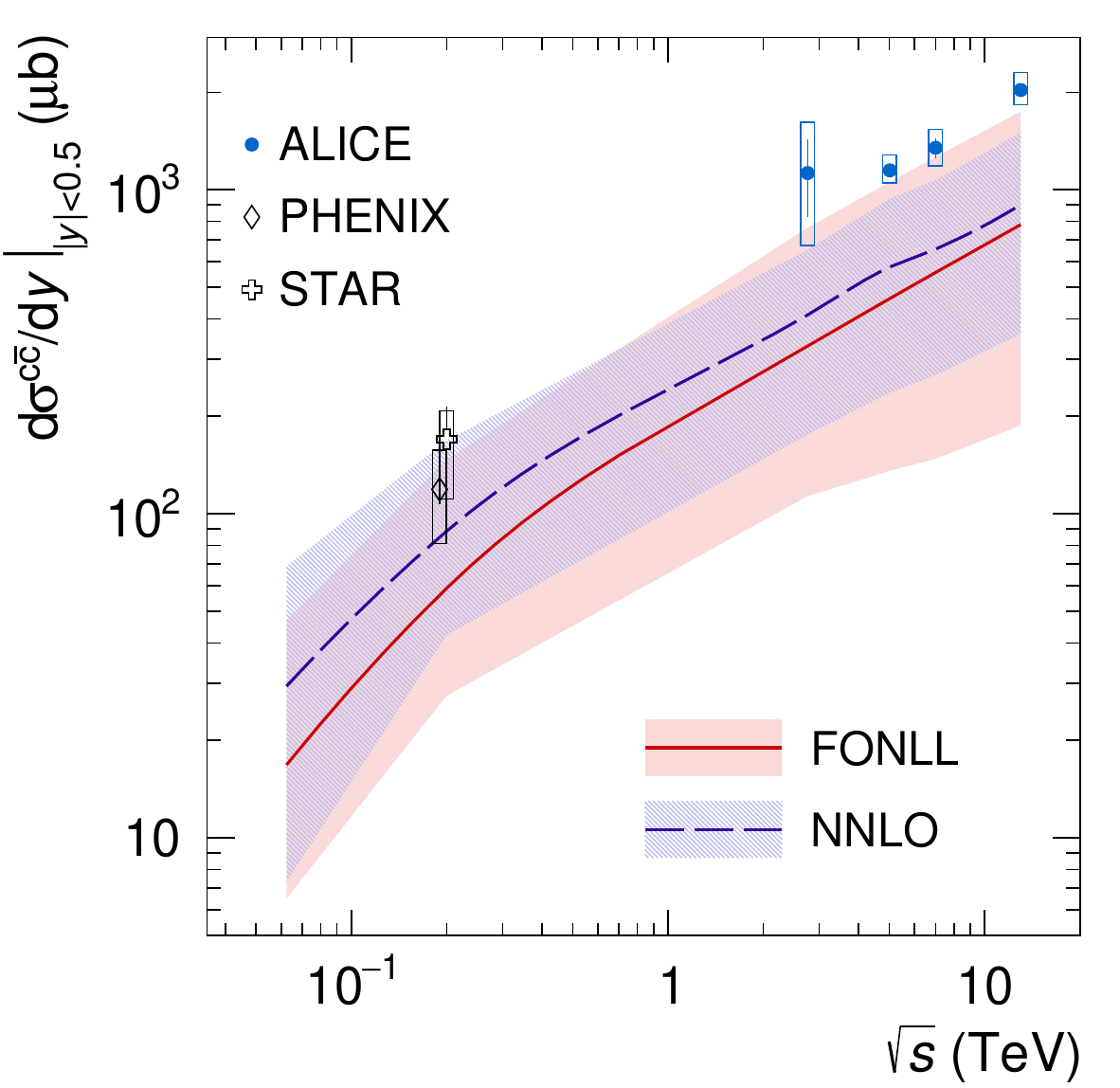} \\
    \caption{Left: charm-quark fragmentation fractions at midrapidity $(|y|<0.5)$ in pp collisions at $\s=5.02$~TeV and $\s=13$~TeV compared with results in $\ee$ and $\ep$ collisions~\cite{Lisovyi:2015uqa}. The fragmentation fractions $f(\rm c\to \hc)$ of $\Jpsi$ mesons are multiplied by a factor 20 for better visibility. Right: $\ccbar$ production cross section per unit of rapidity at midrapidity ($|y|<0.5$) in \pp collisions as a function of $\s$. The measurements are compared with predictions from FONLL~\cite{Cacciari:1998it, Cacciari:2012ny} and NNLO~\cite{dEnterria:2016ids, dEnterria:2016yhy, Czakon:2013goa} calculations. The statistical and systematic uncertainties are reported as vertical bars and boxes, respectively.}
    \label{fig:FF_ccbar}
\end{figure}

The charm-quark fragmentation fractions $f(\charm\to\hc)$ at midrapidity in \pp collisions at $\s=13$~TeV are shown in the left panel of Fig.~\ref{fig:FF_ccbar}. For each hadron species, the production cross section was normalised by the sum of the $\pt$-integrated production cross sections of the measured production cross sections of $\Dzero$, $\Dplus$, $\Ds$, $\Jpsi$, $\Lambdac$, $\XicZero$, and $\XicPlus$. The dashed vertical line separates the fragmentation fractions of the $\Dstar$ mesons and the $\SigmacZeroPlusPlusPlus$ baryons from those of the other charm-hadron species. These two hadrons were not considered in the denominator because they strongly decay into $\Dzero$ and $\Dplus$ mesons and to $\Lambdac$ baryons, respectively, which are already included in the sum.

In this measurement, the systematic uncertainties related to the tracking efficiency and the prompt fraction correction were assumed to be fully correlated among all the particle species, while the uncertainties of the signal extraction and the statistical uncertainty were treated as fully uncorrelated. The extrapolation uncertainty was propagated as partially correlated depending on the adopted techniques for each species. In addition, the possible contribution from $\OmegacZero$-baryon production at midrapidity in \pp collisions at $\s=13$~TeV was taken into account in the systematic uncertainties. According to Ref.~\cite{ALICE:2022cop}, the $\sigma(\OmegacZero)\times\text{ BR}(\OmegacZero\to\Om\pip) / \sigma(\XicZero)$ ratio is around 0.005 in the interval $2< \pt < 12$~GeV$/c$. Scaling the ratio by the theoretical value of the branching ratio $\text{BR}(\OmegacZero\to\Om\pip)=0.51\%^{+2.19\%}_{-0.31\%}$ would imply that the $\OmegacZero$ baryons are produced as abundantly as the $\XicZero$ baryon in this $\pt$ range. However, the branching ratio $\text{BR}(\OmegacZero\to\Om\pip)$ has never been experimentally measured and the one quoted above corresponds to the envelope (uncertainties included) of the values calculated in Refs.~\cite{Hsiao:2020gtc, Gutsche:2018utw, Cheng:1996cs, Hu:2020nkg, Solovieva:2008fw}. Given the large uncertainty of the branching ratio, the $\OmegacZero$-baryon measurement was used only to define an asymmetric systematic uncertainty for the sum of the charm-hadron cross sections used to normalise the fragmentation fraction, which accounts for $\sigma(\OmegacZero) = \sigma(\XicZero)$.

\begin{table}[]
    \renewcommand*{\arraystretch}{1.4}
    \centering
    \caption{Charm-quark fragmentation fractions in \pp collisions at $\s=5.02$~TeV 
      and 13~TeV. The values for the $\XicPlus$ baryon at $\s = 5.02$~TeV 
      are assumed to be the same as the ones of the $\XicZero$ baryon at the same centre-of-mass energy. The values published in Ref.~\cite{ALICE:2021dhb} were updated considering recent measurements of prompt $\Lambdac$ baryon down to $\pt=0$ and of prompt $\Jpsi$ mesons, as mentioned in the text. The \enquote{syst.} uncertainty also includes the contribution of the extrapolation uncertainty.}
    \begin{tabular}{|lll|}
         \hline
         $f(\charm\to \hc)$ & \pp, $\s=5.02$~TeV (\%) & \pp, $\s=13$~TeV (\%) \\
         \hline
         %
         $\Dzero$
         & 39.6 $\pm$ 1.7 (stat.) $^{+2.6}_{-3.8}$ (syst.)
         & 38.2 $\pm$ 1.3 (stat.) $^{+2.3}_{-4.3}$ (syst.) \\
         %
         $\Dplus$
         & 17.5 $\pm$ 1.8 (stat.) $^{+1.7}_{-2.1}$ (syst.)
         & 19.1 $\pm$ 1.4 (stat.) $^{+1.5}_{-2.3}$ (syst.) \\
         %
         $\Ds$
         & 7.4 $\pm$ 1.0 (stat.) $^{+1.9}_{-1.1}$ (syst.)
         & 6.1 $\pm$ 0.5 (stat.) $^{+1.2}_{-0.9}$ (syst.) \\
         %
         $\Lambdac$
         & 18.9 $\pm$ 1.3 (stat.) $^{+1.5}_{-2.0}$ (syst.)
         & 16.8 $\pm$ 0.8 (stat.) $^{+1.5}_{-2.1}$ (syst.) \\
         %
         $\XicZero$
         & 8.1 $\pm$ 1.2 (stat.) $^{+2.5}_{-2.5}$ (syst.)
         & 9.9 $\pm$ 1.3 (stat.) $^{+2.3}_{-2.4}$ (syst.) \\
         %
         $\XicPlus$
         & Assumed to be the same as $\XicZero$
         & 9.6 $\pm$ 1.2 (stat.) $^{+3.9}_{-4.8}$ (syst.) \\
         %
         $\Jpsi$
         & 0.44 $\pm$ 0.03 (stat.) $^{+0.04}_{-0.06}$ (syst.)
         & 0.37 $\pm$ 0.02 (stat.) $^{+0.04}_{-0.05}$ (syst.)\\
         \hline
         %
         $\Dstar$
         & 15.7 $\pm$ 1.2 (stat.) $^{+4.1}_{-1.9}$ (syst.)
         & 15.6 $\pm$ 0.7 (stat.) $^{+2.5}_{-2.2}$ (syst.) \\
         %
         $\SigmacZeroPlusPlusPlus$
         & $-$
         & 7.2 $\pm$ 1.2 (stat.) $^{+1.6}_{-1.9}$ (syst.) \\
         \hline
         
    \end{tabular}
    \label{tab:FF}
\end{table}

The results in \pp collisions at $\s=13$~TeV are compared in the left panel of Fig.~\ref{fig:FF_ccbar} and in Table~\ref{tab:FF} with those in \pp collisions at $\s=5.02$~TeV. The previous measured values published in Ref.~\cite{ALICE:2021dhb} were updated for this paper considering more recent cross section measurements of prompt $\Lambdac$ baryon down to $\pt=0$~\cite{ALICE:2022ych} and of prompt $\Jpsi$ mesons~\cite{ALICE:2021edd}. As reported in Ref.~\cite{ALICE:2022ych}, the $\pt$-integrated $\Lambdac$-baryon cross section in $|y|<0.5$ decreases by about 10\% compared to the previously published results~\cite{ALICE:2020wfu, ALICE:2020wla}, where the measurement did not extend down to $\pt = 0$~GeV$/c$ and instead relied on an extrapolation. This reduction of the $\Lambdac$ production cross section leads to a reduction of the $f(\rm{c}\to\Lambdac)$ by about 7\%.

To compute the $\XicZeroPlus$ fragmentation fractions in pp collisions at $\s=5.02$~TeV, the $\XicZero$-baryon cross section was considered twice, as done in Ref.~\cite{ALICE:2021dhb}. This was due to the lack of $\XicPlus$-baryon measurements at this collision energy.  
The $\XicPlus$-baryon fragmentation fraction at midrapidity ($|y|<0.5$) in \pp collisions at $\s~=~13$~TeV is compatible with the $\XicZero$-baryon fragmentation fractions in \pp collisions at $\s=5.02$~TeV and $\s=13$~TeV within uncertainties. The uncertainties are dominated by the $\sim44\%$ uncertainty of the branching ratio $\text{BR}(\XicPlus \to \X\pip\pip)$. The measurements of the $\Jpsi$ fragmentation fraction at the two different centre-of-mass energies are also shown in the left panel of Fig.~\ref{fig:FF_ccbar}, where they are scaled by a factor 20 for visibility.

The measurements in \pp collisions at the LHC are compared with those in $\ee$ collisions at LEP and at B factories, as well as those in $\ep$ collisions at HERA~\cite{Lisovyi:2015uqa}. The prompt $\Lambdac$-baryon fragmentation fraction in \pp collisions at $\s=13$~TeV is about three times larger than in $\ee$ and $\ep$ collisions. Each of the $\XicZeroPlus$ baryons accounts for about 10\% of the total charm hadron production at midrapidity, while their production was considered to be negligible in $\ee$ and $\ep$ collisions. Since the fragmentation fractions sum to unity, this enhancement of baryon production implies an overall reduction of the relative D-meson abundance by about a factor 1.5 relative to $\ee$ and $\ep$ collisions.

In Table~\ref{tab:FF} the first measurement of the $\SigmacZeroPlusPlusPlus$-baryon fragmentation fraction in \pp collisions at the LHC is also provided. The charm-quark fragmentation fraction into $\SigmacZeroPlusPlusPlus$ baryons in $\ee$ collisions can be estimated to be about 1\%, taking into account that $\SigmacZeroPlusPlusPlus/\Dzero\approx0.02$, as reported in Ref.~\cite{Acharya:2021vpo}, calculated with the charm-hadron cross sections reported in Ref.~\cite{Niiyama:2017wpp}, and the fragmentation fraction $f(\rm{c}\to\Dzero)\approx0.59$ from Ref.~\cite{Lisovyi:2015uqa}.
An enhancement of $f(\rm{c}\to\SigmacZeroPlusPlusPlus)$ of about a factor seven is observed at the LHC compared to e$^{+}$e$^{-}$ collisions. The $\SigmacZeroPlusPlusPlus$ production accounts for about 40\% of the prompt $\Lambdac$-baryon production at midrapidity in \pp collisions at $\s=13$~TeV~\cite{Acharya:2021vpo}. This is significantly larger than the $\SigmacZeroPlusPlusPlus/\Lambdac\approx0.17$ measured in $\ee$ collisions by the Belle Collaboration (Table IV in Ref.~\cite{Niiyama:2017wpp}) and the $\approx0.13$ from PYTHIA 8 Monash tune simulations. Therefore, a larger $\Lambdac$ feed-down from $\SigmacZeroPlusPlusPlus$-baryon decays is observed in \pp collisions at the LHC. 

Within the current precision, the measured fragmentation fractions (left panel of Fig.~\ref{fig:FF_ccbar}) do not show any significant energy dependence of the relative charm-hadron production at midrapidity in \pp collisions at the LHC. Therefore, these results confirm that the baryon enhancement at the LHC with respect to $\ee$ collisions is caused by different hadronisation mechanisms at play in the parton-rich environment produced in \pp collisions, regardless of the centre-of-mass energy.

\subsubsection{Production cross section of $\ccbar$ at midrapidity in $\rm{\bold{pp}}$ collisions}
\label{sec:ccbarVsE}

The $\ccbar$ production cross section at midrapidity (${\rm d}\sigma^{\ccbar}/{\rm d}y|_{|y|<0.5}$) in pp collisions at $\s=13$~TeV is shown in the right panel of Fig.~\ref{fig:FF_ccbar}. As for the measurement of the fragmentation fractions, the $\ccbar$ production cross section is calculated from the sum of the production cross sections at midrapidity (${\rm d}\sigma^{\rm H}/{\rm d}y|_{|y|<0.5}$) of the $\Dzero$, $\Dplus$, $\Ds$, $\Jpsi$, $\Lambdac$, $\XicZero$, and $\XicPlus$ hadrons. The possible contribution of $\OmegacZero$ baryons to the total cross section was taken into account as an asymmetric systematic uncertainty, as discussed in Sec.~\ref{sec:FFVsE} for the fragmentation fractions.

As done for the measurement in \pp collisions at $\s=5.02$~TeV~\cite{ALICE:2021dhb} and at $\s=7$~TeV~\cite{ALICE:2017olh}, two correction factors were applied to account for the different shape of the rapidity distributions of charm hadrons, single charm quarks, and $\ccbar$ pairs. The first factor was evaluated with FONLL calculations in the relevant rapidity range and it accounted for the possible differences in the rapidity distributions of hadrons and single charm quarks. This factor was found to be at unity, excluding any relevant difference in the two rapidity distributions. A 2\% uncertainty for this factor was estimated from the difference to PYTHIA~8 simulations. The second correction factor accounted for the possible differences between the rapidity distributions of single charm quarks and $\ccbar$ pairs. This was evaluated to be 1.036 according to NLO pQCD calculations using POWHEG~\cite{Frixione_2007}. An additional uncertainty of 3\% was estimated by varying independently by a factor of two the factorisation and normalisation scales in the calculations, and using different sets of PDFs~\cite{PhysRevD.82.074024, PhysRevD.93.033006, PhysRevD.103.014013, PhysRevD.78.013004, Forte:2020yip}.

The $\ccbar$ production cross section in \pp collisions at $\s=13$~TeV is
\begin{equation}
    \left.\frac{{\rm d}\sigma}{{\rm d}y}\right|_{|y|<0.5}^{\text{\pp}, \s=13\text{~TeV}} = 2031 \pm 61\text{ (stat.)} ^{+135}_{-141} \text{ (syst.)}
    ^{+196}_{-63} \text{ (extrap.)} \pm 97\text{ (BR)} \pm 33 \text{ (lumi.}) \pm 73 \text{ (\emph{y})}
    \text{ } \rm \upmu b 
    \;.
    \label{eq:ccbarpp13TeV}
\end{equation}

In the computation of the systematic uncertainties, those related to the tracking efficiency and the prompt fraction correction were propagated as fully correlated among the different hadron species. The systematic uncertainties related to the branching ratios of the channels chosen for the reconstruction (BR), the collected luminosity (lumi.), the rapidity correction factors ($y$), and the extrapolation factors (extrap.), which also includes the possible $\OmegacZero$-baryon contribution, are quoted separately.

As for the fragmentation fractions, the $\ccbar$ production cross section at midrapidity ${\rm d}\sigma/{\rm d}y|_{|y|<0.5}$ at $\s=5.02$~TeV was updated taking into account the more recent cross section measurements of prompt $\Lambdac$ baryon down to $\pt=0$~\cite{ALICE:2022ych} and of prompt $\Jpsi$ mesons~\cite{ALICE:2021edd}. The updated $\ccbar$ production cross section in \pp collisions at $\s=5.02$~TeV is
\begin{equation}
    \left.\frac{{\rm d}\sigma}{{\rm d}y}\right|_{|y|<0.5}^{\text{\pp}, \s=5.02\text{~TeV}} = 1148 \pm 43\text{ (stat.)} ^{+62}_{-65} \text{ (syst.)}
    ^{+98}_{-36} \text{ (extrap.)} \pm 43\text{ (BR)} \pm 24 \text{ (lumi.}) \pm 41 \text{ (\emph{y})} 
    \text{ } \rm \upmu b 
    \;.
    \label{eq:ccbarpp5TeV}
\end{equation}

The extrapolated fraction of the $\ccbar$ cross section is around 11\% for both the measurements at $\s=5.02$~TeV and $\s = 13$~TeV. This is significantly reduced with respect to the previous measurement in \pp collisions at $\s=5.02$~TeV~\cite{ALICE:2021dhb}, where it reached about 20\% of the $\ccbar$ cross section due to the smaller $\pt$ reach of baryon measurements at low $\pt$. 
The right panel of Fig.~\ref{fig:FF_ccbar} shows the $\ccbar$ production cross section at midrapidity ($|y|<0.5$) in \pp collisions as a function of centre-of-mass energy $\s$. The blue points represent the measurements from the ALICE Collaboration. The result from this work corresponds to the first measurement of the $\ccbar$ production cross section at midrapidity in \pp collisions at $\s=13$~TeV based on the measured cross sections of several charm-hadron species, while previously only estimates based on dilepton measurements were available~\cite{ALICE:2018gev}. In Fig.~\ref{fig:FF_ccbar} the results in \pp collisions at $\s=200$~GeV by the STAR~\cite{STAR:2012nbd} and PHENIX~\cite{PHENIX:2010xji} Collaborations are also shown. The $\ccbar$ cross section was obtained from the measurement of $\Dzero$- and $\Dstar$-meson abundances and from the production cross section of electrons from heavy-flavour hadron decays, as discussed in Ref.~\cite{ALICE:2021dhb}. In these cases, the production cross sections were scaled by fragmentation fractions obtained from $\ee$ collisions.

The measured $\ccbar$ production cross sections at midrapidity in \pp collisions at different centre-of-mass energies are compared with the FONLL and NNLO predictions. The NNLO values were obtained by the authors of~\cite{dEnterria:2016ids, dEnterria:2016yhy} by applying a $K$ scaling factor (NNLO/NLO) to the central value of ${\rm d}\sigma^{\ccbar}/{\rm d}y$ from FONLL, calculated as discussed in Ref.~\cite{ALICE:2021dhb}. 
The results from RHIC are compatible with the upper edge of the theoretical calculations. The results at the LHC are systematically higher as an effect of the measured baryon enhancement at midrapidity. However, within the current experimental precision the measured ${\rm d}\sigma^{\ccbar}/{\rm d}y|_{|y|<0.5}$ at the LHC are compatible with the FONLL predictions within about $1\sigma$ in \pp collisions at $\s=5.02$~TeV and about $1.4\sigma$ in \pp collisions at $\s=13$~TeV. In the end, calculations based on a factorisation approach describe quantitatively the evolution of the $\ccbar$ cross section at midrapidity in \pp collisions as a function of the centre-of-mass energy. The precise measurements from LHC data shown in Fig.~\ref{fig:FF_ccbar} can provide useful constraints to reduce the uncertainties related either to the factorisation and renormalisation scales in pQCD calculations of the hard-scattering cross section, or to the PDFs.

\section{Summary}
\label{sec:summary}

In this article, the measurements of the production cross sections of prompt $\Dzero$, $\Dplus$, $\Ds$, and $\Dstar$ mesons at midrapidity ($|y|<0.5$) in \pp collisions at $\s=13$~TeV with the ALICE detector are reported. The D-meson production cross sections are described within uncertainties by perturbative QCD calculations based on a factorisation approach over the full $\pt$ ranges of the measurements. The $\pt$-differential ratios of the measured charm-hadron cross sections are compared with the same quantities measured in \pp collisions at $\s=5.02$~TeV and 7 TeV.  Within the experimental uncertainties, no significant energy dependence of the D-meson ratios in \pp collisions at the LHC is observed. These results suggest common fragmentation functions of charm quarks to pseudoscalar and vector mesons and to mesons with and without strange quark content in \pp collisions at different LHC energies.

The production cross sections of prompt $\Lambdac$ and $\XicPlus$ baryons in \pp collisions at $\s=13$ TeV are measured down to $\pt=0$ and $\pt=3$ GeV$/c$, respectively. Similarly to the meson-to-meson ratios at the LHC, the measured $\Lambdac/\Dzero$ ratio in \pp collisions at $\s=5.02$~TeV, 7 TeV, and 13 TeV, and the $\XicZeroPlus/\Dzero$ ratios in \pp collisions at $\s=5.02$ TeV and 13 TeV do not show a significant energy dependence. Within the current uncertainties, the measurements do not discriminate among the model calculations that describe charm hadronisation at the LHC employing new mechanisms. However, the extended measurements of the $\Lambdac$ and $\XicPlus$ production confirm the baryon-to-meson enhancement at the LHC compared to $\ee$ results down to lower $\pt$. These results support the scenario of charm-quark hadronisation in \pp collisions at the LHC via mechanisms other than those in $\ee$ collisions.

The $\pt$-differential cross sections of prompt D mesons measured at midrapidity ($|y|<0.5$) are compared with those measured at forward rapidity by the LHCb Collaboration in three intervals $2<y<2.5$, $3<y<3.5$, and $4<y<4.5$ at the same collision energy. The mid-over-forward rapidity ratios show an increasing trend with $\pt$ when going to more forward rapidities and a common trend is shared by all the D-meson species. This behaviour can be explained as a softening of the $\pt$ spectra at forward rapidity, which is described by pQCD-based calculations (FONLL). The measurement of the $\Dzero$-meson mid-to-forward rapidity ratios between \pp collisions at $\s=5.02$ TeV and 13 TeV is provided and compared with FONLL predictions employing the CTEQ6.6 and NNPDF30 sets of PDFs. The perturbative QCD calculations reproduce the magnitude and the  $\pt$ dependence of the ratios. The measurements are precise enough to constrain the gluon PDFs employed in the NNPDF30 set at low $\pt$, given that the uncertainties related to the scale variations in the pQCD calculations are found to be subdominant. A similar comparison would be interesting considering predictions from pQCD-based calculations employing more recent PDF sets.

The D-meson strangeness suppression factor $\gamma_{\rm s}$  in \pp collisions at $\s=13$~TeV is compared with previous measurements from the ALICE, ATLAS, H1, and ZEUS Collaborations. The results are compatible within uncertainties and they agree with the average of LEP measurements.
These results indicate that the production of prompt strange D mesons relative to that of prompt non-strange D mesons in $\ee$, $\ep$ and \pp collisions does not show any significant dependence on the collision system and energy.

The measurement of charm-quark fragmentation fractions in \pp collisions at $\s=13$~TeV is provided. In addition to the charm-quark fragmentation fractions into $\Dzero$, $\Dplus$, $\Ds$, $\Dstar$, $\Lambdac$, and $\XicZero$ hadrons, also the results for the fragmentation into $\Jpsi$ mesons and $\XicPlus$ and $\SigmacZeroPlusPlusPlus$ baryons are reported. The results are compared with those in \pp collisions at $\s=5.02$~TeV and no significant energy dependence is observed according to the current uncertainties. These results confirm that the baryon enhancement at the LHC with respect to $\ee$ collisions is caused by different hadronisation mechanisms at play in the parton-rich environment produced in \pp collisions. Finally, the first measurement in \pp collisions at $\s=13$~TeV of the $\ccbar$ production cross section at midrapidity ($|y|<0.5$) based on the sum of the production cross sections at midrapidity of $\Dzero$, $\Dplus$, $\Ds$, $\Jpsi$, $\Lambdac$, $\XicZero$, and $\XicPlus$ hadrons is reported. This measurement, whose maximum relative uncertainty is about 13\%, is found to be compatible with the upper edge of the FONLL and NNLO predictions within uncertainties. Thanks to their better precision, the measurements at the LHC can provide useful constraints to reduce the theoretical uncertainties on the calculations of $\ccbar$ production at midrapidity in \pp collisions.


\newenvironment{acknowledgement}{\relax}{\relax}
\begin{acknowledgement}
\section*{Acknowledgements}

We gratefully acknowledge Professor Matteo Cacciari for the precious contribution made to validate the FONLL calculations compared with the measurements reported in this manuscript.


The ALICE Collaboration would like to thank all its engineers and technicians for their invaluable contributions to the construction of the experiment and the CERN accelerator teams for the outstanding performance of the LHC complex.
The ALICE Collaboration gratefully acknowledges the resources and support provided by all Grid centres and the Worldwide LHC Computing Grid (WLCG) collaboration.
The ALICE Collaboration acknowledges the following funding agencies for their support in building and running the ALICE detector:
A. I. Alikhanyan National Science Laboratory (Yerevan Physics Institute) Foundation (ANSL), State Committee of Science and World Federation of Scientists (WFS), Armenia;
Austrian Academy of Sciences, Austrian Science Fund (FWF): [M 2467-N36] and Nationalstiftung f\"{u}r Forschung, Technologie und Entwicklung, Austria;
Ministry of Communications and High Technologies, National Nuclear Research Center, Azerbaijan;
Conselho Nacional de Desenvolvimento Cient\'{\i}fico e Tecnol\'{o}gico (CNPq), Financiadora de Estudos e Projetos (Finep), Funda\c{c}\~{a}o de Amparo \`{a} Pesquisa do Estado de S\~{a}o Paulo (FAPESP) and Universidade Federal do Rio Grande do Sul (UFRGS), Brazil;
Bulgarian Ministry of Education and Science, within the National Roadmap for Research Infrastructures 2020-2027 (object CERN), Bulgaria;
Ministry of Education of China (MOEC) , Ministry of Science \& Technology of China (MSTC) and National Natural Science Foundation of China (NSFC), China;
Ministry of Science and Education and Croatian Science Foundation, Croatia;
Centro de Aplicaciones Tecnol\'{o}gicas y Desarrollo Nuclear (CEADEN), Cubaenerg\'{\i}a, Cuba;
Ministry of Education, Youth and Sports of the Czech Republic, Czech Republic;
The Danish Council for Independent Research | Natural Sciences, the VILLUM FONDEN and Danish National Research Foundation (DNRF), Denmark;
Helsinki Institute of Physics (HIP), Finland;
Commissariat \`{a} l'Energie Atomique (CEA) and Institut National de Physique Nucl\'{e}aire et de Physique des Particules (IN2P3) and Centre National de la Recherche Scientifique (CNRS), France;
Bundesministerium f\"{u}r Bildung und Forschung (BMBF) and GSI Helmholtzzentrum f\"{u}r Schwerionenforschung GmbH, Germany;
General Secretariat for Research and Technology, Ministry of Education, Research and Religions, Greece;
National Research, Development and Innovation Office, Hungary;
Department of Atomic Energy Government of India (DAE), Department of Science and Technology, Government of India (DST), University Grants Commission, Government of India (UGC) and Council of Scientific and Industrial Research (CSIR), India;
National Research and Innovation Agency - BRIN, Indonesia;
Istituto Nazionale di Fisica Nucleare (INFN), Italy;
Japanese Ministry of Education, Culture, Sports, Science and Technology (MEXT) and Japan Society for the Promotion of Science (JSPS) KAKENHI, Japan;
Consejo Nacional de Ciencia (CONACYT) y Tecnolog\'{i}a, through Fondo de Cooperaci\'{o}n Internacional en Ciencia y Tecnolog\'{i}a (FONCICYT) and Direcci\'{o}n General de Asuntos del Personal Academico (DGAPA), Mexico;
Nederlandse Organisatie voor Wetenschappelijk Onderzoek (NWO), Netherlands;
The Research Council of Norway, Norway;
Commission on Science and Technology for Sustainable Development in the South (COMSATS), Pakistan;
Pontificia Universidad Cat\'{o}lica del Per\'{u}, Peru;
Ministry of Education and Science, National Science Centre and WUT ID-UB, Poland;
Korea Institute of Science and Technology Information and National Research Foundation of Korea (NRF), Republic of Korea;
Ministry of Education and Scientific Research, Institute of Atomic Physics, Ministry of Research and Innovation and Institute of Atomic Physics and University Politehnica of Bucharest, Romania;
Ministry of Education, Science, Research and Sport of the Slovak Republic, Slovakia;
National Research Foundation of South Africa, South Africa;
Swedish Research Council (VR) and Knut \& Alice Wallenberg Foundation (KAW), Sweden;
European Organization for Nuclear Research, Switzerland;
Suranaree University of Technology (SUT), National Science and Technology Development Agency (NSTDA), Thailand Science Research and Innovation (TSRI) and National Science, Research and Innovation Fund (NSRF), Thailand;
Turkish Energy, Nuclear and Mineral Research Agency (TENMAK), Turkey;
National Academy of  Sciences of Ukraine, Ukraine;
Science and Technology Facilities Council (STFC), United Kingdom;
National Science Foundation of the United States of America (NSF) and United States Department of Energy, Office of Nuclear Physics (DOE NP), United States of America.
In addition, individual groups or members have received support from:
European Research Council, Strong 2020 - Horizon 2020 (grant nos. 950692, 824093), European Union;
Academy of Finland (Center of Excellence in Quark Matter) (grant nos. 346327, 346328), Finland;

\end{acknowledgement}

\bibliographystyle{utphys}   
\bibliography{bibliography}

\newpage

\appendix

%
%

\section{The ALICE Collaboration}
\label{app:collab}
\begin{flushleft} 
\small

S.~Acharya\,\orcidlink{0000-0002-9213-5329}\,$^{\rm 126}$, 
D.~Adamov\'{a}\,\orcidlink{0000-0002-0504-7428}\,$^{\rm 86}$, 
G.~Aglieri Rinella\,\orcidlink{0000-0002-9611-3696}\,$^{\rm 33}$, 
M.~Agnello\,\orcidlink{0000-0002-0760-5075}\,$^{\rm 30}$, 
N.~Agrawal\,\orcidlink{0000-0003-0348-9836}\,$^{\rm 51}$, 
Z.~Ahammed\,\orcidlink{0000-0001-5241-7412}\,$^{\rm 134}$, 
S.~Ahmad\,\orcidlink{0000-0003-0497-5705}\,$^{\rm 16}$, 
S.U.~Ahn\,\orcidlink{0000-0001-8847-489X}\,$^{\rm 71}$, 
I.~Ahuja\,\orcidlink{0000-0002-4417-1392}\,$^{\rm 38}$, 
A.~Akindinov\,\orcidlink{0000-0002-7388-3022}\,$^{\rm 142}$
% ...

\section*{Affiliation Notes}

$^{\rm I}$ Also at: Max-Planck-Institut f\"{u}r Physik, Munich, Germany\\
$^{\rm II}$ Also at: Italian National Agency for New Technologies, Energy and Sustainable Economic Development (ENEA), Bologna, Italy\\
$^{\rm III}$ Also at: Dipartimento DET del Politecnico di Torino, Turin, Italy\\
$^{\rm IV}$ Also at: Department of Applied Physics, Aligarh Muslim University, Aligarh, India\\
$^{\rm V}$ Also at: Institute of Theoretical Physics, University of Wroclaw, Poland\\
$^{\rm VI}$ Also at: An institution covered by a cooperation agreement with CERN\\

\section*{Collaboration Institutes}

$^{1}$ A.I. Alikhanyan National Science Laboratory (Yerevan Physics Institute) Foundation, Yerevan, Armenia\\
$^{2}$ AGH University of Krakow, Cracow, Poland\\
$^{3}$ Bogolyubov Institute for Theoretical Physics, National Academy of Sciences of Ukraine, Kiev, Ukraine\\
$^{4}$ Bose Institute, Department of Physics  and Centre for Astroparticle Physics and Space Science (CAPSS), Kolkata, India\\
$^{5}$ California Polytechnic State University, San Luis Obispo, California, United States\\
$^{6}$ Central China Normal University, Wuhan, China\\
$^{7}$ Centro de Aplicaciones Tecnol\'{o}gicas y Desarrollo Nuclear (CEADEN), Havana, Cuba\\
$^{8}$ Centro de Investigaci\'{o}n y de Estudios Avanzados (CINVESTAV), Mexico City and M\'{e}rida, Mexico\\
$^{9}$ Chicago State University, Chicago, Illinois, United States\\
$^{10}$ China Institute of Atomic Energy, Beijing, China\\
$^{11}$ China University of Geosciences, Wuhan, China\\
$^{12}$ Chungbuk National University, Cheongju, Republic of Korea\\
$^{13}$ Comenius University Bratislava, Faculty of Mathematics, Physics and Informatics, Bratislava, Slovak Republic\\
$^{14}$ COMSATS University Islamabad, Islamabad, Pakistan\\
$^{15}$ Creighton University, Omaha, Nebraska, United States\\
$^{16}$ Department of Physics, Aligarh Muslim University, Aligarh, India\\
$^{17}$ Department of Physics, Pusan National University, Pusan, Republic of Korea\\
$^{18}$ Department of Physics, Sejong University, Seoul, Republic of Korea\\
$^{19}$ Department of Physics, University of California, Berkeley, California, United States\\
$^{20}$ Department of Physics, University of Oslo, Oslo, Norway\\
$^{21}$ Department of Physics and Technology, University of Bergen, Bergen, Norway\\
$^{22}$ Dipartimento di Fisica, Universit\`{a} di Pavia, Pavia, Italy\\
$^{23}$ Dipartimento di Fisica dell'Universit\`{a} and Sezione INFN, Cagliari, Italy\\
$^{24}$ Dipartimento di Fisica dell'Universit\`{a} and Sezione INFN, Trieste, Italy\\
$^{25}$ Dipartimento di Fisica dell'Universit\`{a} and Sezione INFN, Turin, Italy\\
$^{26}$ Dipartimento di Fisica e Astronomia dell'Universit\`{a} and Sezione INFN, Bologna, Italy\\
$^{27}$ Dipartimento di Fisica e Astronomia dell'Universit\`{a} and Sezione INFN, Catania, Italy\\
$^{28}$ Dipartimento di Fisica e Astronomia dell'Universit\`{a} and Sezione INFN, Padova, Italy\\
$^{29}$ Dipartimento di Fisica `E.R.~Caianiello' dell'Universit\`{a} and Gruppo Collegato INFN, Salerno, Italy\\
$^{30}$ Dipartimento DISAT del Politecnico and Sezione INFN, Turin, Italy\\
$^{31}$ Dipartimento di Scienze MIFT, Universit\`{a} di Messina, Messina, Italy\\
$^{32}$ Dipartimento Interateneo di Fisica `M.~Merlin' and Sezione INFN, Bari, Italy\\
$^{33}$ European Organization for Nuclear Research (CERN), Geneva, Switzerland\\
$^{34}$ Faculty of Electrical Engineering, Mechanical Engineering and Naval Architecture, University of Split, Split, Croatia\\
$^{35}$ Faculty of Engineering and Science, Western Norway University of Applied Sciences, Bergen, Norway\\
$^{36}$ Faculty of Nuclear Sciences and Physical Engineering, Czech Technical University in Prague, Prague, Czech Republic\\
$^{37}$ Faculty of Physics, Sofia University, Sofia, Bulgaria\\
$^{38}$ Faculty of Science, P.J.~\v{S}af\'{a}rik University, Ko\v{s}ice, Slovak Republic\\
$^{39}$ Frankfurt Institute for Advanced Studies, Johann Wolfgang Goethe-Universit\"{a}t Frankfurt, Frankfurt, Germany\\
$^{40}$ Fudan University, Shanghai, China\\
$^{41}$ Gangneung-Wonju National University, Gangneung, Republic of Korea\\
$^{42}$ Gauhati University, Department of Physics, Guwahati, India\\
$^{43}$ Helmholtz-Institut f\"{u}r Strahlen- und Kernphysik, Rheinische Friedrich-Wilhelms-Universit\"{a}t Bonn, Bonn, Germany\\
$^{44}$ Helsinki Institute of Physics (HIP), Helsinki, Finland\\
$^{45}$ High Energy Physics Group,  Universidad Aut\'{o}noma de Puebla, Puebla, Mexico\\
$^{46}$ Horia Hulubei National Institute of Physics and Nuclear Engineering, Bucharest, Romania\\
$^{47}$ Indian Institute of Technology Bombay (IIT), Mumbai, India\\
$^{48}$ Indian Institute of Technology Indore, Indore, India\\
$^{49}$ INFN, Laboratori Nazionali di Frascati, Frascati, Italy\\
$^{50}$ INFN, Sezione di Bari, Bari, Italy\\
$^{51}$ INFN, Sezione di Bologna, Bologna, Italy\\
$^{52}$ INFN, Sezione di Cagliari, Cagliari, Italy\\
$^{53}$ INFN, Sezione di Catania, Catania, Italy\\
$^{54}$ INFN, Sezione di Padova, Padova, Italy\\
$^{55}$ INFN, Sezione di Pavia, Pavia, Italy\\
$^{56}$ INFN, Sezione di Torino, Turin, Italy\\
$^{57}$ INFN, Sezione di Trieste, Trieste, Italy\\
$^{58}$ Inha University, Incheon, Republic of Korea\\
$^{59}$ Institute for Gravitational and Subatomic Physics (GRASP), Utrecht University/Nikhef, Utrecht, Netherlands\\
$^{60}$ Institute of Experimental Physics, Slovak Academy of Sciences, Ko\v{s}ice, Slovak Republic\\
$^{61}$ Institute of Physics, Homi Bhabha National Institute, Bhubaneswar, India\\
$^{62}$ Institute of Physics of the Czech Academy of Sciences, Prague, Czech Republic\\
$^{63}$ Institute of Space Science (ISS), Bucharest, Romania\\
$^{64}$ Institut f\"{u}r Kernphysik, Johann Wolfgang Goethe-Universit\"{a}t Frankfurt, Frankfurt, Germany\\
$^{65}$ Instituto de Ciencias Nucleares, Universidad Nacional Aut\'{o}noma de M\'{e}xico, Mexico City, Mexico\\
$^{66}$ Instituto de F\'{i}sica, Universidade Federal do Rio Grande do Sul (UFRGS), Porto Alegre, Brazil\\
$^{67}$ Instituto de F\'{\i}sica, Universidad Nacional Aut\'{o}noma de M\'{e}xico, Mexico City, Mexico\\
$^{68}$ iThemba LABS, National Research Foundation, Somerset West, South Africa\\
$^{69}$ Jeonbuk National University, Jeonju, Republic of Korea\\
$^{70}$ Johann-Wolfgang-Goethe Universit\"{a}t Frankfurt Institut f\"{u}r Informatik, Fachbereich Informatik und Mathematik, Frankfurt, Germany\\
$^{71}$ Korea Institute of Science and Technology Information, Daejeon, Republic of Korea\\
$^{72}$ KTO Karatay University, Konya, Turkey\\
$^{73}$ Laboratoire de Physique Subatomique et de Cosmologie, Universit\'{e} Grenoble-Alpes, CNRS-IN2P3, Grenoble, France\\
$^{74}$ Lawrence Berkeley National Laboratory, Berkeley, California, United States\\
$^{75}$ Lund University Department of Physics, Division of Particle Physics, Lund, Sweden\\
$^{76}$ Nagasaki Institute of Applied Science, Nagasaki, Japan\\
$^{77}$ Nara Women{'}s University (NWU), Nara, Japan\\
$^{78}$ National and Kapodistrian University of Athens, School of Science, Department of Physics , Athens, Greece\\
$^{79}$ National Centre for Nuclear Research, Warsaw, Poland\\
$^{80}$ National Institute of Science Education and Research, Homi Bhabha National Institute, Jatni, India\\
$^{81}$ National Nuclear Research Center, Baku, Azerbaijan\\
$^{82}$ National Research and Innovation Agency - BRIN, Jakarta, Indonesia\\
$^{83}$ Niels Bohr Institute, University of Copenhagen, Copenhagen, Denmark\\
$^{84}$ Nikhef, National institute for subatomic physics, Amsterdam, Netherlands\\
$^{85}$ Nuclear Physics Group, STFC Daresbury Laboratory, Daresbury, United Kingdom\\
$^{86}$ Nuclear Physics Institute of the Czech Academy of Sciences, Husinec-\v{R}e\v{z}, Czech Republic\\
$^{87}$ Oak Ridge National Laboratory, Oak Ridge, Tennessee, United States\\
$^{88}$ Ohio State University, Columbus, Ohio, United States\\
$^{89}$ Physics department, Faculty of science, University of Zagreb, Zagreb, Croatia\\
$^{90}$ Physics Department, Panjab University, Chandigarh, India\\
$^{91}$ Physics Department, University of Jammu, Jammu, India\\
$^{92}$ Physics Program and International Institute for Sustainability with Knotted Chiral Meta Matter (SKCM2), Hiroshima University, Hiroshima, Japan\\
$^{93}$ Physikalisches Institut, Eberhard-Karls-Universit\"{a}t T\"{u}bingen, T\"{u}bingen, Germany\\
$^{94}$ Physikalisches Institut, Ruprecht-Karls-Universit\"{a}t Heidelberg, Heidelberg, Germany\\
$^{95}$ Physik Department, Technische Universit\"{a}t M\"{u}nchen, Munich, Germany\\
$^{96}$ Politecnico di Bari and Sezione INFN, Bari, Italy\\
$^{97}$ Research Division and ExtreMe Matter Institute EMMI, GSI Helmholtzzentrum f\"ur Schwerionenforschung GmbH, Darmstadt, Germany\\
$^{98}$ Saga University, Saga, Japan\\
$^{99}$ Saha Institute of Nuclear Physics, Homi Bhabha National Institute, Kolkata, India\\
$^{100}$ School of Physics and Astronomy, University of Birmingham, Birmingham, United Kingdom\\
$^{101}$ Secci\'{o}n F\'{\i}sica, Departamento de Ciencias, Pontificia Universidad Cat\'{o}lica del Per\'{u}, Lima, Peru\\
$^{102}$ Stefan Meyer Institut f\"{u}r Subatomare Physik (SMI), Vienna, Austria\\
$^{103}$ SUBATECH, IMT Atlantique, Nantes Universit\'{e}, CNRS-IN2P3, Nantes, France\\
$^{104}$ Sungkyunkwan University, Suwon City, Republic of Korea\\
$^{105}$ Suranaree University of Technology, Nakhon Ratchasima, Thailand\\
$^{106}$ Technical University of Ko\v{s}ice, Ko\v{s}ice, Slovak Republic\\
$^{107}$ The Henryk Niewodniczanski Institute of Nuclear Physics, Polish Academy of Sciences, Cracow, Poland\\
$^{108}$ The University of Texas at Austin, Austin, Texas, United States\\
$^{109}$ Universidad Aut\'{o}noma de Sinaloa, Culiac\'{a}n, Mexico\\
$^{110}$ Universidade de S\~{a}o Paulo (USP), S\~{a}o Paulo, Brazil\\
$^{111}$ Universidade Estadual de Campinas (UNICAMP), Campinas, Brazil\\
$^{112}$ Universidade Federal do ABC, Santo Andre, Brazil\\
$^{113}$ University of Cape Town, Cape Town, South Africa\\
$^{114}$ University of Derby, Derby, United Kingdom\\
$^{115}$ University of Houston, Houston, Texas, United States\\
$^{116}$ University of Jyv\"{a}skyl\"{a}, Jyv\"{a}skyl\"{a}, Finland\\
$^{117}$ University of Kansas, Lawrence, Kansas, United States\\
$^{118}$ University of Liverpool, Liverpool, United Kingdom\\
$^{119}$ University of Science and Technology of China, Hefei, China\\
$^{120}$ University of South-Eastern Norway, Kongsberg, Norway\\
$^{121}$ University of Tennessee, Knoxville, Tennessee, United States\\
$^{122}$ University of the Witwatersrand, Johannesburg, South Africa\\
$^{123}$ University of Tokyo, Tokyo, Japan\\
$^{124}$ University of Tsukuba, Tsukuba, Japan\\
$^{125}$ University Politehnica of Bucharest, Bucharest, Romania\\
$^{126}$ Universit\'{e} Clermont Auvergne, CNRS/IN2P3, LPC, Clermont-Ferrand, France\\
$^{127}$ Universit\'{e} de Lyon, CNRS/IN2P3, Institut de Physique des 2 Infinis de Lyon, Lyon, France\\
$^{128}$ Universit\'{e} de Strasbourg, CNRS, IPHC UMR 7178, F-67000 Strasbourg, France, Strasbourg, France\\
$^{129}$ Universit\'{e} Paris-Saclay, Centre d'Etudes de Saclay (CEA), IRFU, D\'{e}partment de Physique Nucl\'{e}aire (DPhN), Saclay, France\\
$^{130}$ Universit\'{e}  Paris-Saclay, CNRS/IN2P3, IJCLab, Orsay, France\\
$^{131}$ Universit\`{a} degli Studi di Foggia, Foggia, Italy\\
$^{132}$ Universit\`{a} del Piemonte Orientale, Vercelli, Italy\\
$^{133}$ Universit\`{a} di Brescia, Brescia, Italy\\
$^{134}$ Variable Energy Cyclotron Centre, Homi Bhabha National Institute, Kolkata, India\\
$^{135}$ Warsaw University of Technology, Warsaw, Poland\\
$^{136}$ Wayne State University, Detroit, Michigan, United States\\
$^{137}$ Westf\"{a}lische Wilhelms-Universit\"{a}t M\"{u}nster, Institut f\"{u}r Kernphysik, M\"{u}nster, Germany\\
$^{138}$ Wigner Research Centre for Physics, Budapest, Hungary\\
$^{139}$ Yale University, New Haven, Connecticut, United States\\
$^{140}$ Yonsei University, Seoul, Republic of Korea\\
$^{141}$  Zentrum  f\"{u}r Technologie und Transfer (ZTT), Worms, Germany\\
$^{142}$ Affiliated with an institute covered by a cooperation agreement with CERN\\
$^{143}$ Affiliated with an international laboratory covered by a cooperation agreement with CERN.\\

\end{flushleft} 
  
\end{document}